\documentclass[12pt]{article}
\usepackage{amsmath}
\usepackage{graphicx,psfrag,epsf}
\usepackage{enumerate}
\usepackage{natbib}
\usepackage{url} 

\usepackage{natbib,bm}
\usepackage{amsmath,amssymb}
\usepackage{graphicx}
\usepackage{epsfig}
\usepackage{epstopdf}
\usepackage{rotating}
\usepackage{multirow}
\usepackage{multicol}
\usepackage[small]{caption}
\usepackage{subfig}
\usepackage{color}
\usepackage{mathtools}
\usepackage{enumerate}
\usepackage{pdflscape}
\usepackage{longtable}
\usepackage{pifont}

\usepackage{setspace}

\newcommand{\nmathbf}{\bm}

\def\bfV{\nmathbf V}

\def\bfX{\nmathbf X}

\def\bfs{\nmathbf s}

\def\bfbeta   {\nmathbf \beta}

\def\bftheta  {\nmathbf \theta}

\def\bflambda {\nmathbf \lambda}
\def\bfmu     {\nmathbf \mu}

\def\bfSigma  {\nmathbf \Sigma}

\newcommand{\bfzero}{{\nmathbf 0}}
\newcommand{\bfone}{{\nmathbf 1}}

\def\boldfacefake#1{\kern-4pt
   \hbox{ \mathsurround=0pt
   \hbox to 0.4pt{$#1$\hss}\hbox to 0.4pt{$#1$\hss}\hbox {$#1$}}}

\newcommand{\btable}{\begin{table}[h]\centering}
\newcommand{\etable}{\end{table}}
\newcommand{\bt}{\begin{parag}\small \let\b=\nsb \let\sb=\nssb \begin{tabular}}
\newcommand{\et}{\end{tabular}\let\b=\nb \let\sb=\nsb\end{parag}}

\newenvironment{parag}{\par}{\par}

\newcommand{\be}{\begin{eqnarray}}
\newcommand{\ee}{\end{eqnarray}}
\newcommand{\ba}{\begin{eqnarray*}}
\newcommand{\ea}{\end{eqnarray*}}

\newcommand{\reals}{\mbox{\rm I\kern-.20em R}}
\newcommand{\sreals}{\mbox{\small \rm I\kern-.20em R}}

\newcommand{\mc}[1]{\multicolumn{1}{c}{#1}}

\newcommand{\blind}{0}

\begin{document}

\def\spacingset#1{\renewcommand{\baselinestretch}%
{#1}\small\normalsize} \spacingset{1}


\if0\blind
{
  \title{\bf Hierarchical models for semi-competing risks data with application to quality of end-of-life care for pancreatic cancer}
  \author{Kyu Ha Lee\thanks{
    This work was supported by National Cancer Institute grant (P01 CA134294-02) and National Institutes of Health grants (ES012044, K18 HS021991, R01 CA181360-01). }\hspace{.2cm}\\
    Epidemiology and Biostatistics Core, The Forsyth Institute\\
    and \\
    Francesca Dominici \\
    Department of Biostatistics, Harvard T.H. Chan School of Public Health\\
    and \\
    Deborah Schrag \\
    Department of Medical Oncology, Dana Farber Cancer Institute \\
    and \\
    Sebastien Haneuse \\
    Department of Biostatistics, Harvard T.H. Chan School of Public Health}
  \maketitle
} \fi

\if1\blind
{
  \bigskip
  \bigskip
  \bigskip
  \begin{center}
    {\LARGE\bf Title}
\end{center}
  \medskip
} \fi

\bigskip
\begin{abstract}
Readmission following discharge from an initial hospitalization is a key marker of quality of health care in the United States. For the most part, readmission has been studied among patients with `acute' health conditions, such as pneumonia and heart failure, with analyses based on a logistic-Normal generalized linear mixed model~\citep{normand1997statistical}. Na\"{i}ve application of this model to the study of readmission among patients with `advanced' health conditions such as pancreatic cancer, however, is problematic because it ignores death as a competing risk. A more appropriate analysis is to imbed such a study within the semi-competing risks framework. To our knowledge, however, no comprehensive statistical methods have been developed for cluster-correlated semi-competing risks data. To resolve this gap in the literature we propose a novel hierarchical modeling framework for the analysis of cluster-correlated semi-competing risks data that permits parametric or non-parametric specifications for a range of components giving analysts substantial flexibility as they consider their own analyses. Estimation and inference is performed within the Bayesian paradigm since it facilitates the straightforward characterization of (posterior) uncertainty for all model parameters, including hospital-specific random effects. Model comparison and choice is performed via the deviance information criterion and the log-pseudo marginal likelihood statistic, both of which are based on a partially marginalized likelihood. An efficient computational scheme, based on the Metropolis-Hastings-Green algorithm, is developed and had been implemented in the \texttt{SemiCompRisks} \texttt{R} package. A comprehensive simulation study shows that the proposed framework performs very well in a range of data scenarios, and outperforms competitor analysis strategies. The proposed framework is motivated by and illustrated with an on-going study of the risk of readmission among Medicare beneficiaries diagnosed with pancreatic cancer. Using data on n=5,298 patients at $J$=112 hospitals in the six New England states between 2000-2009, key scientific questions we consider include the role of patient-level risk factors on the risk of readmission and the extent of variation in risk across hospitals not explained by differences in patient case-mix.
\end{abstract}

\vfill

\newpage
\spacingset{1.45} 
\section{Introduction} \label{sec:intro}

Cancer of the pancreas is one of the most deadly. In 2013, an estimated 38,460 individuals died from pancreatic cancer in the United States making it the fourth most prevalent cause of cancer death~\citep{acs:2013}. Unfortunately, since there are no effective screening tests for pancreatic cancer, most patients are diagnosed at a late stage of the disease, specifically once it has metastasized to other parts of the body. As a result, survival is poor with 1-year and 5-year mortality rates are of 75\% and 94\%, respectively~\citep{shin2012pancreatic}. In practice, since prognosis is poor and mortality rates high, the treatment and management of patients diagnosed with pancreatic cancer generally focuses on palliative care aimed at enhancing quality of end-of-life care~\citep{plos:2012}. Such care is expensive, however, with patients diagnosed with pancreatic cancer accruing an estimated \$165,000 in health care costs in their last year of life~\citep{mariotto2011projections}.

Despite the huge costs, there are currently no comprehensive national efforts to monitor quality of end-of-life care for pancreatic cancer nor for any of a broad range of other `advanced' health conditions for which the management of disease focuses on palliative care. Outside the context of these conditions, however, there is substantial interest in understanding variation in quality of health care. The recent literature, in particular, has focused on readmission as a key marker of quality of care, in part because it is expensive but also because it is thought of as an often-preventable event~\citep{joshua5determinants, warren2011end, brooks2014acute, stitzenberg2014exploring}. In addition, as the nation's largest payer of health care costs in the United States, the Centers for Medicare and Medicaid Services (CMS) uses hospital-specific readmission rates as a central component in two programs: the Hospital Inpatient Quality Reporting Program, which requires hospitals to annually  report, among other measures, readmission rates for pneumonia, heart failure and myocardial infarction in order to receive a full update to their reimbursement payments~\citep{2013hiqr}; and, the Readmission Reduction Program, which requires CMS to reduce payments to hospitals with excess readmissions~\citep{2013readmRed}.

Across all of these efforts, investigations of readmission in the literature have invariably used a logistic-Normal generalized linear mixed model (LN-GLMM) to analyze patients clustered within hospitals~\citep{normand1997statistical, ash2012statistical}. While reasonable for health conditions with effective treatment options and low mortality, direct application of this model to investigate variation in risk of readmission following a diagnosis of pancreatic cancer is inappropriate because of the strong force of mortality. Consider, for example, $n$=5,298 Medicare beneficiaries diagnosed with pancreatic cancer at $J$=112 hospitals in six New England states between 2000-2009 and suppose interest lies in understanding determinants of readmission 90 days post-discharge. While additional detail is given below, we note at the outset that 1,257 patients (24\%) died within 30 days of discharge without experiencing a readmission event; furthermore, 1,912 patients (36\%) died within 90 days of discharge without experiencing a readmission event. Na\"{i}ve application of a standard LN-GLMM to these data ignores the fact that a substantial portion of the patients are not at risk to experience the event of `readmission by 90 days' for much of the timeframe. Such an analysis may lead to bias and, if incorporated into existing CMS programs, could have a major impact on how hospitals are penalized for poor quality of care.

In the statistics literature, data that arise from studies in which primary scientific interest lies with some \textit{non-terminal} event (e.g. readmission) whose observation is subject to a \textit{terminal} event (e.g. death) are referred to as \textit{semi-competing risks data}~\citep{Fine:Jian:2001}. Broadly, published methods for the analysis of semi-competing risks data can be classified into three groups: methods that specify dependence between the non-terminal and terminal events via a copula~\citep{Fine:Jian:2001, peng2007regression, Hsie:Wang:2008}; methods based on multi-state models that induce dependence via a shared patient-specific frailty~\citep{Knei:Henn:2008,Xu:Kalb:2010,zeng2012estimating,han2014bayesian,zhang2014bayesian,Lee2014bscr}; and, methods based on principal stratification~\citep{zhang2003estimation, egleston2007causal, tchetgen2014identification}. Common to all of these methods, however, is that their development has focused exclusively on settings where individual study units are independent. As such, the methods are not design to address scientific questions that arise naturally in the context of cluster-correlated data~\citep{diggle2002analysis, fitzmaurice2012applied}. In the context of readmission following a diagnosis of pancreatic cancer, such questions include: (i) the investigation of between- and within-hospital risks factors for readmission while acknowledging death as a competing force, (ii) characterizing and quantifying between-hospital variation in risk of the terminal event not explained by differences in patient case-mix, and (iii) estimating, and quantifying uncertainty for, hospital-specific effects, as well as ranking. Furthermore, it is well-known that if one is to perform valid inference all potential sources of correlation must be accounted for in the analysis.

To our knowledge, while the literature on the related competing risks problem has considered methods for cluster-correlated data settings~\citep{katsahian2006analysing, chen2008competing, gorfine2011frailty, zhou2012competing, gorfine2014calibrated}, only one paper on the analysis of cluster-correlated semi-competing risks data has been published. Specifically,~\cite{liquet2012investigating} recently proposed a multi-state model that incorporated a hospital-specific random effect to account for cluster-correlation. Estimation and inference was performed within the frequentist paradigm, based on an integrated likelihood that marginalizes over the random effect, implemented in the \texttt{frailtypack} \texttt{R} package~\citep{rondeau2012package}. For our purposes, however, their approach is limited in a number of important ways. First, the analyses presented in~\cite{liquet2012investigating} permit either a patient-specific frailty to account for dependence between $T_1$ and $T_2$ or a hospital-specific random effect to account for cluster-correlation but not both simultaneously. Second, the proposed specification assumed that the hospital-specific random effect for the non-terminal event is independent of the hospital-specific random effect for the terminal event, precluding a potentially important form of dependence. Third, towards understanding variation in risk of readmission, the hospital-specific random effects are themselves key parameters of scientific interest and not nuisance parameters to be marginalized over. Finally, evaluation of the integrated likelihood requires the specification of a parametric distribution for the hospital-specific random effects. While estimation and inference for regression parameters is generally robust to misspecification of random effects distributions in GLMMs, misspecification is known to adversely impact the shape of the estimated distribution of the random effects themselves~\citep{mcculloch2011misspecifying, mcculloch2011prediction, neuhaus2011estimation}. This is particularly important in quality of health care studies where identifying a hospital as being in the tail of the distribution can have a substantial impact on their evaluation.

Towards overcoming these limitations, we develop a novel, comprehensive hierarchical multi-state modeling framework for cluster-correlated semi-competing risks data. A key feature of the framework, and its implementation, is that it permits either parametric or non-parametric specifications for a range of model components, including baseline hazard functions and distributions for hospital-specific random effects. This gives analysts substantial flexibility as they consider their own analyses. Estimation and inference is performed within the Bayesian paradigm which facilities the straightforward quantification of uncertainty for all model parameters, including hospital-specific random effects and variance components. The remainder of this paper paper is organized as follows. Section \ref{sec:data} introduces an on-going study of readmission among patients diagnosed with pancreatic cancer, and provides a description of the available Medicare data. Section \ref{sec:framework} describes the proposed framework, including specification of prior distributions; Section \ref{sec:posterior} provides a brief overview of an efficient computational algorithm for obtaining samples from the joint posterior, its implementation and methods for comparing goodness-of-fit across model specifications. Section \ref{sec:sim} presents a comprehensive simulation study investigating the performance of the proposed framework, including a comparison with the methods of \cite{liquet2012investigating}. Section \ref{sec:application} reports on a detailed analysis of the motiving pancreatic cancer study; sensitivity analyses regarding the specification of certain model parameters are reported in Section \ref{sec:sensitivity}. Finally Section \ref{sec:discussion} concludes the paper with a discussion. Where appropriate, detailed derivations and additional results are provided in an online Supplementary Materials document.

\section{Risk of Readmission Among Patients Diagnosed with Pancreatic Cancer} \label{sec:data}

As mentioned in the Introduction, readmission is a key marker of quality-of-care~\citep{ash2012statistical, 2013hiqr, 2013readmRed}. To-date, however, studies of readmission have focused on health conditions that have relatively good prognosis and/or low mortality including heart failure, myocardial infarction and pneumonia~\citep{krumholz1997readmission, krumholz2011administrative, joynt2011thirty, epstein2011relationship}. Beyond these conditions, however, little is known about variation in risk of readmission for patients diagnosed with terminal conditions such as pancreatic cancer. We are therefore currently engaged in a collaboration investigating readmission among Medicare enrollees diagnosed with pancreatic cancer. The overarching goals of the study are to improve end-of-life quality of care for these patients by first understanding patient-level risk factors associated with readmission and second understanding variation in risk at the level of the hospital (i.e. that not explained by differences in patient case-mix). Towards this we identified all $n$=5,298 Medicare enrollees who were diagnosed with pancreatic cancer during a hospitalization at one of $J$=112 hospitals in the six New England states (Connecticut, Maine, Massachusetts, New Hampshire, Rhode Island and Vermont) between 2000-2009. Information on the initial hospitalization and diagnosis, patient characteristics and co-morbid conditions, discharge destination and subsequent readmissions is obtained from the Medicare Fee-For-Service inpatient claims file (Part A). Specific covariates of interest include sex (0/1 = male/female), age, race (0/1 = white/non-white), the patients Charlson/Deyo comorbidity score~\citep{sharabiani2012systematic}, information on entry route for the initial admission (0/1 = from the ER/transfer from some other facility), whether or not the patient underwent a pancreatic cancer-specific procedure (resection, bypass, or stent), the length of hospitalization and the discharge destination. For the latter, patients could have been discharged to their home, their home with care, a hospice, an intermediate care or skilled nursing facility (ICF/SNF) or some other facility (e.g. a rehabilitation facility or to inpatient care). Table \ref{pcData:summary} provides a summary of observed distributions for these covariates.

\begin{table}[h!]
\centering
\caption{Covariate and outcome information for $n$=5,298 Medicare beneficiaries diagnosed with pancreatic cancer in the six New England states during a hospitalization between 2000-2009. Outcome information is considered with administrative censoring applied at 30 and 90 days post-discarge. \label{pcData:summary}}
\begin{tabular}{llcrr}
\hline
& && $n$ & Percent \\
\hline
\multicolumn{5}{l}{\textit{Covariate information}} \\
~~Sex						& Female					&& 3,037 		&  57.3 \\
							& Male					&& 2,261 		& 42.7 \\
~~Age, years					& 65-69					&& 727		& 13.7 \\
							& 70-74					&& 1,052		& 19.9 \\
							& 75-79					&& 1,226		& 23.1 \\
							& 80-84					&& 1,129		& 21.3 \\
							& $\ge$ 85				&& 1,164		& 20.0 \\
~~Race						& White					&& 4,982 		& 94.0 \\
 							& Non-white				&& 316 		& 6.0 \\				
~~Charlson/Deyo comorbidity score	&$\leq 1$					&& 4,854 		& 91.6 \\
							&$> 1$					&& 444		& 8.4 \\
~~Entry route					& Emergency room			&& 2,255		& 42.6 \\
							& Transfer from another facility	&& 3,043		& 57.4 \\
~~Procedure during hospitalization	&Yes					&& 1,291 		& 24.4 \\
							& No						&& 4,007 		& 75.6 \\
~~Length of hospitalization, days	& 1-7					&& 3,170		& 59.8 \\
							& 8-14					&& 1,465		& 27.7 \\
							& $\ge$ 15				&& 663		& 12.5 \\
~~Discharge destination 			& Home					&& 1,823		& 34.4 \\
							& Home with care			&& 1,571 		& 29.7 \\
							& Hospice					&& 419 		& 7.9 \\
							& SNF/ICF				&& 1,219		& 23.0 \\
							& Other facility				&& 266		& 5.0 \\ \\
\multicolumn{5}{l}{\textit{Outcome information with administrative censoring at 30 days}} \\
~~Readmission and death				& 					&& 205		& 3.9 \\
~~Readmission and censored prior to death	& 				&& 853		& 16.1 \\
~~Death without readmission			& 					&& 1,257		& 23.7 \\
~~Censored prior to readmission or death	&					&& 2,983		& 56.3 \\
\multicolumn{5}{l}{\textit{Outcome information with administrative censoring at 90 days}} \\
~~Readmission and death				& 					&& 608		& 11.5 \\
~~Readmission and censored prior to death	& 				&& 930		& 17.6 \\
~~Death without readmission			& 					&& 1,912		& 36.1 \\
~~Censored prior to readmission or death	&					&& 1,848		& 34.9 \\
\hline
\end{tabular}
\end{table}

Also provided in Table \ref{pcData:summary} is a summary of the observed outcome information at 30 and 90 days post-discharge. Specifically, each patient is classified into one of four groups: (1) they experienced a readmission event and were subsequently observed to die; (2) they experienced a readmission event but were censored prior to death; (3) they were observed to die without having experienced a readmission event; and, (4) they were censored prior to experiencing either a readmission or death event. The administrative censoring at 30 and 90 days is driven by a several important factors. First, scientific and health policy interest regarding readmission has generally focused on a patient's experience in the immediate months following discharge~\citep{2013hiqr}. The primary rationale for this is that post-discharge management for patients diagnosed with pancreatic cancer generally focuses on palliative care, with a specific emphasis on pain management. As patients and their health care providers coordinate this care, the early phases are particularly important for long-term success and are therefore of key interest. A second consideration is that readmission events that occur soon after a patient is discharged are more likely to be directly related to their diagnosis and subsequent care. Readmission events that occur a long time after diagnosis are less likely to be directly related to the quality of care they receive in the immediate aftermath of the diagnosis and, arguably, should not count against a hospitals performance.

\begin{figure}[h!]
\centering
\includegraphics[width=5in, angle=270]{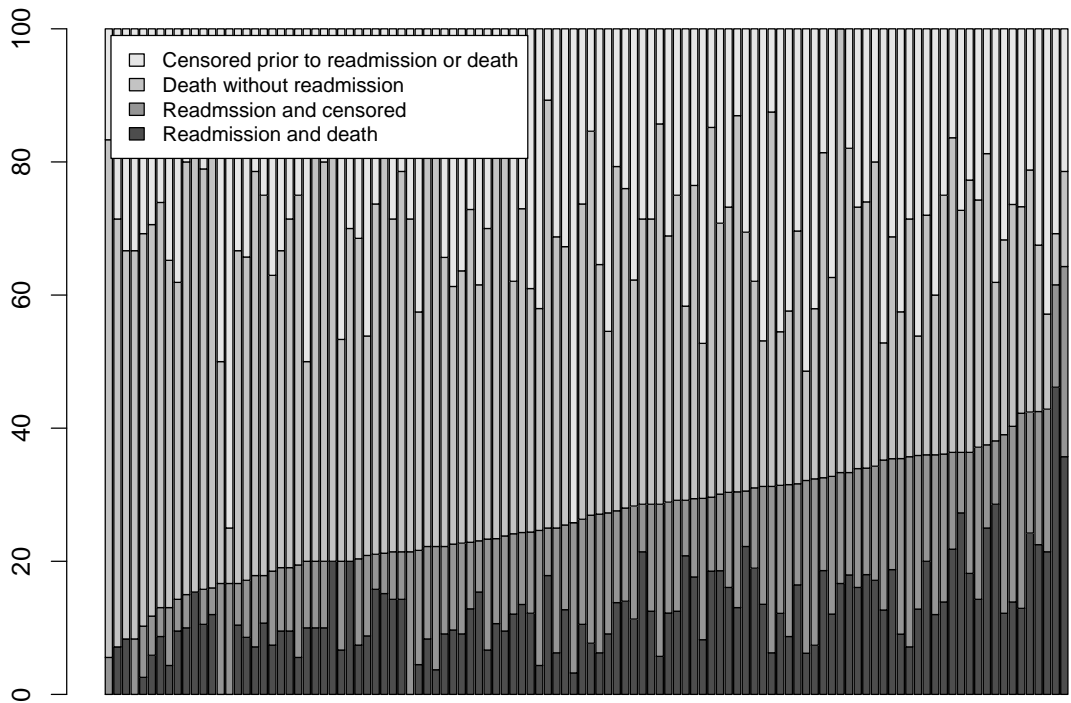}
\caption{Hospital-specific distributions of the outcome information for $n$=5,298 Medicare beneficiaries diagnosed with pancreatic cancer at one of $J$=112 hospitals in the six New England states between 2000-2009. Outcomes have been administratively censored at 90 days and distributions ordered according to the observed percentage of deaths (with and without readmission). \label{pcData:barplot}}
\end{figure}

A central feature of the Medicare data is that the $n$=5,298 patients are clustered within $J$=112 hospitals; cluster sizes vary from 10-420 with a median of 30 patients. The inherent clustering of patients within hospitals is important from both a statistical and a scientific perspective: valid inference requires acknowledging potential correlation among patients and understanding between-hospital variation in readmission rates is a key scientific goal. Towards the latter, Figure \ref{pcData:barplot} provides a barplot of the hospital-specific distributions of the four outcome groups based on censoring at 90 days. While there are many ways in which the $J$=112 hospitals could be ordered, Figure \ref{pcData:barplot} orders them according to the total percentage of patients readmitted within 90 days (i.e. with or without a subsequent death event). From the figure we see that there is substantial variation in observed readmission rates across hospitals, with the lowest being 5.6\% and the highest being 64.3\%. Moving beyond these raw adjusted rates would need to first account for case-mix differences across the hospitals, second account for death as a competing risk and third account for the cluster-correlation.

\section{A Bayesian Framework for Cluster-Correlated Semi-competing Risks Data}\label{sec:framework}

\subsection{Model specification}\label{sec:framework:model}

Viewing `discharge', `readmission' and `death' as three states, the underlying data generating mechanism that gave rise to these data can be represented by a multi-state model, specifically an \textit{illness-death model}~\citep{gill1997statistical,putter2007tutorial,Xu:Kalb:2010}. Letting $T_1$ denote the time to non-terminal event and $T_2$ the time to the terminal event, the illness-death model is characterized by three hazard functions that govern the rates at which patients transitions between the states:
\be
\label{def.haz1}
	h_1(t_1) & = & \lim_{\Delta\rightarrow 0}\frac{1}{\Delta}\Pr[T_1\in [t_1, t_1+\Delta)|\ T_1\geq t_1, T_2 \geq t_1],\ \hspace*{0.25in} t_1 > 0 \\
\label{def.haz2}
	h_2(t_2) & = & \lim_{\Delta\rightarrow 0}\frac{1}{\Delta}\Pr[T_2\in [t_2, t_2+\Delta)|\ T_1\geq t_2, T_2 \geq t_2],\ \hspace*{0.25in} t_2 > 0 \\
\label{def.haz3}
	h_3(t_2|t_1) & = & \lim_{\Delta\rightarrow 0}\frac{1}{\Delta}\Pr[T_2\in [t_2, t_2+\Delta)|\ T_1= t_1, T_2 \geq t_2],\ \hspace*{0.25in} t_2 > t_1.
\ee
In practice, analyses based on the illness-death model characterized by (\ref{def.haz1})-(\ref{def.haz3}) proceeds by placing structure on these functions, specifically as a function of covariates and frailties/random effects. Towards this, let $T_{ji1}$ and $T_{ji2}$ denote the time to the non-terminal event and time to the terminal event for the $i^{\textrm{th}}$ patient in the $j^{\textrm{th}}$ cluster, respectively, for $i = 1, \ldots, n_j$ and $j = 1, \ldots, J$. Furthermore, let $\bfX_{jig}$ be a vector of time-invariant covariates for the $i^{\textrm{th}}$ patient in the $j^{\textrm{th}}$ cluster that will be considered in the model for the $g^{\textrm{th}}$ transition, $g$=1,2,3. Consider the following general modeling specification:
\be
\label{model1}
	h_1(t_{ji1}; \gamma_{ji}, \bfX_{ji1}, V_{j1}) &=& \gamma_{ji}\ h_{01}(t_{ji1})\ \mbox{exp}\{\bfX_{ji1}^T\bfbeta_1\ +\ V_{j1}\}, \hspace*{0.25in}  t_{ji1} > 0 \\
\label{model2}
	h_2(t_{ji2}; \gamma_{ji}, \bfX_{ji2}, V_{j2}) &=& \gamma_{ji}\ h_{02}(t_{ji2})\ \mbox{exp}\{\bfX_{ji2}^T\bfbeta_2\ +\ V_{j2}\}, \hspace*{0.25in}  t_{ji2} > 0 \\
\label{model3}
	h_3(t_{ji2}|t_{ji1}; \gamma_{ji}, \bfX_{ji3}, V_{j3}) &=&\gamma_{ji} \ h_{03}(t_{ji2}| t_{ji1})\ \mbox{exp}\{\bfX_{ji3}^T\bfbeta_3\ +\ V_{j3}\}, \hspace*{0.25in}  t_{ji2} > t_{ji1},
\ee
where $\gamma_{ji}$ is a shared patient-specific frailty, $\bfV_j = (V_{j1}, V_{j2}, V_{j3})$ is a vector of cluster-specific random effects, each specific to one of the three possible transitions, and $\bfbeta_g$ is a transition-specific vector of fixed-effect log-hazard ratio regression parameters. As described by ~\cite{Xu:Kalb:2010}, model (\ref{model3}) is often simplified in practice by either assuming that $h_{03}(t_{ji2}| t_{ji1}) =  h_{03}(t_{ji2})$ or that $h_{03}(t_{ji2}| t_{ji1}) =  h_{03}(t_{ji2} - t_{ji1})$. Given the former specification, the model is referred to as being \textit{Markov} in the sense that the hazard for death given readmission does not depend on the actual time of readmission; under the latter specification, the model is referred to as \textit{semi-Markov}. For simplicity we focus the exposition in this section on Markov models although note that the methods and computational algorithms have also been developed and implemented for the semi-Markov model; the analyses in Sections \ref{sec:application} and \ref{sec:sensitivity} also consider both models.

\subsection{The observed data likelihood}\label{sec:framework:likelihood}

To complete the notation developed so far, let $C_{ji}$ denote the right censoring time for the $i^{\textrm{th}}$ patient in the $j^{\textrm{th}}$ cluster. Furthermore, let $Y_{ji1}$ = min($T_{ji1}$, $T_{ji2}$, $C_{ji}$), $\Delta_{ji1}$ = 1 if $Y_{ji1}$ = $T_{ji1}$ (i.e. a readmission event is observed) and 0 otherwise, $Y_{ji2}$ = min($T_{ji2}$, $C_{ji}$) and, $\Delta_{ji2}$ = 1 if $Y_{ji2}$ = $T_{ji2}$ (i.e. a death event is observed) and 0 otherwise. Finally, let $\mathcal{D}_{ji}$ = $\{y_{ji1}, \delta_{ji1}, y_{ji2}, \delta_{ji2}\}$ denote the observed outcome data for the $i^{\textrm{th}}$ patient in the $j^{\textrm{th}}$ cluster and $H_{0g}(\cdot)$ the cumulative baseline hazard function corresponding to $h_{0g}(\cdot)$. Let $\vec{\gamma}$ and $\vec{\bfV}$ denote the collections of the $\gamma_{ji}$ and $\bfV_j$, respectively. Following~\cite{putter2007tutorial}, for a given specification of (\ref{model1})-(\ref{model3}), the observed data likelihood as a function of the unknown parameters $\Phi$ = $\{\bfbeta_1, \bfbeta_2, \bfbeta_3, h_{01}, h_{02}, h_{03}, \vec{\gamma}, \vec{\bfV}\}$, is given by:
\be
	\mathcal{L}(\mathcal{D}|\Phi) &=& \prod_{j=1}^J\prod_{i=1}^{n_j}\mathcal{L}(\mathcal{D}_{ji} | \bfbeta_1, \bfbeta_2, \bfbeta_3, h_{01}, h_{02}, h_{03}, \gamma_{ji},  \bfV_j) \nonumber \\
	&=& \prod_{j=1}^J\prod_{i=1}^{n_j} \left\{\gamma_{ji}h_{01}(y_{ji1})\eta_{ji1}\right\}^{\delta_{ji1}(1-\delta_{ji2})} \left\{\gamma_{ji}^{2} h_{01}(y_{ji1})\eta_{ji1} h_{03}(y_{ji2})\eta_{ji3}\right\}^{\delta_{ji1}\delta_{ji2}} \nonumber \\
	&& \hspace*{0.5in} \times \left\{\gamma_{ji}h_{02}(y_{ji2})\eta_{ji2}\right\}^{(1-\delta_{ji1})\delta_{ji2}} \exp\left\{-\gamma_{ji} r(y_{ji1}, y_{ji2})\right\}, \label{lh1}
\ee
where $\eta_{jig} = \exp\{\bfX_{jig}^T\bfbeta_{g} + V_{jg}\}$ and $r(t_{ji1}, t_{ji2}) = [H_{01}(t_{ji1})\eta_{ji1} + H_{02}(t_{ji1})\eta_{ji2} + \left\{H_{03}(t_{ji2}) - \right.$ $\left.H_{03}(t_{ji1})\right\}\eta_{ji3}]$.

In the remainder of this section, we complete the specification of the Bayesian model by providing detail on a range of possible choices for specification of the baseline hazard functions in (\ref{model1})-(\ref{model3}), the population distribution for the hospital-specific random effects and, finally, prior distributions. To facilitate the exposition, Table \ref{tab:four:models} provides a summary of four possible specifications of the model along with the hyperparameters that require specification by the analyst.

\begin{table}[h!]
\caption{Summary of four models based on parametric and non-parametric specifications of the baseline hazard functions and hospital-specific random effects distributions. Hyperparameters that require specification by the analyst are provided in parenthesis. Note, $(a_{\theta}, b_{\theta})$, the hyperparameters for the patient-specific frailty variance component, require specification in all of four models. \label{tab:four:models}}
\begin{center}
\begin{tabular}{lcc}
\hline
									& \multicolumn{2}{c}{\underline{Hospital-specific random effects, $\bfV_j$}} \\
									& 	MVN				& DPM$^\dag$ 		\\
Baseline hazard functions, $h_{0g}(\cdot)$	&	$(\Psi_v, \rho_v)$	& $(\Psi_0, \rho_0, \tau)$ \\
\hline
Weibull 	$(a_{\alpha, g}, b_{\alpha, g}, a_{\kappa, g}, b_{\kappa, g})$ &	Weibull-MVN 	&		Weibull-DPM \\
PEM$^\dag$ 	$(\alpha_{K_g}, a_{\sigma, g}, b_{\sigma, g})$ 				& PEM-MVN	& 		PEM-DPM\\
\hline
\multicolumn{3}{l}{\footnotesize $^\dag$ PEM: piecewise exponential model; DPM: Dirichlet process mixture}
\end{tabular}
\end{center}
\end{table}

\subsection{Baseline hazard functions}\label{sec:framework:baseline}

Within the frequentist paradigm estimation and inference for time-to-event models is often based on a partial likelihood which conditions on risk sets, removing the need for analysts to specify baseline hazard functions. In the Bayesian paradigm, however, one is required to specify these functions. Here, we consider two strategies. The first assumes that the underlying transition times follow Weibull($\alpha_{w,g}$, $\kappa_{w,g}$) distributions, parameterized so that $h_{0g}(t)$ = $\alpha_{w,g} \kappa_{w,g} t^{\alpha_{w,g}-1}$. While such a parametric specification is appealing due to its computational simplicity, especially in small-sample settings, the Weibull is somewhat restrictive in that the corresponding hazard function is strictly monotone. As an alternative, we consider a non-parametric specification based on taking each of the log-baseline hazard functions to be a flexible mixture of piecewise constant functions~\citep{mckeague2000bayesian}. Briefly, let $s_{g, \textrm{max}}$ denote the maximum observed time for transition $g$ and partition (0, $s_{g, \textrm{max}}$] into $K_g+1$ intervals: $\bfs_g$ = $\{s_{g,0}, s_{g,1},\ \ldots,\ s_{g, K_g+1}\}$, with $s_{g,0}$ $\equiv$ 0 and $s_{g, K_g+1}$ $\equiv$ $s_{g, \textrm{max}}$. Given the partition ($K_g, \bfs_g$), we assume
\be
	\log h_{0g}(t)\ =\ \lambda_g(t)\ =\ \sum_{k=1}^{K_g+1}\ 1_{[s_{g, k-1} < t \le s_{g, k}]}\lambda_{g,k},
\ee
where $\lambda_{g,k}$ is the (constant) height of the log-baseline hazard function on the interval $(s_{g, k-1}, s_{g, k}]$. We refer to this specification as a \textit{piecewise exponential model} (PEM) for the baseline hazard function. Note, while numerous options are available for specifying these functions~\citep[e.g.][]{Ibra:Chen:2001}, a key benefit of this structure is that it balances flexibility and computational convenience, since the integrals in the likelihood (specifically for the cumulative hazard functions) are replaced with summations~\citep{Lee2014bscr}.

\subsection{Hospital-specific random effects}\label{sec:framework:Vj}

As with specification of the baseline hazard functions, we consider two options for the specification of the population distribution for the $J$ hospital-specific vectors of random effects. First, motivated by the current standard for analyses of readmission (i.e. a LN-GLMM), we consider a specification in which the $\bfV_j$ arise as i.i.d draws from a mean-zero multivariate Normal distribution with variance-covariance matrix $\bfSigma_V$. The diagonal elements of the 3$\times$3 matrix $\bfSigma_V$ characterize variation across hospitals in risk for readmission, death and death following readmission, respectively, that is not explained by covariates included in the linear predictors. Crucially, that each random effect has its own variance component allows the characterization of differential variation across hospitals for each of the three transitions. In addition, the off-diagonals of $\bfSigma_V$ permit covariation between the three random effects across the hospitals giving researchers the ability to characterize, for example, whether or not hospitals with high mortality rates tend to have low readmission rates. 

While conceptually simple and computationally convenient, the Normal distribution it is often criticized as being a strong assumption. As an alternative we consider the use of a so-called Bayesian nonparametric specification for the population distribution of $\bfV_j$, specifically a Dirichlet process mixture of multivariate Normal distributions (DPM-MVN)~\citep{ferguson1973bayesian, bush1996semiparametric, walker1997hierarchical}. One representation of this model is as follows:
\be
	\bfV_j | \bfmu_j, \Sigma_j &\sim& \mbox{MVN$_3$}(\bfmu_j, \Sigma_j), \nonumber \\
	\bfmu_j, \Sigma_j | G &\sim& G,  \nonumber \\
	G &\sim& DP(G_0, \tau),
\ee
where $\bfmu_j, \Sigma_j$ are the cluster-specific latent mean and variance of $\bfV_j$, which are taken to be draws from some (unknown) distribution $G$ to which a Dirichlet process prior is assigned. Finally the Dirichlet process is indexed by $G_0$, the so-called \textit{centering distribution}, and $\tau$, the so-called \textit{precision} parameter. 

\subsection{Hyperparameters and prior distributions}\label{sec:framework:priors}

The proposed Bayesian framework is completed with the specification of prior distributions for unknown parameters introduced in Sections \ref{sec:framework:model}-\ref{sec:framework:Vj}.

\subsubsection{Stage one parameters}\label{sec:framework:priors:one}

For each of the transition-specific regression parameters, $\bfbeta_g$ for $g$=1,2,3, a non-informative flat prior on the real line is adopted. For the shared patient-specific frailties, $\gamma_{ji}$, we assume that they arise from a Gamma$(\theta^{-1}, \theta^{-1})$ distribution, parameterized so that $E[\gamma_{ji}] = 1$ and $V [\gamma_{ji}] = \theta$. In the absence of prior knowledge on the frailty variance component, a Gamma$(a_{\theta}, b_{\theta})$ hyperprior for the precision $\theta^{-1}$ is adopted.

\subsubsection{Baseline hazard functions}

For the parametric Weibull baseline hazard functions, since the hyperparameters have support on (0, $\infty$), we complete the specification of this model by adopting gamma prior distributions for both; that is, we take $\alpha_{w,g} \sim$ Gamma$(a_{\alpha,g},\ b_{\alpha,g})$ and $\kappa_{w,g} \sim$ Gamma$(a_{\kappa,g},\ b_{\kappa,g})$, $g$=1,2,3.

To complete the non-parametric PEM model specification, we specify that the $K_g+1$ heights arise from a multivariate Normal distribution. Specifically, letting $\bflambda_g$ = ($\lambda_{g,1}$,\ldots, $\lambda_{g,K_g}$, $\lambda_{g, K_g+1}$) denote the transition-specific heights, we assume that $\bflambda_g$ $\sim$ MVN($\mu_{\lambda_g}\bfone$, $\sigma_{\lambda_g}^2\bfSigma_{\lambda_g}$), where $\mu_{\lambda_g}$ is the overall mean, $\sigma^2_{\lambda_g}$ is a common variance component for the $K_g+1$ elements and $\bfSigma_{\lambda_g}$ is a correlation matrix. To induce \textit{a priori} smoothness in the baseline hazard functions we view the components of $\bflambda_g$ in terms of a one-dimensional spatial problem, so that adjacent intervals can `borrow' information from each other. To do this we specify $\bfSigma_{\lambda_g}$ via a Gaussian intrinsic conditional autoregression (ICAR)\citep{Besa:Koop:1995}. Additional technical details regarding the corresponding MVN-ICAR are provided in Supplementary Materials A. Finally, we specify a series of hyperpriors for the additional parameters introduced in the MVN-ICAR. In particular, we adopt a flat prior on the real line for $\mu_{\lambda_g}$ and a conjugate Gamma$(a_{\sigma,g}, b_{\sigma,g})$ distribution for the precision $\sigma^{-2}_{\lambda_g}$. For the ICAR specification, we avoid reliance on a fixed partition of the time scales by permitting the partition ($K_g$, $\bfs_g$) to vary and be updated via a reversible jump MCMC scheme~\citep{green1995reversible}. Towards this we first adopt a Poisson($\alpha_{K_g}$) prior for the number of splits in the partition, $K_g$. Conditional on the number of splits, we take the locations to be \textit{a priori} distributed as the even-numbered order statistics:
\be
	 \pi(\bfs_g | K_g) &\propto& \frac{(2K_g+1)! \prod_{k=1}^{K_g+1}(s_{g,k}-s_{g, k-1})}{(s_{g,K_g+1})^{(2K_g+1)}},
\ee
Jointly, these choices form a time-homogeneous Poisson process prior for the partition $(K_g$, $\bfs_g)$ so that \textit{a posteriori}, after mixing over partitions as they arise in the MCMC scheme, the value of $\lambda_g(t)$ in any given small interval of time is characterized as a smooth exponentiated mixture of piecewise constant functions~\citep{arjas1994nonparametric, mckeague2000bayesian, sebastien2008separation}.

\subsubsection{Hospital-specific random effects}

For the parametric specification of a single MVN$_3$($\bfzero$, $\bfSigma_V$) distribution, we adopt a conjugate inverse-Wishart($\Psi_v$, $\rho_v$) prior for the variance-covariance matrix $\bfSigma_V$. Completion of the non-parametric DPM-MVN model requires specification of prior choices for the centering distribution and the precision parameter. Here we take $G_0$ to be a multivariate Normal/inverse-Wishart (NIW) distribution for which the probability density function can be expressed as the product:
\be
	f_{\mathcal{NIW}}(\mu, \Sigma | \Psi_0, \rho_0) &=& f_{MVN}(\bfmu | \bfzero, \Sigma) \times f_{inv-Wishart}(\Sigma | \Psi_0, \rho_0), \nonumber
\ee 
where $f_D(\cdot | \bftheta_D)$ is the density function for a distribution $D$ indexed by $\bftheta_D$. This choice is appealing in that one can exploit prior-posterior conjugacy in the MCMC scheme~\citep{neal2000markov}. Finally, we treat hyperparameter the precision parameter in DPM-MVN specification, $\tau$, as unknown and assign a Gamma$(a_{\tau}, b_{\tau})$ hyperprior~\citep{escobar1995bayesian}. 

\section{Posterior Inference and Model Comparison}\label{sec:posterior}

\subsection{Markov Chain Monte Carlo}\label{sec:posterior:mcmc}

To perform estimation and inference for each of the models in Table \ref{tab:four:models} we use a random scan Gibbs sampling algorithm to generate samples from their joint posterior distributions. In the corresponding Markov chain Monte Carlo (MCMC) scheme, parameters are updated by either exploiting conjugacies inherent to the model structure or using a Metropolis-Hastings step. For models that adopt a PEM specification for the baseline hazard functions, updating the partition $(K_g, \bfs)$ requires a change in the dimension of the parameter space and a Metropolis-Hastings-Green step is used~\citep{green1995reversible}. A detailed description of proposed computational scheme is given in Supplementary Materials B; as mentioned in Section \ref{sec:framework:model}, the computation scheme has been developed for both the Markov and semi-Markov models for $h_{03}(\cdot)$.

Finally, we note that the algorithms are implemented in the \texttt{SemiCompRisks} package for \texttt{R}~\citep{R:man:2014}. Given the complexity of the proposed models, and the numerous updates in the MCMC scheme, \texttt{C} has been used as the primary computational engine to ensure that analyses can be conducted within a reasonable timeframe.

\subsection{Model comparison}\label{sec:posterior:comparison}

In practice, analysts have to balance model complexity with the realities of sample size and availability of information. While each of models in Table \ref{tab:four:models} has its own merit and utility, it may be of interest to directly compare their goodness of fit to the observed data. To this end, we consider two model assessment metrics: the deviance information criterion~\citep[DIC; ][]{spiegelhalter2002bayesian}, and the log-pseudo marginal likelihood statistic~\citep[LPML; ][]{geisser1979predictive,gelfand1995bayesian}. Although DIC is often the default choice for model comparison in the Bayesian paradigm, its use in the context of complex hierarchical models requires care~\citep{celeux2006deviance}. Specifically for models that condition on latent parameters, such as the patient-specific $\gamma_{ji}$ in models (\ref{model1})-(\ref{model3}), DIC computed on the basis of a likelihood that is marginalized with respect to these parameters performs more reliably as a metric for comparison than DIC computed on the basis of a likelihood that conditions on them~\citep{millar2009comparison}. For our purposes, since the $\bfV_j$ random effects are of intrinsic scientific interest, we propose to evaluate DIC and LPML on the basis of a partially marginalized likelihood, one that integrates solely over the distribution of the patient-specific frailties:
\be
	\mathcal{L}^*(\mathcal{D}|\Phi^*) &=& \int \prod_{j=1}^J\prod_{i=1}^{n_j}\mathcal{L}(\mathcal{D}_{ji} | \bfbeta_1, \bfbeta_2, \bfbeta_3, h_{01}, h_{02}, h_{03}, \gamma,  \bfV_j) f_\theta(\gamma; \theta)d\gamma
\label{like:marginal}
\ee
where $\Phi^*$ = $\{\bfbeta_1, \bfbeta_2, \bfbeta_3, h_{01}, h_{02}, h_{03}, \theta, \vec{\bfV}\}$, $\mathcal{L}(\mathcal{D}_{ji} | \cdot)$ is given by expression (\ref{lh1}) and $f(\cdot; \theta)$ is the density of a Gamma$(\theta^{-1}, \theta^{-1})$ distribution (see Section \ref{sec:framework:priors:one}).

Given expression (\ref{like:marginal}), we therefore compute DIC as:
\be
	\textrm{DIC} = \textrm{D}(\overline{\Phi^*}) + 2p_D,
\ee
where $D(\Phi^*)$ = $\sum\limits_{j,i} -2\log L^*(\mathcal{D}_{ji} | \Phi^*)$ is the (marginal) deviance and $\overline{\Phi^*}$ is the posterior mean of $\Phi^*$. The penalty term, $p_D$, is given by $\overline{D}(\Phi^*) - D(\overline{\Phi^*})$, where $\overline{D}(\Phi^*)$ is the posterior mean of $D(\Phi^*)$. Note, a model with smaller DIC indicates a better fit of the model for the data.

The LPML statistic is computed as $\sum\limits_{j,i} \log(\textrm{CPO})_{ji}$, the sum of the logarithms of the patient-specific conditional predictive ordinate (CPO)~\citep{geisser1993predictive}, each defined as:
\be
	\textrm{CPO}_{ji} = L^*(\mathcal{D}_{ji} | \mathcal{D}^{(-ji)}) = \int L^*(\mathcal{D}_{ji} | \Phi^*) \pi(\Phi^* | \mathcal{D}^{(-ji)})d\Phi^*,
\ee
where $\mathcal{D}^{(-ji)}$ denotes the data with the observation from the $i^{\textrm{th}}$ patient in the $j^{\textrm{th}}$ cluster removed. Intuitively, the CPO$_{ji}$ is the posterior probability of the observed outcome for $i^{\textrm{th}}$ patient in the $j^{\textrm{th}}$ cluster, i.e. $(y_{ji1}, \delta_{ji1}, y_{ji2}, \delta_{ji2})$, on the basis of a model fit to a dataset that excludes that particular patient. Thus, large values of CPO$_{ji}$ attribute high posterior probability to the observed data and, therefore, indicate a better fit. Although a closed form expression for CPO$_{ji}$ is not available for our proposed models, following~\cite{shao2000monte} we approximate CPO$_{ji}$ via a Monte Carlo estimator:
\be
	\left(\frac{1}{Q}\sum_{q = 1}^Q L^*(\mathcal{D}_{ji} | \Phi^{*(q)})^{-1}\right)^{-1},
\ee
where $\{\Phi^{*(q)}; q= 1,2,\ldots, Q\}$ are MCMC samples drawn from the (marginal) joint posterior distribution of $\Phi^*$.

\section{Simulation Studies}\label{sec:sim}

The performance of the proposed models is investigated through a series of simulation studies. The overarching goals of the simulation studies are to investigate the small sample operating characteristics of the models summarized in Table \ref{tab:four:models} under a variety of scenarios as well as to compare their performance with the methods of \cite{liquet2012investigating}.

\subsection{Set-up and data generation}

Towards developing a comprehensive understanding of the performance of the proposed methods we consider six data scenarios that vary in terms of the true underlying baseline hazard distributions, the true distribution of the cluster-specific random effects and the true extent of variation in the patient-specific frailties. Table \ref{tab:simSetting} provides a summary. In scenarios 1-5, the baseline hazard functions are set to correspond to the hazard of a Weibull distribution so that the event rates in the simulated data are similar to those in the observed Medicare data when the outcomes are administratively censored at $t$=90; specifically, we set ($\alpha_{w,1}$, $\kappa_{w,1}$)=(0.8, 0.05), ($\alpha_{w,2}$, $\kappa_{w,2}$)=(1.1, 0.01), and ($\alpha_{w,3}$, $\kappa_{w,3}$)=(0.9, 0.01). To evaluate the performance of the model when the baseline hazard functions do not correspond to a Weibull, scenario 6 takes them to be piecewise linear functions: $h_{0g}(t)$ = \{($k_g$-$b_g$)$t$/40+$b_g$\}$I$($t$$\leq$40) + \{(3$k_g$-$b_g$)/2-($k_g$-$b_g$)$t$/80\}$I$($t$$>$40), with $b_1$=0.1, $b_2$=0.05, $b_3$=0.15, and $k_1$=$k_2$=$k_3$=0.0005 specified so that the true baseline hazard functions are not monotone increasing or decreasing functions like a Weibull.

\begin{table}[ht]
\centering
\caption{Summary of six simulation scenarios explored in Section \ref{sec:sim}.\label{tab:simSetting}}
\begin{tabular}{c c c c c}
\hline
Scenario		&Distribution of &Distribution of &&\\
 & baseline hazard functions & cluster-specific random effects, $\bfV_j$ & $\theta$\\
\hline
1 & Weibull & MVN($\bfzero$, 0.25$\cdot I$) & 0.50 \\
2 & Weibull & MVN($\bfzero$, 0.25$\cdot I$) & 1.00 \\
3 & Weibull & MVN\Big($\bfzero$, \renewcommand{\arraystretch}{0.5} $\begin{bmatrix}  ~~0.25 & -0.10 & -0.10 \\ -0.10 & ~~0.25 & ~~0.20 \\ -0.10 & ~~0.20 & ~~0.25 \end{bmatrix}$\Big)& 0.50 \\
4 & Weibull & MVN($\bfzero$, 0.25$\cdot I$) & 0.00 \\
5 & Weibull & 0.5$\cdot$MVN($\bfzero$, $I$)+0.5$\cdot$MVN($\bfzero$, 0.01$\cdot I$) & 0.50 \\
6 & Piecewise linear & MVN($\bfzero$, 0.25$\cdot I$) & 0.50 \\
\hline
\end{tabular}
\end{table}

With regard to the `true' distribution of the cluster-specific random effects, scenarios 1, 2, 4 and 6 consider a multivariate Normal distribution in which the components are independent.  Scenario 3 expands on this by considering the impact of covariation across the $\bfV_j$, while Scenario 5 examines the performance of the models when the true distribution is a mixture of two multivariate Normal distributions.

Finally, with regard to the `true' variance of the patient-specific frailties, scenarios 1, 3, 5 and 6 consider a base value of $\theta$=0.5. This value was chosen as a compromise across the posterior medians from the fits of the four models in Table \ref{tab:four:models} to the Medicare data (see Table \ref{tab:results:var} below). Scenario 2 considers the impact of greater variation in the patient-specific frailties, while Scenario 4 corresponds to a misspecification of the proposed model with the `true' $\theta$ set to 0.

For each of the six scenarios we generated $R$=500 simulated datasets under the semi-Markov illness-death model described in Section \ref{sec:framework:model}. Across all simulated datasets, we set the number of clusters and cluster-specific sample sizes to be those observed in the Medicare data. Furthermore, we specified that each of the three transition-specific hazard functions depended on three covariates: $X_{jig, 1}$ and $X_{jig, 2}$ both Normal(0, 1) random variables and $X_{jig, 3}$ a Bernoulli(0.5) random variable. The regression coefficients are set to $\bfbeta_1$=$\bfbeta_2$=(0.5, 0.8, -0.5) and $\bfbeta_3$=(1.0, 1.0, -1.0), so that the covariate effects on the risk of the terminal event depend on whether or not the non-terminal event has occurred. Finally, we note that the function used to simulate the semi-competing risks data is available in the \texttt{SemiCompRisks} package.

\subsection{Analyses}\label{sec:sim:details}

For each of the $R$=500 datasets under each of the six scenarios we fit each of the four models in Table \ref{tab:four:models}. For the proposed models in which the baseline hazard function was specified via a Weibull distribution, we set ($a_{\alpha,g}$, $b_{\alpha,g}$) = (0.5, 0.01) and ($a_{\kappa,g}$, $b_{\kappa,g}$) = (0.5, 0.05) for the transition-specific shape and rate parameters. For models in which a non-parametric PEM specification was adopted for the baseline hazard function, we set the prior Poisson rate on the number of intervals to be $\alpha_g$ = 10. For the precision parameter in the MVN-ICAR specification, we set ($a_{\sigma,g}$, $b_{\sigma,g}$) = (0.7, 0.7) so that the induced prior for $\sigma_{\lambda_g}^2$ had a median of 1.72 and 95\% central mass between 0.23 and 156.

For the variance component associated with the patient-specific frailties, we set ($a_\theta$, $b_\theta$) = (0.7, 0.7); that is the same prior was used for the precision $\theta^{-1}$ for the $\gamma_{ji}$ frailties as the precision component in the MVN-ICAR specification for the PEM model. For the hospital-specific random effects variance components, given a MVN specification, we set ($\Psi_v$, $\rho_v$) = ($I_3$, 5) so that the induced prior on $\Sigma_V$ has a prior mean given by the 3$\times$3 identity matrix. The same prior was adopted for the variance-covariance matrix of the centering distribution of the DPM-MVN specification, $G_0$; that is, we set ($\Psi_0$, $\rho_0$) = ($I_3$, 5). Finally, for the precision parameter in the DPM-MVN specification we set ($a_{\tau}$, $b_{\tau}$) = (1.5, 0.0125) so that \textit{a priori} $\tau$ had a mode of 40 and standard deviation of 98. Given the prior specifications, two independent chains were run for a total of six million scans each; the Gelman-Rubin potential scale reduction (PSR) statistic~\citep{gelman2003bayesian} was used to assess convergence, specifically requiring the PSR to be less than 1.05 for all model parameters.

In addition to the models in Table \ref{tab:four:models}, we analyzed each simulated dataset using the methods of \cite{liquet2012investigating}. Specifically, we considered the `shared frailty' (SF) model implemented in the \texttt{frailtypack} package for \texttt{R} \citep{rondeau2012package} and summarized using notation consistent with that adopted in this manuscript in Supplementary Materials Section C. Briefly, this model adopts a Cox-type regression structure for each transition-specific hazard, as we do in expressions (\ref{model1})-(\ref{model3}). For the baseline hazard functions, two options are available: one that corresponds to a Weibull distribution and another where each $h_g(\cdot)$ is specified via a flexible penalized smoothing spline. To distinguish these models, we refer to them as the \textit{Weibull-SF} and \textit{Spline-SF} models, respectively. In contrast to the specification in (\ref{model1})-(\ref{model3}), the SF model introduces a cluster-specific frailties as a multiplicative factors for each transition-specific hazard. Two options are available for the distribution of these factors across the clusters; either they arise from three independent gamma distributions or they arise from three independent log-Normal distributions. For either option, estimation and inference is performed within the frequentist paradigm specifically based on an integrated likelihood that marginalizes out the cluster-specific frailties; estimation of the latter is performed via empirical Bayes. In this paper, we present the results from the SF models that adopt independent gamma distributions for cluster-specific frailties while we provide those from the SF models with independent log-Normal frailties in Supplementary Materials D. Finally, we note that in contrast to the specification in expressions (\ref{model1})-(\ref{model3}), the SF model does not account for within-patient correlation. That is there is no quantity that corresponds to the patient-specific $\gamma_{ji}$ terms in the proposed models. 

\subsection{Results}\label{sec:sim:results}

\subsubsection{Baseline survivor functions}

\begin{figure}[h!]
\centering
\includegraphics[width = 6in]{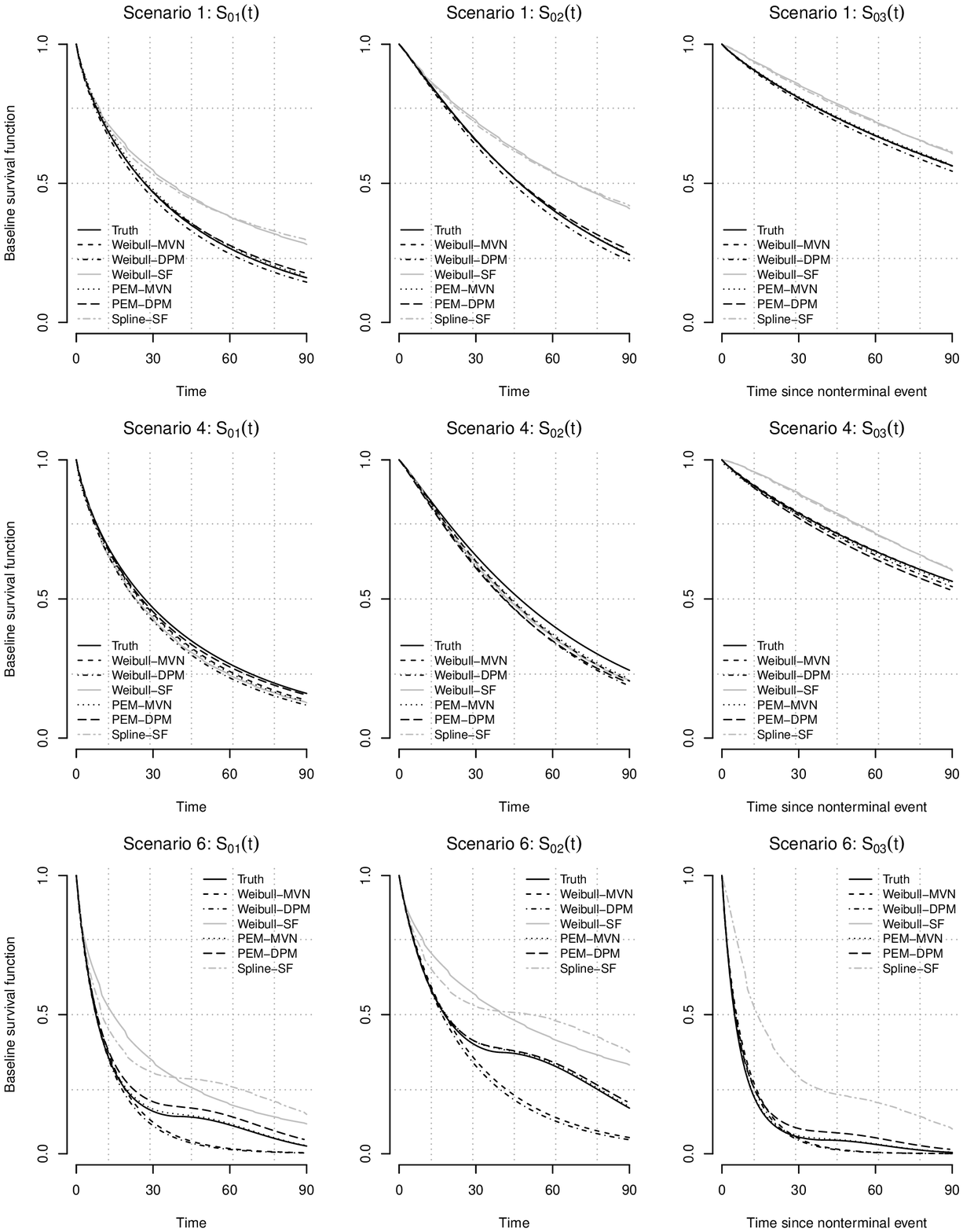}
\caption{Estimated transition-specific baseline survival functions, $S_{0g}(\cdot)$ = $\exp(-H_{0g}(\cdot))$, for each six analyses described in Section \ref{sec:sim:details} under simulation scenarios 1, 4 and 6. \label{fig:sim1BS}}
\end{figure}

Figure \ref{fig:sim1BS} presents the mean estimated transition-specific baseline survival functions under scenarios 1, 4 and 6 across the six models. Under scenarios 1 and 4, for which the baseline hazard functions are Weibull, all four of the proposed models estimate the three baseline survivor functions very well. In contrast the two SF models only perform well in scenario 4 for which $\theta$=0. This is to be expected since, as described in detail in Supplementary Materials Section C, the SF model does not include patient-specific frailties; effectively, it assumes that $\theta$=0 even when it is not. In scenario 6, for which the baseline hazard functions are not Weibull, the proposed PEM-MVN and PEM-DPM specifications capture the true shape of the baseline survivor functions well; all four of the models that assume the baseline hazard function to be a Weibull, however, are unable to capture the shape.

\subsubsection{Regression parameters and $\theta$}

Focusing on scenarios 1-3, each corresponding to a `true' Weibull-MVN model, Table \ref{tab:simReg} indicates that all four of the proposed models in Table \ref{tab:four:models} perform very well in terms of estimation and inference for $\bfbeta_1$ and $\theta$. Across the board, we find that percent bias is no larger than 3.2\% and the estimated coverage probabilities are all close to the nominal 0.95. In contrast, both the Weibull-SF and Spline-SF models yield point estimates of $\bfbeta_1$ that are significantly biased and, as such, have poor coverage probabilities. The poor performance of the SF models is likely tied to the fact that they do not account for within-patient correlation; hence $\theta$ is not estimated by these models. The results for these models, however, is dramatically improved under scenario 4 for which the true value of $\theta$ is zero (i.e. the scenario they explicitly accommodate). Interestingly, the four proposed models each exhibit a small amount of bias under this scenario (up to approximately 5\%). In addition the coverage probabilities for $\beta_{11}$ and $\beta_{12}$ are poor, particularly for the two models that adopt a PEM specification for the baseline hazard function. In scenario 5, we again see that all four of the proposed models perform well. Finally, under scenario 6 we see that the PEM-MVN and PEM-DPM models perform very well in terms of bias and coverage. In contrast, the Weibull-MVN and Weibull-DPM models perform poorly, particularly with respect to estimation of $\theta$, illustrating the potential danger of adopting a parametric Weibull baseline hazard function when the truth is not a Weibull.

\renewcommand{\arraystretch}{1}
\begin{sidewaystable}
\centering
\caption{Estimated percent bias and coverage probability for $\bfbeta_1$ and $\theta$ for six analyses described in Section \ref{sec:sim:details}, across six simulation scenarios given in Table \ref{tab:simSetting}. Throughout values are based on results from $R$=500 simulated datasets. \label{tab:simReg}}
\scalebox{0.85}{
\begin{tabular}{ccrc  cccccc  ccccccc}
  \hline
	 	&&   && \multicolumn{6}{c}{Percent Bias} && \multicolumn{6}{c}{Coverage Probability} \\\cline{5-10}\cline{12-17}   
Scenario	 	&& 	\mc{True}	&& Weibull & Weibull & Weibull & PEM & PEM & Spline && Weibull & Weibull & Weibull & PEM & PEM & Spline \\
	 && \mc{value}	&& -MVN & -DPM &-SF & -MVN& -DPM & -SF && -MVN & -DPM &-SF & -MVN& -DPM & -SF \\	 
  \hline
	 &$\beta_{11}$ & 0.50 && 0.1 & 0.2 & -19.8 & 0.4 & 0.4 & -21.0 && 0.96 & 0.96 & 0.01 & 0.95 & 0.96 & 0.00 \\ 
 1	 & $\beta_{12}$& 0.80 && 0.2 & 0.3 & -19.7 & 0.5 & 0.4 & -21.0 && 0.95 & 0.95 & 0.00 & 0.96 & 0.97 & 0.00 \\ 
 	& $\beta_{13}$& -0.50 && 0.3 & 0.3 & -19.8 & 0.3 & 0.3 & -21.2 && 0.97 & 0.96 & 0.31 & 0.96 & 0.96 & 0.25 \\ 
 	 & $\theta$	& 0.50 && 1.0 & 1.3 &  & 1.4 & 1.2 &  		&& 0.95 & 0.95 &  & 0.93 & 0.94 &  \\ 
   \hline	 
	 &$\beta_{11}$ & 0.50 && -0.1 & -0.0 & -31.8 & 0.1 & 0.1 & -32.8 && 0.94 & 0.94 & 0.00 & 0.94 & 0.93 & 0.00 \\ 
 2	 & $\beta_{12}$& 0.80 && 0.1 & 0.2 & -31.7 & 0.4 & 0.3 & -32.7 && 0.97 & 0.97 & 0.00 & 0.94 & 0.95 & 0.00 \\ 
 	& $\beta_{13}$& -0.50 && 1.2 & 1.3 & -31.1 & 1.1 & 1.1 & -32.2 && 0.94 & 0.95 & 0.05 & 0.94 & 0.94 & 0.04 \\ 
 	 & $\theta$	& 1.00 && 0.4 & 0.7 &  & 0.7 & 0.6 &  		&& 0.94 & 0.95 &  & 0.94 & 0.95 &  \\ 
   \hline  
	 &$\beta_{11}$ & 0.50 && 0.3 & 0.3 & -19.9 & 0.7 & 0.7 & -21.0 && 0.94 & 0.94 & 0.00 & 0.93 & 0.94 & 0.00 \\ 
 3	 & $\beta_{12}$& 0.80 && 0.4 & 0.4 & -19.8 & 0.8 & 0.8 & -20.9 && 0.94 & 0.94 & 0.00 & 0.94 & 0.94 & 0.00 \\ 
 	& $\beta_{13}$& -0.50 && 0.4 & 0.3 & -20.1 & 0.5 & 0.6 & -21.2 && 0.96 & 0.96 & 0.31 & 0.95 & 0.96 & 0.27 \\ 
 	 & $\theta$	& 0.50 && 2.0 & 2.1 &  & 3.2 & 3.2 &  		&& 0.96 & 0.96 &  & 0.93 & 0.95 &  \\ 
\hline	    
	 &$\beta_{11}$ & 0.50 && 3.7 & 3.7 & ~~0.2 & 4.7 & 4.6 & ~~0.3 && 0.87 & 0.86 & 0.96 & 0.81 & 0.83 & 0.96 \\ 
 4	 & $\beta_{12}$& 0.80 && 3.6 & 3.6 & ~-0.0 & 4.5 & 4.5 & ~~0.1 && 0.80 & 0.79 & 0.95 & 0.69 & 0.70 & 0.95 \\ 
 	& $\beta_{13}$& -0.50 && 4.0 & 4.0 & ~~0.2 & 4.8 & 4.7 & ~~0.2 && 0.93 & 0.94 & 0.94 & 0.93 & 0.93 & 0.93 \\ 
 	 & $\theta$	& 0.00 &&  &  &  & &  &  		&&  &  &  &  &  &  \\ 
   \hline  
	 &$\beta_{11}$ & 0.50 && -0.3 & 0.1 & -20.3 & 0.0 & 0.3 & -21.1 && 0.94 & 0.95 & 0.00 & 0.96 & 0.96 & 0.00 \\ 
 5	 & $\beta_{12}$& 0.80 && 0.0 & 0.3 & -20.0 & 0.3 & 0.6 & -20.9 && 0.95 & 0.95 & 0.00 & 0.96 & 0.96 & 0.00 \\ 
 	& $\beta_{13}$& -0.50 && -0.2 & 0.2 & -20.4 & -0.2 & 0.2 & -21.3 && 0.94 & 0.94 & 0.29 & 0.94 & 0.94 & 0.25 \\ 
 	 & $\theta$	& 0.50 && -0.2 & 1.0 &  & 0.4 & 1.3 &  			&& 0.95 & 0.95 &  & 0.95 & 0.96 &  \\ 
   \hline  
	 &$\beta_{11}$ & 0.50 && 9.3 & 9.4 & -22.1 & 0.4 & 0.3 & -25.9 && 0.58 & 0.57 & 0.00 & 0.94 & 0.94 & 0.00 \\ 
 6	 & $\beta_{12}$& 0.80 && 9.7 & 9.8 & -22.0 & 0.5 & 0.5 & -25.8 && 0.20 & 0.20 & 0.00 & 0.94 & 0.95 & 0.00 \\ 
 	& $\beta_{13}$& -0.50 && 10.2 & 10.2 & -21.6 & 0.8 & 0.7 & -26.1 && 0.81 & 0.80 & 0.21 & 0.93 & 0.94 & 0.10 \\ 
 	 & $\theta$	& 0.50 && 52.8 & 53.0 &  & 1.8 & 1.7 &  			&& 0.00 & 0.00 &  & 0.95 & 0.96 &  \\ 
   \hline
\end{tabular}}
\end{sidewaystable}

\begin{table}[ht]
\centering
\caption{Average relative width of 95\% credible/confidence intervals for $\bfbeta_1$ and $\theta$, with the Weibull-MVN model taken as the referent, across six simulation scenarios given in Table \ref{tab:simSetting}. Throughout values are based on results from $R$=500 simulated datasets. \label{tab:simWidths}}
\scalebox{0.9}{
\begin{tabular}{cc cccccc}
  \hline
Scenario	&& Weibull & Weibull & Weibull & PEM & PEM & Spline \\
	 	&& -MVN & -DPM &-SF & -MVN& -DPM & -SF \\	 
  \hline
	& $\beta_{11}$ 	& 1.00 & 1.00 & 0.81 & 1.02 & 1.02 & 0.81 \\ 
 1	& $\beta_{12}$ & 1.00 & 1.00 & 0.77 & 1.04 & 1.04 & 0.77 \\ 
 	& $\beta_{13}$ 	& 1.00 & 1.00 & 0.84 & 1.00 & 1.01 & 0.83 \\ 
 	& $\theta$	 	& 1.00 & 1.00 &  & 1.10 & 1.12 &  \\ 
    \hline  
	& $\beta_{11}$ 	 & 1.00 & 1.00 & 0.73 & 1.02 & 1.02 & 0.73 \\ 
 2	& $\beta_{12}$ & 1.00 & 1.00 & 0.69 & 1.03 & 1.04 & 0.69 \\ 
 	& $\beta_{13}$ 	& 1.00 & 1.00 & 0.76 & 1.00 & 1.00 & 0.76 \\ 
 	& $\theta$	 	& 1.00 & 1.00 &  & 1.12 & 1.14 &  \\ 
    \hline  
	& $\beta_{11}$ 	& 1.00 & 1.00 & 0.81 & 1.02 & 1.02 & 0.81 \\ 
 3	& $\beta_{12}$  & 1.00 & 1.00 & 0.76 & 1.04 & 1.04 & 0.77 \\ 
 	& $\beta_{13}$ 	& 1.00 & 1.00 & 0.83 & 1.00 & 1.01 & 0.83 \\ 
 	& $\theta$	 	& 1.00 & 1.00 &  & 1.10 & 1.13 &  \\ 
	\hline    
	& $\beta_{11}$ 	& 1.00 & 1.00 & 0.95 & 1.02 & 1.01 & 0.96 \\ 
 4	& $\beta_{12}$ & 1.00 & 1.00 & 0.94 & 1.03 & 1.03 & 0.95 \\ 
 	& $\beta_{13}$ 	& 1.00 & 1.00 & 0.96 & 1.01 & 1.01 & 0.96 \\ 
 	& $\theta$	 	& 1.00 & 1.00 &  & 1.09 & 1.09 &  \\ 
    \hline  
	& $\beta_{11}$ 	& 1.00 & 1.00 & 0.81 & 1.02 & 1.02 & 0.81 \\ 
 5	& $\beta_{12}$ & 1.00 & 1.00 & 0.77 & 1.03 & 1.03 & 0.77 \\ 
 	& $\beta_{13}$ 	& 1.00 & 1.00 & 0.83 & 1.00 & 1.00 & 0.83 \\ 
 	& $\theta$	 	& 1.00 & 1.00 &  & 1.09 & 1.09 &  \\ 
    \hline  
	& $\beta_{11}$ 	& 1.00 & 1.00 & 0.74 & 0.94 & 0.95 & 0.73 \\ 
 6	& $\beta_{12}$ & 1.00 & 1.00 & 0.72 & 0.96 & 0.97 & 0.71 \\ 
 	& $\beta_{13}$ 	& 1.00 & 1.00 & 0.76 & 0.93 & 0.93 & 0.75 \\ 
 	& $\theta$	 	& 1.00 & 1.00 &  & 0.89 & 0.90 &  \\ 
   \hline
\end{tabular}}
\end{table}

While Table \ref{tab:simReg} explores estimation and the (valid) quantification of uncertainty, Table \ref{tab:simWidths} examines the relative merits of the various analysis approaches in terms of efficiency. Specifically, we computed the average relative width of 95\% credible/confidence intervals for $\bfbeta_1$ and $\theta$ under each analysis with the Weibull-MVN model taken as a common referent. Comparing the Weibull-DPM to the Weibull-MVN as well as the results between the PEM-MVN and PEM-DPM we see that there is no loss of efficiency for any of the regression parameters, and minimal loss for $\theta$, if one adopts the flexible DPM specification for the cluster-specific random effects, even if the true distribution is a MVN. Under all five scenarios for which the true baseline hazard functions were Weibull hazard functions, the two models that adopt a PEM specification have somewhat wider credible intervals particularly for $\theta$. However, as expected, the 95\% credible intervals for the two PEM models under scenario 6 are somewhat tighter indicating improved efficiency when the true baseline hazard functions are not Weibull hazard functions. Finally, across all scenarios, the estimated 95\% confidence intervals for the two SF models are substantially tighter than those for any of the proposed analyses, although this must be balanced with the high bias shown in Table \ref{tab:simReg}.

\subsubsection{Cluster-specific random effects}

Finally, Table \ref{tab:simMSEP} investigates the relative performance of the various analyses with respect to estimation of the cluster-specific random effects. Specifically, we calculated the mean squared error of prediction (MSEP) given by:
\be
	\frac{1}{RJ}\sum_{r=1}^R\sum_{j=1}^J (\hat{V}_{rjg} - V_{rjg})^2,
\ee
where $V_{rjg}$ is the cluster-specific random effect for the $j^{\textrm{th}}$ cluster in the transition $g$ for the $r^{\textrm{th}}$ simulated data set, $r$=1,\ldots,$R$. For each of the four proposed models, $\hat{V}_{rjg}$ was taken as the corresponding posterior median. For the two SF models, $\hat{V}_{rjg}$ was taken as a the log of the empirical Bayes estimates of the transition/cluster-specific frailties (see Supplementary Materials C for details). We note, however, that for some of the simulated datasets, the empirical Bayes estimates returned by the current implemented in the \texttt{frailtypack} package were zero. Since taking the log of these estimates would yield $\hat{V}_{rjg}$ = $-\infty$, we calculated MSEP over the random effects for which the empirical Bayes estimate was non-zero; to place these values in context, Table \ref{tab:simMSEP} also reports the percentage of instances where a frailty was estimated to be zero.

\begin{table}[ht]
\centering
\caption{Mean squared error of prediction ($\times 10^{-2}$) for cluster-specific random effects based on six analyses described in Section \ref{sec:sim:details}, across six data scenarios given in Table \ref{tab:simSetting}. Throughout values are based on results from $R$=500 simulated datasets. \label{tab:simMSEP}}
\scalebox{0.90}{
\begin{tabular}{cc r r rr r r rr}
  \hline
Scenario	 & 		& \mc{Weibull} & \mc{Weibull} & \multicolumn{2}{c}{Weibull}& \mc{PEM} 	& \mc{PEM} & \multicolumn{2}{c}{Spline}\\ 
		 & 		& \mc{-MVN} & \mc{-DPM} & \multicolumn{2}{c}{-SF} & \mc{-MVN} 	& \mc{-DPM} &\multicolumn{2}{c}{-SF}\\ 
		 & 		&&  & & \%F$^{\dag}$ && & & \%F\\ 		 
  \hline
	 &$V_1$ 		& 5.25 & 5.27 & 6.40 &  						& 5.27 & 5.27 & 6.39 &  \\
1	 & $V_2$		& 7.66 & 7.70 & 8.70 & 17.8  					& 7.67 & 7.72 & 8.68 & 0.2   \\
	& $V_3$ 		& 9.91 & 9.95 & 12.13 &   						& 9.91 & 9.96 & 12.11 &    \\
\hline  
	 &$V_1$ 		& 6.36 & 6.41 & 8.10 &   						& 6.37 & 6.41 & 8.09 &    \\
2	 & $V_2$		& 8.76 & 8.85 & 10.23 & 10.4  					 & 8.77 & 8.86 & 10.20 & 0.0   \\
	& $V_3$ 		& 11.13 & 11.19 & 13.85 &   					& 11.13 & 11.19  & 13.91 &    \\
\hline  
	 &$V_1$ 		& 5.03 & 5.04 & 6.27 &   						& 5.04 & 5.04  & 6.22 &    \\
3	 & $V_2$		 & 6.34 & 6.34 & 8.28 & 15.8  					 & 6.36 & 6.36 & 8.24 & 0.0   \\
	& $V_3$ 		& 7.55 & 7.49 & 11.66 &   					& 7.57 & 7.55 & 11.69 &    \\
\hline  
	 &$V_1$ 		& 3.84 & 3.85 & 4.99 &  						& 3.87 & 3.87 	& 5.01 &    \\
4	 & $V_2$		& 6.25 & 6.27 & 7.19 & 12.8  					& 6.25 & 6.27 & 7.12 & 0.4   \\
	& $V_3$ 		& 7.89 & 7.90 & 9.57 &   						& 7.90 & 7.91 & 9.52 &    \\
\hline  
	 &$V_1$ 		& 6.95 & 6.26 & 10.87 &   						& 6.96 & 6.27 & 10.86 &    \\
5	 & $V_2$		& 11.52 & 10.50 & 14.95 & 12.8  				& 11.50 & 10.52 & 14.92 & 0.2   \\
	& $V_3$ 		& 15.46 & 14.66 & 25.04 &   					& 15.46 & 14.72  & 24.94 &    \\
\hline  
	 &$V_1$ 		& 5.05 & 5.01 & 6.34 &   						 & 4.89 & 4.85  & 6.26 &    \\
6	 & $V_2$		& 7.58 & 7.55 & 8.60 & 5.4  					& 7.41 & 7.39 & 8.49 & 1.4   \\
	& $V_3$ 		& 6.72 & 6.65 & 13.42 &   						& 6.44 & 6.40 & 13.70 &    \\
\hline	
\multicolumn{10}{l}{\footnotesize $^\dag$ \% of times SF models yield at least one of $\hat{\bfV}_j$ being $-\infty$, resulting in MSEP being $\infty$}
\end{tabular}}
\end{table}

From Table \ref{tab:simMSEP} we see that under scenarios 1-4, for which the true model is a Weibull-MVN model, the Weibull-MVN analysis generally performs the best. Comparing the Weibull-MVN and PEM-MVN results across these scenarios, we see that over-specification of the baseline hazard functions (i.e. adoption of the more flexible PEM specification) does not meaningfully impact MSEP. In addition, comparing the Weibull-MVN and Weibull-DPM results we see that over-specification of the random effects structure (i.e. adoption of the more flexible DPM specification) does not adversely affect MSEP either. When the true distribution of the random effects is not a multivariate Normal distribution, however, as in scenario 5, both the Weibull-DPM and PEM-DPM models outperform their MVN counterparts, illustrating the potential benefit of the more flexible DPM specification. Furthermore, when the true baseline hazard functions do not correspond to a Weibull distribution, the MSEP for the two PEM models are, as expected, smaller than the corresponding values for the two Weibull models, illustrating the potential benefit of the more flexible PEM specification. Finally, we find that the empirical Bayes estimates of the cluster-specific random effects from the SF models perform relatively poorly when compared to the corresponding estimates from the proposed methods. For example, the Spline-SF model yields approximately 14\% to 55\% higher MSEP than our proposed PEM-MVN model across the six scenarios.

\section{Analysis of Medicare Data}\label{sec:application}

\subsection{Analysis details and prior specifications}\label{sec:application:details}

Returning to the motivating application of readmissions following a diagnosis of pancreatic cancer, we fit each of the four models summarized in Table \ref{tab:four:models} to the Medicare data under both the Markov and semi-Markov assumption for $h_3(\cdot)$ (see Section \ref{sec:framework:model}). Based on the rationale provided in Section \ref{sec:data}, we administratively censored observation time at 90 day. Given the results from the simulation studies, specifically with respect to estimation of the cluster-specific random effects, we decided not to fit the shared frailty models of \cite{liquet2012investigating}. We did, however, perform an analysis based on a LN-GLMM model since this model is the current standard for analyzing variation in the risk of readmission and we believed it would be instructive to examine the potential impact of ignoring death as a competing force. Towards this, let $Y^*_{ji}$ = 0/1 be a binary indicator of whether or not the $i^{th}$ patient in the $j^{th}$ hospital readmitted within 90 days of discharge. Note, if a patient died prior to readmission within 90 days their outcome was set to $Y^*_{ji}$ = 0. The LN-GLMM is then given by:
\be
\label{model:ln:glmm}
	\mbox{logit} \Pr(Y^*_{ji} = 1| \bfX^*_{ji}, V^*_j)\ =\ \bfX^{*T}_{ji}\bfbeta^* + V^*_j
\ee
\noindent where $V^*_j$ is a hospital-specific random effect for readmission taken to be Normally distributed with mean zero and a constant variance, $\sigma_v^2$. To complete the Bayesian specification of this model, we adopted a Gamma(0.7, 0.7) prior for the precision $\sigma_v^{-2}$. For the four proposed models, the hyperparameters outlined in Table \ref{tab:four:models} are specified as in Section \ref{sec:sim:details}

Throughout the analyses, to ensure the baseline hazard functions in the proposed models and the (overall) intercept in the LN-GLMM retained reasonable interpretations, age was standardized so that `zero' corresponded to age 77 years and a one-unit increment corresponded to a 10-year contrast. Furthermore, length of stay during the initial hospitalization was also standardized so that `zero' corresponded to 10 days and a one-unit increment corresponded to a 7-day contrast. 

\subsection{MCMC}\label{sec:application:MCMC}

Towards obtaining summaries of the joint posterior distributions we ran 3 independent chains of the proposed MCMC scheme, each for a total of 6 million scans. Convergence was evaluated by inspection of trace plots as well as calculation of the PSR statistic; an MCMC scheme was determined to have converged if the PSR statistic was less than 1.05 for all parameters in the model (see Supplementary Materials E). Although the hierarchical models are complex and include a large number of parameters, the proposed algorithm achieved an overall acceptance rate of 35\% across the various Metropolis-Hastings and Metropolis-Hastings-Green steps. To provide a sense of computational time, the most complex of our proposed models (i.e. the PEM-DPM model), the implementation in our \texttt{R} package is able to generate 1 million scans in 30 minutes on a 2.5 GHz Intel Core i7 MacBook Pro; for the least complex of the models (i.e. the Weibull-MVN model), the implementation is able to generate 1 million scans in 10 minutes on the same machine.

\subsection{Results}\label{sec:application:results}

\subsubsection{Overall model fit}

Table \ref{tbl:dic} provides DIC and LPML for the eight model fits. For the DIC measure, a general rule of thumb for model comparison is to consider differences of less than 2 to be negligible, differences between 2 and 6 to indicative of positive support for the model with the lower value and differences greater than 6 to be strong support in value of the model with the lower value~\citep{spiegelhalter2002bayesian, millar2009comparison}. For LPML, one can compute the so-called pseudo Bayes factor (PBF) for two models by exponentiating difference in their LPML values~\citep{hanson2006inference}. While the conventional Bayes factor \citep{kass1995bayes} tends to find which model explains the observed data best, predictive methods such as PBF attempt to find which model gives the best predictions for future observations when the same process as the original data is used to generate the observations \citep{kadane2004methods}

\begin{table}[ht]
\centering
\caption{DIC and LPML for eight models fit to the New England Medicare data.\label{tbl:dic}}
\begin{tabular}{llcc}
\hline
			&				& DIC 		& LPML 				\\
\hline			
			&	Weibull-MVN	& 46184.3		&  -23101.6 \\
Markov 		& 	Weibull-DPM	& 46174.1 	&  -23101.2 \\
			&	PEM-MVN	& 45609.2		&  -22812.6 \\
			&	PEM-DPM 	& 45606.8 	&  -22810.7 \\
\hline			
			&	Weibull-MVN	& 46163.7 	&  -23088.8 	\\
semi-Markov 	& 	Weibull-DPM    & 46153.0 	&  -23086.7 	\\
			&	PEM-MVN	& 45574.1 	&  -22790.9 	\\
			&	PEM-DPM 	& 45569.0		&  -22789.3	 \\			
\hline
\end{tabular}
\end{table}

Based on pairwise comparisons of the values in Table \ref{tbl:dic} we draw a number of conclusions. First, each of the models in which a semi-Markov specification is made for $h_{03}(\cdot)$ has a substantially better fit to the data than the corresponding model in which a Markov assumption is made for $h_{03}(\cdot)$; differences in DIC and the PBF range between 20.6-37.8 and the order of $10^5$-$10^9$, respectively. Second, both DIC and LPML indicate that models for which a PEM specification was adopted for the baseline hazard functions have substantially better fit to the data than models for which a Weibull hazard function was adopted; differences in DIC and the PBF range between 567.3-589.6 and the order of $10^{125}$-$10^{129}$, respectively. Finally, although DIC indicates a somewhat better fit for models that adopt a DPM for the random effects distribution compared to a MVN specification, the LPML values are less convincing in this regard; differences in DIC and the PBF range between 2.4-10.7 and 1.5-8.2, respectively. 

\subsubsection{Baseline survival functions}

\begin{figure}[h!]
\centering
\includegraphics[width = 6in]{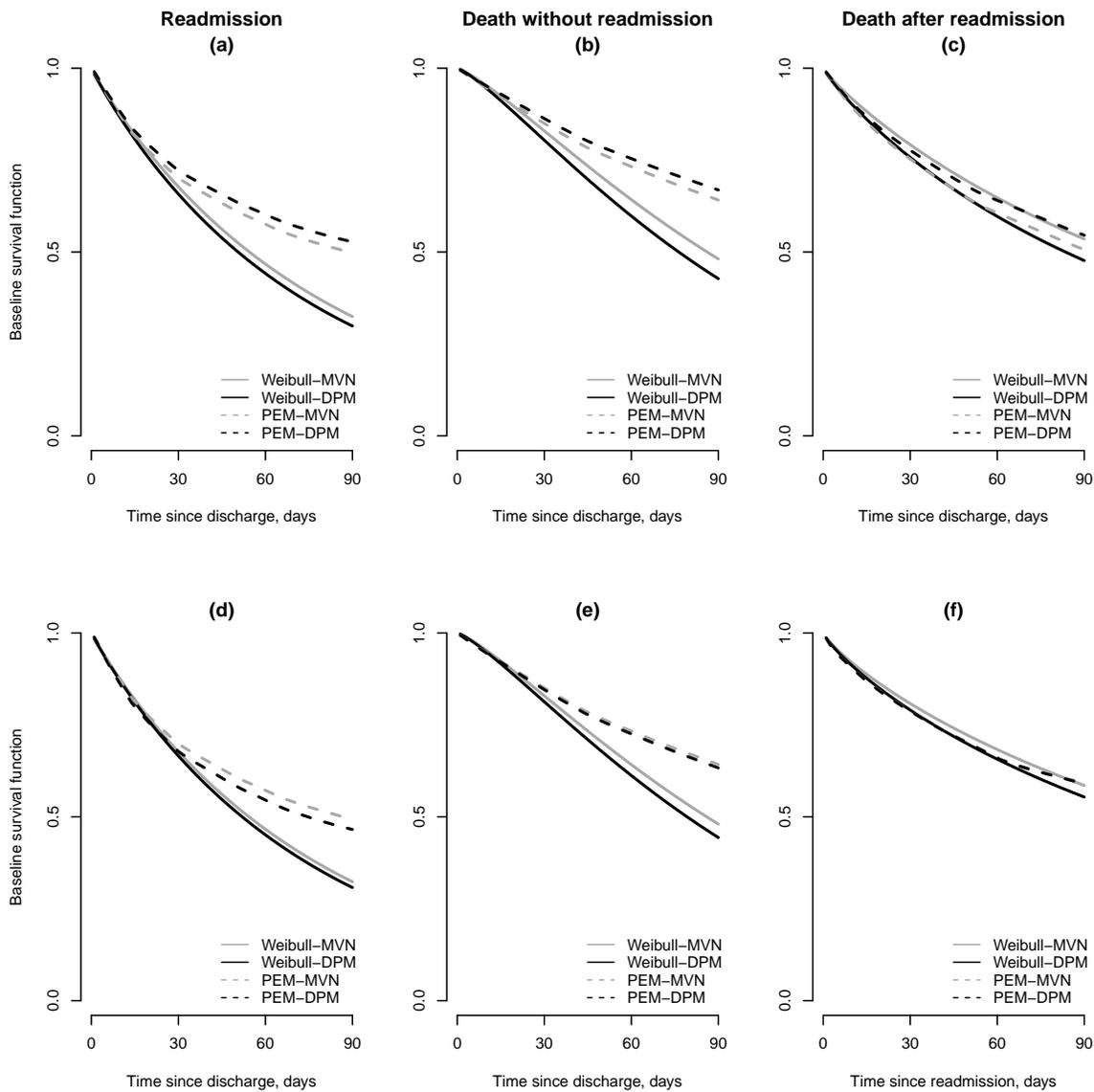}
\caption{Estimated transition-specific baseline survival functions, $S_{0g}(\cdot)$ = $\exp(-H_{0g}(\cdot))$, for the eight models fit to the New England Medicare data (see Table \ref{tbl:dic}). Panels (a)-(c) are estimates based on a Markov specification for $h_{03}(\cdot)$; panels (d)-(f) are estimates based on a semi-Markov specification. Note the difference in time scale between panels (c) and (d) due the differing specifications for $h_{03}(\cdot)$. \label{fig:bs}}
\end{figure}

Since hazard functions are notoriously difficult to interpret, Figure \ref{fig:bs} provides estimates of the corresponding baseline survival functions. Specifically, they provide pointwise time-specific posterior medians for $S_{0g}(\cdot)$ for a 77-year old white female patient who had a Charlson/Deyo comorbidity score of 0 or 1, whose initial hospitalization lasted 10 days and during which they had no pancreatic cancer-related procedures, and were eventually discharged to their own home. In panels (a)-(c) results are presented for models for which a Markov assumption was adopted for $h_{03}(\cdot)$; panels (d)-(f) present results for models for which a semi-Markov assumption was adopted.

From panels (a) and (d) we see that all eight models indicate similar risk of readmission within the first 30 days. After 30 days, however, models with a parametric Weibull specification for the baseline hazard functions indicate substantially higher overall risk for readmission. Note, most of the observed readmission events occur relatively soon after discharge with a median of 18 days and 75\% of observed events occurring within 40 days. As such, the posterior mass is being assigned to values of the two Weibull hyperparameters, ($\alpha_g$, $\kappa_g$), that fit the early time periods well to the detriment of fitting late periods relatively poorly. From panels (b) and (e) a similar phenomenon is observed for the baseline survival function for death without readmission for which the median event time is 20 days and, again, approximately 75\% of observed events occurring within 40 days. In contrast, since the distribution of time to death following readmission is more spread out (median=43 days, IQR=40 days) the estimated baseline survival functions under the Weibull and PEM are more similar (see panels (c) and (f)).

\subsubsection{Regression parameters}

Posterior summaries for the vector of hazard ratio (HR) parameters for readmission, $\exp(\bfbeta_1)$, are presented in Table \ref{tab:results:fixed}. For brevity, based in part on the conclusions drawn from Table \ref{tbl:dic}, results are only presented for models for which a semi-Markov specification was adopted for $h_{03}(\cdot)$; additional results, particularly for $\exp(\bfbeta_2)$ and $\exp(\bfbeta_3)$ are provided in Supplementary Materials E. In addition, posterior summaries for the vector of odds ratio (OR) parameters, $\exp(\bfbeta^*)$ from model (\ref{model:ln:glmm}), are also presented.

\begin{sidewaystable}
\centering
\caption{Posterior summaries (medians and 95\% credible intervals (CI)) for hazard ratio (HR) parameters for readmission, $\exp(\bfbeta_1)$, from semi-competing risks data analyses and odds ratios (OR) based on a LN-GLMM. For the latter, results are reported for models for which a semi-Markov specification was adopted for $h_{03}(\cdot)$ \label{tab:results:fixed}}
\scalebox{0.9}{
\begin{tabular}{ l c c c c c}
  \hline
  	& LN-GLMM 	&	 Weibull-MVN 	& Weibull-DPM &  PEM-MVN & PEM-DPM \\
	& OR (95\% CI)  		 & HR (95\% CI) 	& HR (95\% CI)  & HR (95\% CI)  & HR (95\% CI)  \\ 
  \hline
Sex 					&&& &&\\
~~~Male 				& 1.00 & 1.00& 1.00 & 1.00 & 1.00\\
~~~Female 			& 0.91 (0.80, 1.03) & 0.80 (0.70, 0.90) & 0.79 (0.70, 0.90) & 0.85 (0.76, 0.96) & 0.85 (0.76, 0.95) \\ 
Age$^\dag$				& 0.87 (0.83, 0.92) & 0.90 (0.86, 0.95) & 0.90 (0.86, 0.95) & 0.91 (0.87, 0.94) & 0.91 (0.87, 0.95) \\ 
Race 				&&&&&\\
~~~White				& 1.00 & 1.00& 1.00 & 1.00 & 1.00\\  
~~~Non-white			& 1.17 (0.90, 1.51) & 1.12 (0.86, 1.45) & 1.12 (0.86, 1.46) & 1.11 (0.89, 1.40) & 1.12 (0.89, 1.38) \\ 
Source of entry to initial hospitalization && &&&\\
~~~Emergency room	& 1.00 & 1.00& 1.00 & 1.00 & 1.00\\
~~~Other facility		& 0.99 (0.86, 1.14) & 1.18 (1.03, 1.36) & 1.19 (1.03, 1.35) & 1.12 (1.00, 1.27) & 1.13 (1.00, 1.28) \\ 
Charlson/Deyo score 	&& &&&\\
~~~$\leq 1$ 			& 1.00 & 1.00& 1.00 & 1.00 & 1.00\\
~~~$> 1$ 				& 1.31 (1.05, 1.64) & 1.50 (1.19, 1.85) & 1.49 (1.20, 1.85) & 1.41 (1.15, 1.68) & 1.40 (1.15, 1.70) \\ 
Procedure during hospitalization			&&&&&\\
~~~No 		 		& 1.00 & 1.00& 1.00 & 1.00 & 1.00\\
~~~Yes 				& 0.73 (0.61, 0.86) & 0.45 (0.38, 0.53) & 0.44 (0.37, 0.54) & 0.56 (0.48, 0.66) & 0.57 (0.48, 0.66) \\ 
Length of stay$^*$ 	& 1.14 (1.06, 1.22) & 1.15 (1.07, 1.24) & 1.15 (1.07, 1.24) & 1.12 (1.05, 1.18) & 1.12 (1.05, 1.19) \\ 
Discharge location 	&& &&&\\
~~~Home without care 	& 1.00  & 1.00& 1.00 & 1.00 & 1.00\\
~~~Home with care 		& 0.68 (0.58, 0.79) & 0.95 (0.82, 1.11) & 0.96 (0.82, 1.11) & 0.90 (0.79, 1.02) & 0.89 (0.78, 1.02) \\ 
~~~Hospice 			& 0.06 (0.04, 0.10) & 0.39 (0.22, 0.62) & 0.39 (0.22, 0.62) & 0.27 (0.15, 0.45) & 0.27 (0.16, 0.43) \\ 
~~~ICF/SNF	 		& 0.44 (0.36, 0.53) & 0.88 (0.72, 1.07) & 0.88 (0.73, 1.08) & 0.76 (0.64, 0.90) & 0.76 (0.65, 0.90) \\ 
~~~Other 				& 0.56 (0.41, 0.76) & 1.04 (0.75, 1.42) & 1.06 (0.75, 1.44) & 0.91 (0.70, 1.18) & 0.90 (0.68, 1.18) \\ 
\hline
   	\multicolumn{6}{l}{\footnotesize $^\dag$ Standardized so that 0 corresponds to an age of 77 years and so that a one unit increment corresponds to 10 years}\\
   	\multicolumn{6}{l}{\footnotesize $^*$ Standardized so that 0 corresponds to 10 days and so that a one unit increment corresponds to 7 day} \\
\end{tabular}}
\end{sidewaystable}

Recognizing that the interpretation of the HR and OR parameters differ (due to the different set of frailties/random effects that are conditioned upon), the results in Table \ref{tab:results:fixed} indicate that the LN-GLMM qualitatively identifies a different set of risk factors for readmission than the results based on the proposed framework. For instance, while there is evidence of lower risk for readmission among females diagnosed with pancreatic cancer under the semi-competing risks approach (e.g. HR 0.80; 95\% CI 0.70, 0.90 in Weibull-MVN), one cannot draw the same conclusion based on the LN-GLMM (OR 0.91; 95\% CI 0.80, 1.03). In addition, under the LN-GLMM model there is no evidence of a relationship between source of entry to the initial hospitalization (OR 0.99; 95\% CI 0.86, 1.14) while under each of the semi-competing risks analysis models there is evidence that patients who enter the initial hospitalization via some route other than the emergency room are at higher risk of readmission (e.g. HR 1.12; 95\% CI 1.00, 1.28). Conflicting results are also found with respect to discharge destination. In particular, under the LN-GLMM model patients who are discharged to home with care, a hospice, a ICF/SNF or some `other' facility (e.g. a rehabilitation center) have statistically significant lower estimates odds of readmission than patients discharged to their home without care. In contrast, results from the semi-competing risks analyses fail to indicate differences between patients discharged to home without care and those discharged to home with care (e.g. HR 0.90; 95\% CI 0.79, 1.02) or to some other facility. Furthermore, while patients discharged to either a hospice or ICF/SNF have significantly lower odds of readmission, the estimated effects are substantially attenuated (e.g. compare OR=0.06 under the LN-GLMM to HR=0.27 under the PEM-MVN model). Finally, consistent with the assessment of model fit in Table \ref{tbl:dic}, Table \ref{tab:results:fixed} indicates that for estimation and inference for regression parameters differs somewhat between models based on a Weibull baseline hazard specification and models based on a PEM specification. Comparing the Weibull-MVN model to the PEM-MVN model, for example, estimates for gender, Charlson/Deyo score and whether or not the patient underwent a procedure during the hospitalization are all attenuated; in contrast estimates for discharge location are generally strengthened under the PEM-MVN model, in some cases achieving statistical significance.

\subsubsection{Variance components}\label{subsec:varcomp}

Table \ref{tab:results:var} provides posterior summaries for the standard deviation of the patient-specific frailty distribution, $\sqrt{\theta}$, as well as components of the variance-covariance matrix for the hospital-specific $\bfV = (V_1, V_2, V_3)$ from models in which a semi-Markov specification was adopted for $h_{03}(\cdot)$. For the latter, the summaries are directly with respect to the components of $\bfSigma_V$ under a MVN specification; under the two DPM specifications, posterior summaries are reported for the marginal total variance-covariance matrix obtained by applying the law of total cumulance: $\sum_{j=1}^J$$\big\{\big(\bfmu_{m_j}-\bar{\bfmu}\big)\big(\bfmu_{m_j}-\bar{\bfmu}\big)^{\top} + \Sigma_{m_j}\big\}/J$, where $\bar{\bfmu} =\sum_{j=1}^J\bfmu_{m_j}/J$ \citep{ohlssen2007flexible}. From the Table we see that the components of variation (particularly the standard deviation components) are generally smaller in magnitude for models in which a PEM specification for the baseline hazard functions was adopted. For example, under the Weibull-MVN model the posterior median of $\sqrt{\theta}$ is 1.03, whereas the corresponding posterior median under the PEM-MVN model is 0.61. This is likely due to the $\gamma_{ji}$ patient-specific frailties not only representing patient-level heterogeneity but also accounting, in part, for misspecification of the Weibull model when the underlying baseline hazard functions are not Weibull. Qualitatively, across all four model specifications, we find that there is less variation across hospitals in the random effects specific to readmission compared to the random effects for mortality (either prior to or post-readmission); compare the posterior summaries for SD($V_{j1}$) to those of SD($V_{j2}$) and SD($V_{j2}$). Furthermore, while there is no evidence of correlation between hospital-specific random effects for readmission and corresponding random effects for mortality, there is some evidence of a positive correlation between hospital-specific random effects for mortality pre- and post-readmission, although the 95\% CIs each cover 0.  

\begin{table}[ht]
\centering
\caption{Posterior summaries (medians (PM) and 95\% credible intervals (CI)) for standard deviations (SD) of the underlying population distributions for the patient-specific frailties and hospital-specific random effects. Estimates of population correlation components, between hospital-specific random effects, are also provided. \label{tab:results:var}}
\begin{tabular}{l c c c c c}
\hline
  	& Weibull-MVN & Weibull-DPM &PEM-MVN & PEM-DPM \\
	& PM (95\% CI)  & PM (95\% CI) & PM (95\% CI)  & PM (95\% CI)  \\ 
  \hline
Patient-specific &&&& \\
frailties &&&& \\
~~$\sqrt{\theta}$ & 1.03 (0.94, 1.12) & 1.03 (0.95, 1.12) & 0.61 (0.50, 0.71) & 0.61 (0.49, 0.71) \\
Hospital-specific &&&& \\
random effects &&&& \\
~~SD($V_{j1}$)  &~0.26 (~0.20, 0.34) & ~0.27 (~0.21, 0.35) & ~0.25 (~0.19, 0.32) & ~0.25 (~0.20, 0.32) \\ 
~~SD($V_{j2}$)  &~0.37 (~0.28, 0.47) & ~0.37 (~0.28, 0.47) & ~0.32 (~0.25, 0.41) & ~0.32 (~0.25, 0.42) \\ 
~~SD($V_{j3}$)  &~0.37 (~0.27, 0.50) & ~0.37 (~0.27, 0.50) & ~0.33 (~0.25, 0.44) & ~0.33 (~0.25, 0.45) \\ 
~~corr($V_{j1}, V_{j2}$) &-0.04 (-0.40, 0.33)  & -0.04 (-0.40, 0.33) & -0.12 (-0.44, 0.23) & -0.12 (-0.45, 0.24) \\ 
~~corr($V_{j1}, V_{j3}$)  &~0.06 (-0.32, 0.42) & ~0.06 (-0.32, 0.43) & ~0.03 (-0.32, 0.38) & ~0.03 (-0.33, 0.39) \\ 
~~corr($V_{j2}, V_{j3}$) &~0.39 (-0.02, 0.67) & ~0.37 (-0.03, 0.67) & ~0.28 (-0.12, 0.59) & ~0.29 (-0.11, 0.59) \\ 
  \hline
\end{tabular}
\end{table}

\subsubsection{Hospital-specific random effects}

\begin{figure}
\centering
\includegraphics[width = 6in]{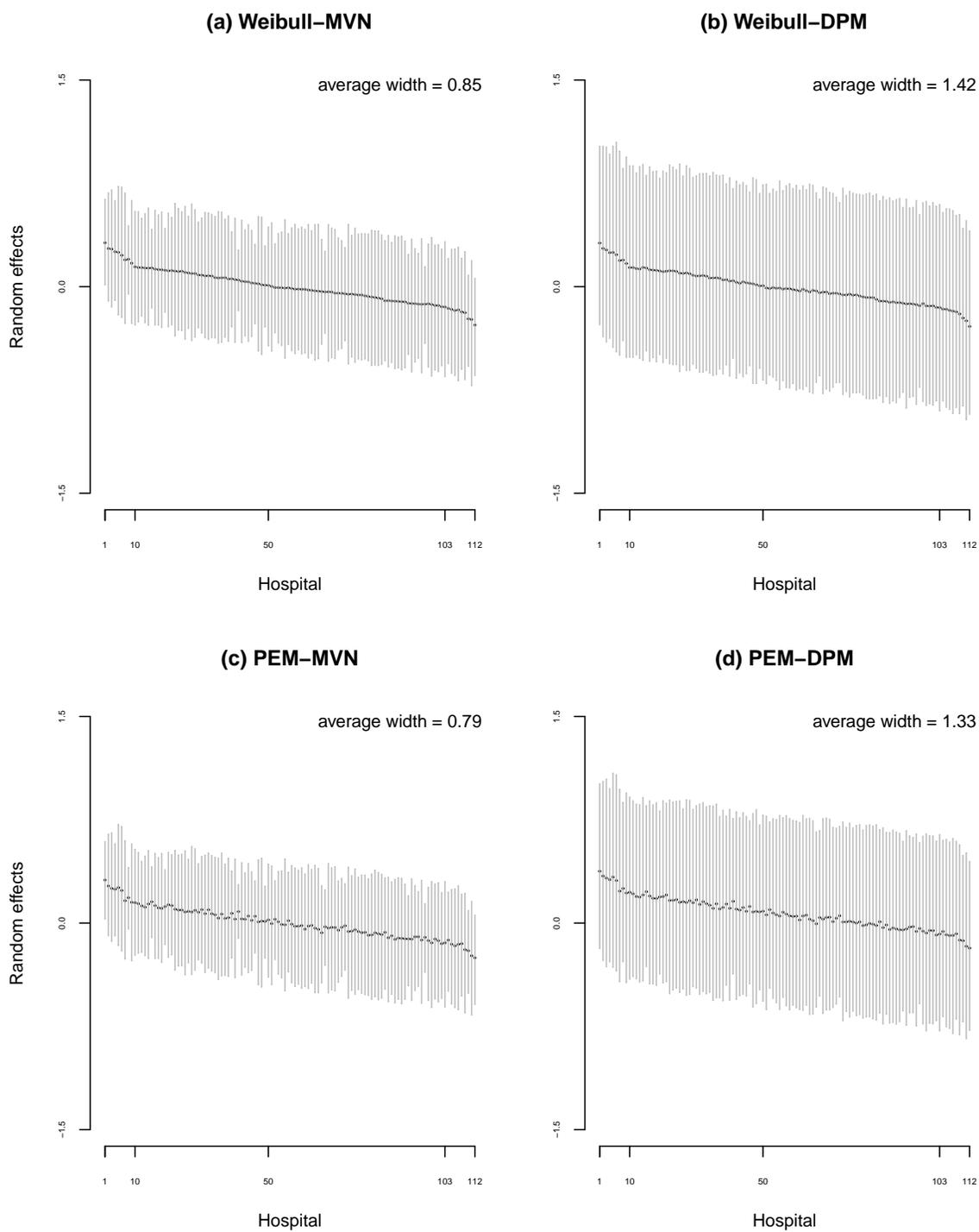}
\caption{Posterior summaries (median and 95\% credible interval) for the hospital-specific random effects for readmission, $V_{1j}$, under four models for which a semi-Markov specification for $h_{03}(\cdot)$ was adopted. In each panel, the hospitals are ordered according to the posterior median under the Weibull-MVN model.}\label{fig:RE}
\end{figure}

As noted, a key advantage of embedding the analysis of cluster-correlated semi-competing risks data in the Bayesian framework is the relatively straightforward nature of obtaining posterior summaries for the hospital-specific random effects themselves. Figure \ref{fig:RE} provides posterior medians and 95\% CIs for $V_{1j}$, $j$ = 1, $\ldots$, 112, based on the four models in which a semi-Markov specification is adopted for $h_{03}(\cdot)$. Note, across the four panels, the ordering of the hospitals is based on the magnitude of the posterior median under the Weibull-MVN model. Comparing the panels we see that posterior uncertainty for the $V_{1j}$ is generally greater under models that adopt a DPM for the hospital-specific $\bfV$ compared to those that adopt a MVN specification. This may not be surprising given the additional complexity of the DPM specification, although we do find that the more `complex' PEM specification for the baseline hazard functions yields lower posterior uncertainty than the Weibull specification. 

\subsubsection{Hospital-specific ranks}\label{subsec:ranks}

\begin{figure}
\centering
\includegraphics[width = 6in]{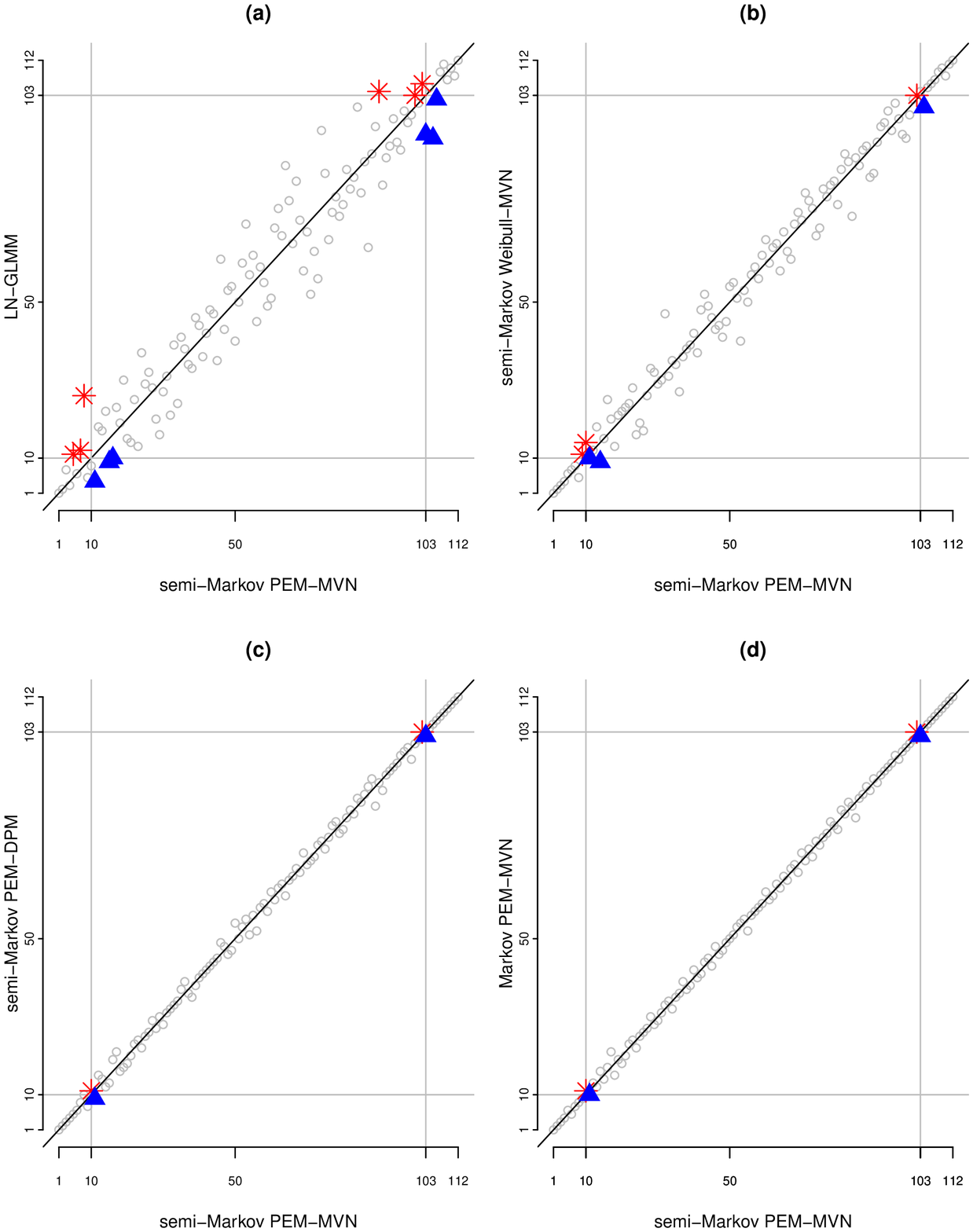}
\caption{Comparison of ranks of $J$=112 hospitals in the Medicare data on the basis of the posterior median for hospital-specific random effects for readmission, $V_{1j}$. Panels compare the ranks on the basis of one of four models to a referent set of ranks based on a PEM-MVN model with a semi-Markov specification for $h_{03}(\cdot)$ (see Section \ref{subsec:ranks} for details). Hospitals marked with a \ding{83} suffer under the given model compared to the referent under the referent semi-Markov PEM-MVN model; hospitals marked with a \ding{115} benefit.}\label{fig:rank}
\end{figure}

In addition to examining the absolute values of the hospital-specific $V_{1j}$, we also considered their rank ordering. Figure \ref{fig:rank} compares the ranks of the $J$=112 hospitals according to the posterior median of $V_{1j}$ under the PEM-MVN model with a semi-Markov specification for $h_{03}(\cdot)$ to the corresponding ranks based on four other models: (a) LN-GLMM; (b) Weibull-MVN with a semi-Markov specification for $h_{03}(\cdot)$; (c) PEM-DPM with a semi-Markov specification for $h_{03}(\cdot)$; and, (d) PEM-MVN model with a Markov specification for $h_{03}(\cdot)$. In each panel, the grey horizontal and vertical lines mark the `top 10' hospitals (i.e. ranks 1-10) and `bottom 10' hospitals (i.e. ranks 103-112).

From panel (a) we see that the correspondence between the ranks under a semi-Markov PEM-MVN model and those under a LN-GLMM it is far from exact. Crucially, from the lower-left portion of the panel, three hospitals that would have been ranked in the top 10 under the semi-Markov PEM-MVN model are ranked outside the top 10 under the LN-GLMM (specifically, those marked with a \ding{83}). Correspondingly there are three hospitals (marked with a \ding{115}) who are indicated as being in the top 10 under the LN-GLMM while the semi-competing risks analysis under the semi-Markov PEM-MVN would have ranked them outside the 10 top. Furthermore, from the top-right portion of the panel, three hospitals ranked in the bottom 10 under the semi-Markov PEM-MVN model are ranked above the bottom 10 under the LN-GLMM model (i.e. those marked with a \ding{115}).

From panels (b)-(d) we find that there is greater correspondence in the ranks of the 112 hospitals across the models within the proposed hierarchical framework. Comparing the ranks under the semi-Markov PEM-MVN specification to the semi-Markov Weibull-MVN specification in panel (b) we see that twos hospital that would have been ranked in the top 10 is now outside the top 10; there is also one hospital that is ranked in the bottom 10 under the semi-Markov PEM-MVN specification but outside the bottom 10 when a more restrictive Weibull model is used for the baseline hazard functions. Panels (c) and (d) are qualitatively similar in that the same two hospitals switch ranks at the lower end and the same two switch at the upper end; more generally, consistent with the conclusions we draw from Table \ref{tbl:dic}, there is very close correspondence in the ranks across the three models represented in these two panels.
 
\section{Sensitivity Analyses}\label{sec:sensitivity}

As outlined in Section \ref{sec:framework:priors} and Table \ref{tab:four:models}, the proposed Bayesian framework requires the specification of a number of hyperparameters. In practice comprehensive sensitivity analyses should be conducted to examine the extent to which conclusions are robust with respect to this specification, especially across key targets of estimation and inference. Here we focus our attention on the choice of hyperparameters for the prior distribution of $\Sigma_V$, the variance-covariance matrix of underlying population distributions for hospital-specific random effects, and their influence on estimation/inference for the random effects as well as $\tau$, the precision parameter in the DPM specification of the baseline hazard function. Towards this, we conducted sensitivity analyses based on the semi-Markov PEM-MVN and PEM-DPM models, specifying a range of values for ($\Psi_v$, $\rho_v$) and ($\Psi_0$, $\rho_0$) such that $\Psi_v$=$\Psi_0$=$\Psi^*(\rho^*-4) I_3$ and $\rho_v$=$\rho_0$=$\rho^*$, where, $\psi^*$=$0.01, 0.1, 1, 10$ and $\rho^*$=$5, 10, 50, 100$. Note, these specifications correspond to a prior distribution for $\Sigma_V$ with a mean of $\psi^* I_3$ and a variance of diagonal elements of $2\psi^{*2}/(\rho^*-6)$. 

Table \ref{tab:sen} presents the results. First, we focus Case I-IV, where $\psi^*$=1, $\rho^*$ = $100, 50, 10, 5$ which correspond to prior distributions of $\Sigma_V$ having a mean of $I_3$ and a standard deviation of diagonal elements of 0.15, 0.21, 0.71, 3.16; for Case IV we note that the (induced) prior standard deviation was calculated from 100,000 random draws from the prior. From the results we see that when the prior distribution is centered around the identity matrix the posterior assigns mass to smaller values of SD($V_{j1}$) as one increases the prior variance (dictated by decreasing $\rho^*$). This is likely due to the discrepancy between the actual variation on the cluster-specific random effects for $h_1()$ and the choice the identity matrix, $I_3$, as the prior mean for $\Sigma_V$ (since $\psi^*$=1) together with the strength attributed to that choice (i.e. the prior variance for $\Sigma_V$ dictated by $\rho^*$). When a prior mean of $I_3$ is chosen for $\Sigma_V$ together with a high value of $\rho^*$ the overall prior overcomes the information in the data such that the posterior for SD($V_{j1}$) is pushed `closer' to 1.0. As $\rho^*$ decreases, however, and less prior mass is given to $\Sigma_V$ = $I_3$, the likelihood is able to overcome the less informative prior so that the posterior is able to move away from the prior mean. Interestingly, based on both the DIC and LPML measures, we find that the overall fit of the data across Cases I-IV improves as the prior variance increases. We therefore interpret these results collectively as indicating that the variation across the (true underlying) $V_{j1}$ is meaningful but relatively small. Turning to Cases V and VI, we note that the induced prior distributions of $\Sigma_V$ are centered around relatively small values, specifically 0.01$I_3$ and 0.1$I_3$, with induced prior standard deviations of diagonal elements of 0.07 and 0.22, respectively. From the DIC and LPML values, these specifications of ($\Psi^*$, $\rho^*$) further improve the overall fit of the model from Case IV, we the posterior summaries for SD($V_{j1}$) again indicating that the variation in the $V_{j1}$ is relatively small.

\begin{table}[ht]
\centering
\caption{Sensitivity analyses for prior specification for hospital-specific random effects. We obtain the fits from PEM-MVN and in PEM-DPM conditioning on a range of values for ($\Psi_v $, $\rho_v$) and ($\Psi_0 $, $\rho_0$): $\Psi_v$=$\Psi_0$=$\Psi^*(\rho^*-4) I $ and $\rho_v$=$\rho_0$=$\rho^*$, where, $\psi^*$=$0.01, 0.1, 1, 10$ and $\rho^*$=$5, 10, 50, 100$. For each model DIC, LPML, and posterior medians (PM) and 95\% credible intervals (CI) for standard deviations (SD) of the underlying population distributions for hospital-specific random effects are reported, as well as the average posterior standard deviation of hospital-specific random effects estimates on readmission ($\bar{\sigma}_{V_{j1}}$). Also presented are the PM for $\tau$ under the PEM-DPM model. \label{tab:sen}}
\begin{tabular}{c c c c c c c c c c}
  \hline
\multirow{2}{*}{Case} & \multirow{2}{*}{($\Psi^*$, $\rho^*$)}	& \multirow{2}{*}{Model}	& \multirow{2}{*}{DIC} & \multirow{2}{*}{LPML} & SD($V_{j1}$) &  \multirow{2}{*}{$\bar{\sigma}_{V_{j1}}$} & $\tau$ \\
	&	&	&	&	& PM (95\% CI) &	& PM \\
  \hline
\multirow{2}{*}{I}	&\multirow{2}{*}{(1, 100)} 	&	MVN		& 45784.6 & -22905.0 & 0.77 (0.69, 0.85) & 0.36 &\\ 
			 	&					&	DPM		& 45779.3 & -22902.4 & 0.77 (0.70, 0.85) & 0.70 & 0.14 \\ \\
\multirow{2}{*}{II}	&\multirow{2}{*}{(1, 50)}	&	MVN		& 45758.2 & -22890.1 & 0.66 (0.59, 0.75) & 0.34 &\\ 
				&					&	DPM		& 45754.4 & -22888.0 & 0.66 (0.59, 0.75) & 0.72 & 0.14 \\ \\
\multirow{2}{*}{III}	&\multirow{2}{*}{(1, 10)} 	&	MVN		& 45642.5 & -22828.0 & 0.39 (0.33, 0.47) & 0.27 &\\ 
				&					&	DPM		& 45644.0 & -22828.7 & 0.39 (0.33, 0.47) & 0.47 & 0.14 \\ \\
\multirow{2}{*}{IV}	&\multirow{2}{*}{(1, 5)} 	&	MVN		& 45574.1 & -22790.9 & 0.25 (0.19, 0.32) & 0.20 &\\ 
				& 					&	DPM		& 45569.0 & -22789.3 & 0.25 (0.20, 0.32) & 0.32 & 0.15 \\ \\
\multirow{2}{*}{V}	&\multirow{2}{*}{(0.01, 5)}	&	MVN		& 45549.4 & -22777.8 & 0.08 (0.04, 0.16) & 0.08 &\\ 
				&					&	DPM		& 45550.6 & -22778.8 & 0.09 (0.04, 0.18) & 0.12 & 0.31 \\ \\
\multirow{2}{*}{VI}	&\multirow{2}{*}{(0.1, 5)} 	&	MVN		& 45545.7 & -22776.7 & 0.14 (0.09, 0.21) & 0.13 &\\ 
				&					&	DPM		& 45540.6 & -22774.0 & 0.15 (0.10, 0.24) & 0.18 & 0.23 \\ 
   \hline
\end{tabular}
\end{table}

While there are clear differences in the posterior summaries for SD($V_{j1}$) across Cases I-VI, within each case we see that there is little difference in the corresponding summaries between the MVN and DPM specifications; that is the conclusions one draws regarding the variation of the true underlying $V_{j1}$ are robust to this choice. However we do find that there are substantial differences in the average posterior standard deviations for the $J$ cluster-specific $V_{j1}$. In Case II, for example, $\bar{\sigma}_{V_{j1}}$ is 0.34 under the MVN specification and 0.70 under the DPM specification. Generally, this ordering is consistent across the six cases, as well as with the results presented in Figure \ref{fig:RE}. When combined with the posterior summaries for SD($V_{j1}$), the results suggest that for our application the trade-off of using the more flexible DPM specification is somewhat detrimental to the analyses; use of the DPM specification rather than the MVN does not serve to change our conclusions regarding the variation across the true $V_{j1}$ but has, rather, served to increase the posterior uncertainty regarding any given specific $V_{j1}$.

Finally, the last column of Table \ref{tab:sen} presents the posterior median for $\tau$, the precision parameter in the DPM specification. If one interprets the DPM specification as a mixture of MVN distributions (see Supplementary Materials Section B), $\tau$ dictates, in part, the number of mixture components and, hence, the complexity of the overall specification. From the results, however, we see that the posterior median of $\tau$, which takes on values in (0, $\infty$), tends towards quite small values and is generally robust to the specification of ($\Psi^*$, $\rho^*$). To further investigate the role of $\tau$ in our analyses, we conducted a series of additional analyses where $\tau$ was fixed at values ranging from 0.1 to 100 (i.e. we did not adopt a gamma hyperprior as described in Section 3.5.3). Although details are not reported here, we found that results of our analyses to be very robust to the specific value of $\tau$, again indicating few gains associated with use of the more flexible DPM specification for our application.

\section{Discussion}\label{sec:discussion}

In this paper, we propose a comprehensive, unified Bayesian framework for the analysis of cluster-correlated semi-competing risks data. The framework is flexible in that it lets researchers take advantage of the numerous benefits afforded by the Bayesian paradigm including the natural incorporation of prior information and the straightforward quantification of uncertainty for all parameters including hospital-specific random effects. The framework is also flexible in that it gives researchers choice in adopting parametric and/or semi-parametric specifications for various model components, a key consideration in practice when small sample size may require pragmatism during the analysis. To facilitate model choice, we have also developed DIC and LMPL measures for model comparison within the proposed framework. Finally, computationally efficient algorithms have been developed and implemented, and are readily-available in a freely-available \texttt{R} package.

The work in this paper was motivated by an on-going collaboration investigating variation in risk of readmission following a diagnosis of pancreatic cancer. Towards this, we applied the framework to a sample of 5,298 Medicare enrollees diagnosed with pancreatic cancer at one of 112 hospitals between 2000-2009. The results from our analysis indicate a number of important determinants of risk of readmission including gender, age, co-morbidity status (as measured by the Charlson/Deyo score), whether or not they under went a procedure during the index hospitalization, the length of stay of the index hospitalization and the location to which the patients was eventually discharged. The analyses also revealed that there is substantially less between-hospital variation in risk of readmission than the risk of death (either prior to or post-readmission), after accounting for patient case-mix. To our knowledge these are the first reported results of this kind in the literature and we are currently expanding our analyses to consider patients across the entire U.S.

More generally, in the clinical and health policy literature, the standard analysis approach for investigating risk of readmission is based on a LN-GLMM \citep{normand1997statistical, ash2012statistical}. In the specific context of our application, compared with results based on the proposed framework, such an analysis yields meaningfully different conclusions regarding which patient-level characteristics are associated with risk of readmission, the magnitude and statistical significance of those associations and the ranks of hospitals. Given the relative robustness across models within the proposed framework, the fact that a LN-GLMM yields different conclusions is likely related to the fact that death is completely ignored as a competing risk. As a concrete example consider the hospital in Figure \ref{fig:rank}(a) that is ranked 8th under the semi-Markov PEM-MVN model and 26th under the LN-GLMM. Closer inspection of the raw data reveals that very few patients diagnosed at this hospital died within the 90-day window we consider. At other hospitals, the force of mortality is stronger and patients die at higher rates within the 90-day window; that these patients die is overlooked by the LN-GLMM which assumes that they remain `at risk' to experience a readmission event. Hence their estimated readmission rates are too small in the LN-GLMM (since the denominator is erroneously inflated). Unfortunately the hospital ranked 8th under the semi-Markov PEM-MVN model suffers from their low mortality rate in the sense that they do not benefit from erroneous inflation of the readmission rate denominator, as other hospitals do. Hence the change in rank.

As indicated, results across models within the proposed framework were relatively robust in our main application. We did find, however, that models which adopted the flexible PEM specification for the baseline hazard functions had substantially better fit to the data than models that adopted a Weibull specification. While models based on a  semi-Markov specification for death following readmission generally had better fit to the data than models based on a Markov specification we note that this choice does not affect the interpretation of the model for readmission (i.e. model (\ref{model1})) the investigation of which was our primary scientific goal. With this in mind, we have not reported on the results for the two models for death (i.e. models (\ref{model2}) and (\ref{model3})) although they are available in the Supplementary Materials E. In practice, researchers may be interested in readmission and death jointly in which case the choice of specification for $h_{03}(\cdot)$ will become critical from a scientific perspective \citep{Lee2014bscr}. In our main application, since the data are relatively rich in terms of sample size and the event rates, we have taken the PEM-MVN and PEM-DPM models as our primary models for comparison of ranks of hospitals and sensitivity analyses. In other less-rich data settings, however, analysts may be in a position where structure is needed either in the forms of the baseline hazard functions or for the random effects. Finally, we note that in our application a MVN specification for the population distribution of the vector of hospital-specific random effects, $\bfV_j$, appeared to be adequate. That is, the so-called Bayesian non-parametric DPM specification did not yield any additional insight into our understanding of variation in risk of readmission nor did it change meaningfully the ranking of hospitals. In other applications this may, of course, not be the case and the proposed framework gives researchers important choice in this regard. 

In Section 5, we show that incorrect assumption of the underlying distribution for cluster-specific random effects or baseline hazard functions result in lower efficiency of the incorrect parametric estimators. In addition, the computational efficiency of proposed models  with non-parametric specification of parameters heavily depends on underlying distributions of model parameters. For PEM models, if the underlying hazard function has an intricate shape, the model estimates a posterior distribution $\alpha_g$ to be centered around a larger value, resulting in expensive computation due to more parameters ($\lambda_{g,k}$'s) to be estimated. For DPM models, if data suggest a larger value of $\tau$, the model will introduce more latent classes in the mixture, implying more parameters to be estimated. 

Our analysis focuses on readmission 90 days post-discharge. However, we note that the computational performance of our proposed approach would not be challenged in the cases when an administrative censoring is not imposed. In particular, the proposed PEM model is flexible in that it allows the time scale for each of three hazard functions to be different for each transition. Following \cite{mckeague2000bayesian} and \cite{sebastien2008separation}, we suggest the last observed event time points be the upper bound in general problems where an administrative censoring is not imposed. In our application, however, since most of patients diagnosed with pancreatic cancer die within 1-year period, we would expect the estimates of baseline hazard functions have a relatively greater uncertainty in the late periods if the administrative censoring is not considered. In addition, in the context of our study, patients can experience multiple readmission events prior to death. The literature on recurrent event semi-competing risks would likely be useful for this setting and thus the development of methods that can accommodate recurrent non-terminal events in the cluster-correlated data setting is a promising area for future development.

The proposed hierarchical models assume constant hazard ratios over time conditional on the cluster-specific and patient-specific random effects. Since the primary interest of our analysis is the study of readmission event within `short time frame' (30 days or 90 days after discharge), the proportionality of hazards is quite reasonable assumption in our application. In the literature of multi-state models, more rigorous diagnosis can be most naturally done by considering a more flexible multi-state model such as a stratified model or inclusion of non-proportional covariate effects \citep{hougaard2000analysis}. Expanding the scope of the proposed models to include deviation from proportional hazards as well as time-varying covariates is our future work. In this paper, we considered a gamma distribution for the within-patient frailty because of its computational tractability. When the frailty distribution is misspecified, the resulting estimator is not guaranteed to be consistent, with the extent of asymptotic bias depending on the discrepancy between the assumed and true frailties distributions. However, \cite{hsu2007robustness} studied the effect of mis-specification of frailty distribution on the marginal regression estimates and hazard functions when gamma distribution is assumed. Their results show that the biases are generally low, even when the true frailty distribution is substantially different from the assumed gamma distribution. Therefore, if the regression parameters and hazard function are of primary interest, the gamma frailty model can be a reasonable choice in practice. 

Finally, we conclude by emphasizing that the proposed framework significantly improves and expands the set of statistical tools researches have to study quality of end-of-life care. While our focus has been on pancreatic cancer, the proposed framework is broadly applicable to all `advanced' health conditions for which current treatment options are limited and the force of mortality is strong. Such studies will be of paramount importance in the near-future because many of these conditions, including other cancers as well as neurodegenerative conditions such as Alzheimers' disease, directly affect large segments of an increasingly aging population. In addition, although it has not been in the focus of this paper, the proposed framework will also be critical in helping policy-makers understand and ultimately control the increasing costs of health care delivery in the U.S. In particular, the proposed framework provides CMS appropriate statistical tools with which to expand the scope of the Hospital Inpatient Quality Reporting Program and the Readmission Reduction Program to include to conditions with strong forces of mortality.

\bigskip
\begin{center}
{\large\bf SUPPLEMENTARY MATERIALS}
\end{center}

\begin{description}

\item[Title:] In online Supplementary Materials, we provide a detail description of Metropolis-Hastings-Green algorithm to fit our proposed models. Additional details regarding the Medicare data and results from the application are also provided. (pdf file)

\item[R-package `\texttt{SemiCompRisks}':] R-package \texttt{SemiCompRisks} contains codes to implement proposed Bayesian framework described in the article. The package is currently available in CRAN.

\end{description}

\bibliographystyle{apalike}

\bibliography{ref.bcscr}
\end{document}


\begin{center}
{\large \bf Supplemental Materials to: \\
``Hierarchical models for clustered semi-competing risks data with application to pancreatic cancer"}
\end{center}

\smallskip

\begin{center}

Kyu Ha Lee \\
{\small Epidemiology and Biostatistics Core, The Forsyth Institute, Cambridge, Massachusetts, U.S.A.} \\
{\small \emph{klee@hsph.harvard.edu}} \\
\vspace{0.2in}
Francesca Dominici \\
{\small Department of Biostatistics, Harvard T.H. Chan School of Public Health, Boston, Massachusetts, U.S.A.} \\
\vspace{0.2in}
Deborah Schrag \\
{\small Department of Medical Oncology, Dana Farber Cancer Institute, Boston, Massachusetts, U.S.A.} \\
\vspace{0.2in}    
Sebastien Haneuse \\
{\small Department of Biostatistics, Harvard T.H. Chan School of Public Health, Boston, Massachusetts, U.S.A.}

\end{center}

\bigskip

\newpage

\renewcommand\thesection{\Alph{section}}


\begin{center}
{\Large \bf Introduction}
\end{center}

This document supplements the main paper titled ``Hierarchical models for clustered semi-competing risks data with application to pancreatic cancer''. In Section \ref{MVNICAR}, we provide technical details regarding the MVN-ICAR specification for baseline hazard functions in PEM models. In Section \ref{webComp}, we provide a detailed description of the Metropolis-Hastings-Green algorithm to implement our proposed Bayesian framework (Weibull-MVN, Weibull-DPM, PEM-MVN, PEM-DPM). In Section \ref{existing}, we examine the potential use of methods proposed in \cite{gorfine2011frailty} and \cite{liquet2012investigating} in the context of the motivating application. In Section \ref{addSim}, we provide results from simulation studies that were not presented in the main paper. Finally, Section \ref{additional} supplements our main paper with some additional results from analyses of Medicare data from New England and a visual assessment of convergence of the proposed MCMC schemes using potential scale reduction factor.

In order to distinguish the two documents, alpha-numeric labels are used for sections, tables, figures, and equations in this document while numeric labels are used in the main paper.

\newpage

\section{MVN-ICAR Specification for $\bflambda_g$}\label{MVNICAR}

In MVN-ICAR, the specification of a prior for the components of $\bflambda_g$ is considered as a one-dimensional spatial problem. The dependence between neighboring intervals are modeled via a Gaussian intrinsic conditional autoregression (ICAR) \citep{Besa:Koop:1995}. Let $\bflambda_g^{(-k)}$ denote the vector given by $\bflambda_g$ with the $k^{th}$ element removed. The full conditional prior for $\lambda_{g,k}$ is then taken to be the following normal distribution:
\be
	\lambda_{g,k} | \bflambda_g^{(-k)} \sim \textrm{Normal}(\nu_{g,k}, \sigma_{g,k}^2), \label{priorLamCond}
\ee
where the conditional mean, $\nu_{g,k}$ = $\mu_{\lambda_g} + \sum_{j\neq k} W_{kj}^g(\lambda_{g,j} - \mu_{\lambda_g})$, is a marginal mean plus a weighted sum of the deviations of the remaining intervals. Let $\bar{\Delta}_{k}^g = s_{g,k} - s_{g,k-1}$ denote the length of the $I_{g,k}$ interval. We determine the weights for the intervals adjacent to the $k^{th}$ intervals based on these lengths as follows:
\be
	W_{k(k-1)}^g = \frac{c_{\lambda_g}(\bar{\Delta}_{k-1}^g + \bar{\Delta}_{k}^g)}{\bar{\Delta}_{k-1}^g + 2\bar{\Delta}_{k}^g + \bar{\Delta}_{k+1}^g}, ~~~~ W_{k(k+1)}^g = \frac{c_{\lambda_g}(\bar{\Delta}_{k}^g + \bar{\Delta}_{k+1}^g)}{\bar{\Delta}_{k-1}^g + 2\bar{\Delta}_{k}^g + \bar{\Delta}_{k+1}^g}, \label{weight}
\ee
where the constant $c_{\lambda_g} \in [0, 1]$ dictates the extent to which $\lambda_{g,k}$ is influenced by adjacent intervals \citep{sebastien2008separation}. The remaining weights corresponding to intervals which are not directly adjacent to the $k^{th}$ interval are set to zero. The conditional variance $\sigma_{g,k}^2$ in (\ref{priorLamCond}) is given by  $\sigma_{\lambda_g}^2 Q_k^g$. The $\sigma_{\lambda_g}^2 $ is an overall measure of variation across the elements of $\bflambda_{g}$ and the diagonal matrix $Q_k^g$ is given by
\be
	\frac{2}{\bar{\Delta}_{k-1}^g + 2\bar{\Delta}_{k}^g + \bar{\Delta}_{k+1}^g}. \label{Qvalue}
\ee
Given (\ref{priorLamCond}), (\ref{weight}), and (\ref{Qvalue}), we can see that $\bflambda_g$ jointly follows a $(K_g+1)$-dimensional multivariate normal (MVN) distribution: 
\be
\textrm{MVN}_{K_g+1}(\mu_{\lambda_g}\bfone,\ \sigma_{\lambda_g}^2\Sigma_{\lambda_g}), \label{ICAR:mvn}
\ee
where $\mu_{\lambda_g}$ is the overall (marginal) mean, $\sigma_{\lambda_g}^2$ the overall variability in elements of $\bflambda_{g}$. The $\Sigma_{\lambda_g}$ is given by $(I - \bfW^g)^{-1}\bfQ^g$, where a $(K_g+1)\times (K_g+1)$ matrix $\bfW^g_{(k,j)}$=$W_{kj}^g$ and a $(K_g+1)\times (K_g+1)$ diagonal matrix $\bfQ^g_{(k,k)}$=$Q_k^g$.

\newpage
\section{Metropolis-Hastings-Green Algorithm}\label{webComp}

\subsection{Weibull models} \label{sec:weib}

Let $\Phi_W= \{\alpha_{w,1}, \alpha_{w,2}, \alpha_{w,1}, \kappa_{w,1}, \kappa_{w,2}, \kappa_{w,3}, \bfbeta_{1}, \bfbeta_{2}, \bfbeta_{3}, \vec{\gamma}, \vec{\bfV}\}$ be a set of parameters in the likelihood function of Weibull models. The observed data likelihood $L_W(\mathcal{D} | \Phi_W)$ is given by
\be
&&\prod_{j=1}^J\prod_{i=1}^{n_j} \left( \gamma_{ji} \alpha_{w,1}\kappa_{w,1} y_{ji1}^{\alpha_{w,1}-1}\eta_{ji1}\right)^{\delta_{ji1}(1-\delta_{ji2})}  \left( \gamma_{ji}^{2} \alpha_{w,1}\kappa_{w,1} y_{ji1}^{\alpha_{w,1}-1} \eta_{ji1} \alpha_{w,3}\kappa_{w,3} y_{ji2}^{\alpha_{w,3}-1}\eta_{ji3}  \right)^{\delta_{ji1}\delta_{ji2}} \nonumber \\
&& \hspace*{0.5in} \times \left( \gamma_{ji} \alpha_{w,2}\kappa_{w,2} y_{ji2}^{\alpha_{w,2}-1} \eta_{ji2} \right)^{\delta_{ji2}(1-\delta_{ji1})}  \exp\left\{-r_W(y_{ji1}, y_{ji2})\right\}, \label{LH_M}
\ee
where $\eta_{jig} = \exp\left(\bfx_{jig}^{\top}\bfbeta_g + V_{jg} \right)$ and
\be
 r_W(t_{ji1}, t_{ji2})\hspace*{4.5in}  \nonumber
\ee
\[
=
\begin{cases}
   \gamma_{ji}\left\{\kappa_{w,1}t_{ji1}^{\alpha_{w,1}}\eta_{ji1} + \kappa_{w,2}t_{ji1}^{\alpha_{w,2}}\eta_{ji2}+ \left(\kappa_{w,3}t_{ji2}^{\alpha_{w,3}} - \kappa_{w,3}t_{ji1}^{\alpha_{w,3}}\right) \eta_{ji3}\right\}, & \text{for Markov model}\\
    \gamma_{ji}\left[\kappa_{w,1}t_{ji1}^{\alpha_{w,1}}\eta_{ji1} + \kappa_{w,2}t_{ji1}^{\alpha_{w,2}}\eta_{ji2} + \left\{\kappa_{w,3}(t_{ji2}-t_{ji1})^{\alpha_{w,3}}\right\} \eta_{ji3}\right],  & \text{for semi-Markov model}
\end{cases}
\]

For Weibull models, we use a random scan Gibbs sampling scheme, randomly selecting and updating a (vector of) model parameter at each iteration.

\subsubsection{Updating $\bfbeta_g$} \label{subsec:beta}
Let $\Phi^{-(\beta)}$ denote a set of parameters $\Phi$ with $\beta$ removed.
The full conditional posterior distribution of $\bfbeta_{ 1}$ can be obtained by
\be
	\pi(\bfbeta_{ 1} | \Phi_W^{-(\bfbeta_{ 1})}, \theta, \Sigma_V)&\propto& L_{W}(D | \Phi_W). \nonumber\\
	&\propto& \prod_{j=1}^J \prod_{i=1}^{n_j} \exp\left(\delta_{ji1} \bfx_{ji1}^{\top}\bfbeta_{ 1}- \gamma_{ji} \kappa_{w,1}y_{ji1}^{\alpha_{w,1}} e^{\bfx_{ji1}^{\top}\bfbeta_{ 1} + V_{j1}}\right). \nonumber
\ee	
Analogously, the full conditionals of $\bfbeta_{ 2}$ and $\bfbeta_{ 3}$ are given by
\be
	\pi(\bfbeta_{ 2} | \Phi_W^{-(\bfbeta_{ 2})}, \theta, \Sigma_V)&\propto& \prod_{j=1}^J \prod_{i=1}^{n_j} \exp\left\{\delta_{ji2}(1-\delta_{ji1}) \bfx_{ji2}^{\top}\bfbeta_{ 2} - \gamma_{ji}\kappa_{w,2} y_{ji1}^{\alpha_{w,2}} e^{\bfx_{ji2}^{\top}\bfbeta_{ 2} + V_{j2}}\right\},\nonumber \\
	\pi(\bfbeta_{ 3} | \Phi_W^{-(\bfbeta_{ 3})}, \theta, \Sigma_V)&\propto& \prod_{j=1}^J \prod_{i=1}^{n_j} \exp\left\{\delta_{ji1}\delta_{ji2} \bfx_{ji3}^{\top}\bfbeta_{ 3} - \gamma_{ji}\kappa_{w,3}\Big(y_{ji2}^{\alpha_{w,3}} - y_{ji1}^{\alpha_{w,3}}\Big) e^{\bfx_{ji3}^{\top}\bfbeta_{ 3} + V_{j3}}\right\}.\nonumber 
\ee
Since the full conditionals do not have standard forms, we use Metropolis Hastings (MH) algorithm to update each element of $\bfbeta_g$, $\beta_{g,1}, \ldots, \beta_{g,p_1}$. In our algorithm, the conventional random walk MH is improved in convergence speed by taking some meaningful function of the current value $\beta_{g, k}^{(t-1)}$ for the mean and variance of Normal proposal density. Specifically, let $D_1(\beta_{g, k})$ and $D_2(\beta_{g, k})$ denote the first and second gradients of log-full conditional of $\bfbeta_g$ with respect to $\beta_{g,k}$, then a proposal $\beta^*$ is drawn from a Normal proposal density that is centered at $\mu(\beta_{g, k}^{(t-1)})=\beta_{g, k}^{(t-1)}-D_1(\beta_{g, k}^{(t-1)})/D_2(\beta_{g, k}^{(t-1)})$, the updated value from the Newton-Raphson algorithm, with a variance $\sigma^2(\beta_{g, k}^{(t-1)})=-2.4^2/D_2(\beta_{g, k}^{(t-1)})$, based on the inverse Fisher information evaluated with $\beta_{g, k}^{(t-1)}$ \citep{roberts2001optimal, gelman2003bayesian}. Therefore, the acceptance probability for $\beta_{g,k}$ is given by
\be
	\frac{\pi(\bfbeta_g^* | \Phi_W^{-(\bfbeta_g)}, \theta, \Sigma_V)\textrm{Normal}\left(\beta_{g, k}^{(t-1)}|\mu(\beta^*), \sigma^2(\beta^*)\right)}{\pi(\bfbeta_g^{(t-1)} | \Phi_W^{-(\bfbeta_g)}, \theta, \Sigma_V)\textrm{Normal}\left(\beta^*|\mu(\beta_{g, k}^{(t-1)}), \sigma^2(\beta_{g, k}^{(t-1)})\right)},
\ee
where $\bfbeta_g^{(t-1)}$ is a sample of $\bfbeta_g$ at current iteration and $\bfbeta_g^*$ is the $\bfbeta_g$ with $k$-th element replaced by $\beta^*$.

\subsubsection{Updating $\alpha_{w,g}$}

The full conditional posterior distribution of $\alpha_{w,1}$ is given by
\be
	&&\pi(\alpha_{w,1} | \Phi_W^{-(\alpha_{w,1})}, \theta, \Sigma_V)\nonumber \\
	&\propto& L_{W}(D | \Phi_W) \times \pi(\alpha_{w,1}) \nonumber\\
	&\propto& \alpha_{w,1}^{a_{\alpha, 1}-1}e^{- b_{\alpha, 1}\alpha_{w,1}}	\prod_{j=1}^J \prod_{i=1}^{n_j} \left(\alpha_{w,1}y_{ji1}^{\alpha_{w,1}}\right)^{\delta_{ji1}} \exp\left( \gamma_{ji} \kappa_{w,1}y_{ji1}^{\alpha_{w,1}} \eta_{ji1}\right).\nonumber
\ee	
Analogously, the full conditionals of $\alpha_{w,2}$ and $\alpha_{w,3}$ are given by
\be
\pi(\alpha_{w,2} | \Phi_W^{-(\alpha_{w,2})}, \theta, \Sigma_V)&\propto& \alpha_{w,2}^{a_{\alpha, 2}-1}e^{- b_{\alpha, 2}\alpha_{w,2}}	\prod_{j=1}^J \prod_{i=1}^{n_j} \left(\alpha_{w,2}y_{ji2}^{\alpha_{w,2}}\right)^{\delta_{ji2}(1-\delta_{ji1})} \exp\left(- \gamma_{ji} \kappa_{w,2} y_{ji1}^{\alpha_{w,2}} \eta_{ji2}\right),\nonumber \\
\pi(\alpha_{w,3} | \Phi_W^{-(\alpha_{w,3})}, \theta, \Sigma_V)&\propto& \alpha_{w,3}^{a_{\alpha, 3}-1}e^{- b_{\alpha, 3}\alpha_{w,3}}	\prod_{j=1}^J \prod_{i=1}^{n_j} \left(\alpha_{w,3}y_{ji2}^{\alpha_{w,3}}\right)^{\delta_{ji1}\delta_{ji2}} \nonumber \\
	&& \hspace*{0.1in} \times \exp\left\{- \gamma_{ji} \kappa_{w,3}\Big(y_{ji2}^{\alpha_{w,3}} - y_{ji1}^{\alpha_{w,3}}\Big) \eta_{ji3}\right\}. \nonumber
\ee
In MH algorithm to update $\alpha_{w,g}$, we generate a proposal $\alpha^*$ from a Gamma distribution with $\textrm{Gamma}\left((\alpha_{w,g}^{(t-1)})^2/k_0, \alpha_{w,g}^{(t-1)}/k_0\right)$ which corresponds to a distribution with a mean of $\alpha_{w,g}^{(t-1)}$ and a variance of $k_0$. The value of $k_0$ is specified such that the MH step for $\alpha_{w,g}$ achieves an acceptance rate of $25\%\sim30\%$. Finally the acceptance probability to update $\alpha_{w,g}$ can be written as
\be
	\frac{\pi(\alpha^* | \Phi_W^{-(\alpha_{w,g})}, \theta, \Sigma_V)\mathcal{G}\left(\alpha_{w,g}^{(t-1)} | (\alpha_{w,g}^*)^2/k_0, \alpha_{w,g}^*/k_0\right)}{\pi(\alpha_{w,g}^{(t-1)} | \Phi_W^{-(\alpha_{w,g})}, \theta, \Sigma_V)\mathcal{G}\left(\alpha_{w,g}^* | (\alpha_{w,g}^{(t-1)})^2/k_0, \alpha_{w,g}^{(t-1)}/k_0\right)}. \nonumber
\ee

\subsubsection{Updating $\kappa_{w,g}$}

\noindent The full conditional posterior distribution of $\kappa_{w,g}$ can be obtained by
\be
\pi(\kappa_{w,g} | \Phi_W^{-(\kappa_{w,g})}, \theta, \Sigma_V) \propto L_{W}(D | \Phi_W) \times \pi(\kappa_{w,g}). \nonumber
\ee
We see that the full conditionals of $\kappa_{w,g}$ are gamma distributions and the samples can be drawn from following distributions:

\be
\kappa_{w,1} | \Phi_W^{-(\kappa_{w,1})}, \theta, \Sigma_V &\sim&	\textrm{Gamma}\left(\sum_{j=1}^J\sum_{i=1}^{n_j}\delta_{ji1} + a_{\kappa, 1}, \ \sum_{j=1}^J\sum_{i=1}^{n_j}\gamma_{ji}y_{ji1}^{\alpha_{w,1}}\eta_{ji1} + b_{\kappa, 1} \right), \nonumber\\
\kappa_{w,2} | \Phi_W^{-(\kappa_{w,2})}, \theta, \Sigma_V &\sim& 	\textrm{Gamma}\left(\sum_{j=1}^J\sum_{i=1}^{n_j}\delta_{ji2}(1-\delta_{ji1}) + a_{\kappa, 2}, \ \sum_{j=1}^J\sum_{i=1}^{n_j}\gamma_{ji}y_{ji1}^{\alpha_{w,2}}\eta_{ji2} + b_{\kappa, 2} \right),\nonumber\\
\kappa_{w,3} | \Phi_W^{-(\kappa_{w,3})}, \theta, \Sigma_V &\sim&	\textrm{Gamma}\left(\sum_{j=1}^J\sum_{i=1}^{n_j}\delta_{ji1}\delta_{ji2} + a_{\kappa, 3}, \ \sum_{j=1}^J\sum_{i=1}^{n_j}\gamma_{ji}\left(y_{ji2}^{\alpha_{w,3}} - y_{ji1}^{\alpha_{w,3}}\right)\eta_{ji3} + b_{\kappa, 3} \right).\nonumber
\ee

\subsubsection{Updating $\gamma_{ji}$}
The full conditional posterior distribution of $\gamma_{ji}$ is given by
\be
	&&\pi(\gamma_{ji} |  \Phi_W^{-(\gamma_{ji})}, \theta, \Sigma_V) \nonumber \\
	&\propto&  L_{W}(D | \Phi_W) \times \pi(\gamma_{ji} | \theta) \nonumber \\
	&\propto&	\gamma_{ji}^{\delta_{ji1} + \delta_{ji2} + \theta^{-1} - 1} \exp\left[-r_W(y_{ji1}, y_{ji2}) - \theta^{-1}\gamma_{ji}\right]. \nonumber
\ee	
Therefore, we sample $\gamma_{ji}$ from
\be
	\textrm{Gamma}\left(\delta_{ji1} + \delta_{ji2} + \theta^{-1} , \ r_W(y_{ji1}, y_{ji2}; \gamma_{ji} = 1) + \theta^{-1} \right). \nonumber
\ee

\subsubsection{Updating $\theta$} \label{subsec:theta}

Let $\xi = 1/\theta$ denote the precision parameter of frailty distribution. The full conditional posterior distribution of $\xi$ is given by
\be
	\pi(\xi |  \Phi_W,\Sigma_V) &\propto& \pi(\xi) \prod_{j=1}^J\prod_{i=1}^{n_j} \pi(\gamma_{ji} | \xi)  \nonumber \\
	&\propto& \frac{\xi^{n\xi + b_{\theta} - 1}e^{-\xi(\sum_{j=1}^J\sum_{i=1}^{n_j}\gamma_{ji} + a_{\theta})}}{\{\Gamma(\xi)\}^n}\prod_{j=1}^J\prod_{i=1}^{n_j}\gamma_{ji}^{\xi-1}.\nonumber
\ee
We revise the traditional random walk MH algorithm for updating $\xi$ as done in Section \ref{subsec:beta} for $\bfbeta_g$. Let $\mu_{\xi}(\xi) = \xi - \min\{0, D_{1,\xi}(\xi)/D_{2,\xi}(\xi)\}$ and $\sigma_{\xi}^2(\xi) = -c_0/D_{2, \xi}(\xi)$, where $D_{1,\xi}(\xi)$ and $D_{2,\xi}(\xi)$  are the first and second gradients of $\log\pi(\xi |  \Phi_W^{-(\xi)},\Sigma_V)$ with respect to $\xi$. A proposal $\xi^*$ is generated from the following Gamma distribution
\be
	\textrm{Gamma}\left(\mu_{\xi}(\xi^{(t-1)})^2/\sigma_{\xi}^2(\xi^{(t-1)}), \ \mu(\xi^{(t-1)})/\sigma_{\xi}^2(\xi^{(t-1)})\right). \nonumber
\ee 
The value of $c_0>0$ is specified such that the algorithm achieve the desired acceptance rate. The acceptance probability to update $\xi$ is then given by
\be
	\frac{\pi(\xi^* |  \Phi_W,\Sigma_V)\textrm{Gamma}\left(\xi^* | \mu_{\xi}(\xi^{*})^2/\sigma_{\xi}^2(\xi^{*}), \ \mu(\xi^{*})/\sigma_{\xi}^2(\xi^{*})\right)}    {\pi(\xi^{(t-1)} |  \Phi_W,\Sigma_V) \textrm{Gamma}\left(\xi^* | \mu_{\xi}(\xi^{(t-1)})^2/\sigma_{\xi}^2(\xi^{(t-1)}), \ \mu(\xi^{(t-1)})/\sigma_{\xi}^2(\xi^{(t-1)})\right)}.\nonumber 
\ee

\subsubsection{Updating $\bfV_j$ for Weibull-MVN model}\label{subsec:MVNV}

The full conditional posterior distribution of $V_{j1}$ can be obtained by
\be
\pi(V_{j1} |  \Phi_W^{-(V_{j1})}, \theta, \Sigma_V) 
	&\propto&  L_{W}(D | \Phi_W) \times  \pi(\bfV_j| \Sigma_{V}). \nonumber \\
	&\propto& \exp\left\{\sum_{i=1}^{n_j} \left(V_{j1}\delta_{ji1} - \gamma_{ji}\kappa_{w,1}y_{ji1}^{\alpha_{w,1}}\eta_{ji1} \right)- \frac{1}{2}\bfV_j^{\top}\Sigma_{V}^{-1}\bfV_j\right\}. \nonumber
\ee
Analogously, the full conditionals of $V_{j2}$ and $V_{j3}$ can be written as
\be
	\pi(V_{j2} |  \Phi_W^{-(V_{j2})}, \theta, \Sigma_V)  &\propto& \exp\left\{\sum_{i=1}^{n_j} \left(V_{j2}\delta_{ji2}(1-\delta_{ji1}) - \gamma_{ji}\kappa_{w,2}y_{ji1}^{\alpha_{w,2}}\eta_{ji2} \right)- \frac{1}{2}\bfV_j^{\top}\Sigma_{V}^{-1}\bfV_j\right\}, \nonumber \\
	\pi(V_{j3} |  \Phi_W^{-(V_{j3})}, \theta, \Sigma_V)  &\propto& \exp\left\{\sum_{i=1}^{n_j} \left(V_{j3}\delta_{ji1}\delta_{ji2} - \gamma_{ji}\kappa_{w,3}(y_{ji2}^{\alpha_{w,3}}-y_{ji1}^{\alpha_{w,3}})\eta_{ji3} \right)- \frac{1}{2}\bfV_j^{\top}\Sigma_{V}^{-1}\bfV_j\right\}. \nonumber	
\ee
As done in Section \ref{subsec:beta}, in a MH step for updating $V_{jg}$, we sample a proposal $V^*$ from a Normal distribution that is centered at $\mu_V(V_{jg}^{(t-1)}) = V_{jg}^{(t-1)}-D_{1,V}(V_{jg}^{(t-1)})/D_{2,V}(V_{jg}^{(t-1)})$ and has a variance of $\sigma_V^2(V^{(t-1)}) = -2.4^2/D_{2,V}(V^{(t-1)})$, where $D_{1,V}(V_{jg})$ and $D_{2,V}(V_{jg})$ are the first and the second gradients of $\log\pi(V_{jg} |  \Phi_W^{-(V_{jg})}, \theta, \Sigma_V)$ with respect to $V_{jg}$. Finally, the acceptance probability is given by
\be
	\frac{\pi(V^* |  \Phi_W^{-(V_{jg})}, \theta, \Sigma_V) \textrm{Normal}\left(V_{jg}^{(t-1)} | \mu_V(V^*), \sigma_V^2(V^*)\right)}		{\pi(V_{jg}^{(t-1)} |  \Phi_W^{-(V_{jg})}, \theta, \Sigma_V)\textrm{Normal}\left(V^* | \mu_V(V_{jg}^{(t-1)}), \sigma_V^2(V_{jg}^{(t-1)})\right)}. \nonumber
\ee

\subsubsection{Updating $\Sigma_V$ for Weibull-MVN model} \label{subsec:MVNSigma}
The full conditional posterior distribution of $\Sigma_V$ can be written as
\be
	\pi(\Sigma_V |  \Phi_W,\theta) &\propto& \pi(\Sigma_V) \prod_{j=1}^J \pi(\bfV_j| \Sigma_{V})\nonumber  \\
	&\propto& |\Sigma_V|^{-\frac{J+\rho_v+4}{2}} \exp\left\{-\frac{1}{2}\left(\sum_{j=1}^J\bfV_j\bfV_j^{\top}+\Psi_v\right)\Sigma_V^{-1}\right\}. \nonumber
\ee
Therefore, we update $\Sigma_V$ from the following inverse-Wishart distribution:
\be
\Sigma_V |  \Phi_W,\theta \sim \textrm{inverse-Wishart}\left(\sum_{j=1}^J \bfV_j\bfV_j^{\top} + \Psi_v, \   J+\rho_v\right). \nonumber
\ee

\subsubsection{Updating $\bfV_j$ and $\Sigma_V$ for Weibull-DPM model} \label{subsec:DPM}

Towards developing this model, suppose that, instead of arising from a single distribution, the $\bfV_j$ are draws from a finite mixture of $M$ multivariate Normal distributions, each with their own mean vector and variance-covariance matrix, ($\bfmu_m$, $\bfSigma_m$) for $m = 1, \ldots, M$. Let $m_j \in \{1,\ldots, M\}$ denote the specific component or class to which the $j^{\textrm{th}}$ hospital belongs. Since the class-specific ($\bfmu_m$, $\bfSigma_m$) are not known they are taken to be draws from some distribution, $G_0$. Furthermore, since the `true' class memberships are not known, we denote the probability that the $j^{\textrm{th}}$ hospital belongs to any given class by the vector $\bfp = (p_1,\ldots, p_M)$ whose components add up to 1.0. In the absence of prior knowledge regarding the distribution of class memberships for the $J$ hospitals across the $M$ classes, a natural prior for $\bfp$ is the conjugate symmetric Dirichlet($\tau/M, \ldots, \tau/M$) distribution; the hyperparameter, $\tau$ \citep{walker1997hierarchical}. Jointly, this finite mixture distribution can be summarized by:
\be
	\bfV_j | m_j &\sim& MVN(\bfmu_{m_j}, \Sigma_{m_j}), \nonumber \\
	(\bfmu_m, \Sigma_m) &\sim& G_{0},\ \textrm{for $m = 1, \ldots, M$}, \nonumber \\
	m_{j} | \bfp &\sim& \textrm{Discrete}(m_{j}|\ p_{1}, \ldots,p_{M}),\nonumber \\
	\bfp &\sim& \textrm{Dirichlet}(\tau/M,\ldots, \tau/M),
\ee
Finally, letting $M\rightarrow \infty$ the resulting specification is referred to as a Dirichlet process mixture of multivariate Normal distributions (DPM-MVN)~\citep{ferguson1973bayesian, bush1996semiparametric}. When $M\rightarrow \infty$, we cannot explicitly represent the infinite number of ($\bfmu_m$, $\Sigma_m$). Instead, following \cite{neal2000markov}, we represent and implement the MCMC sampling for only those ($\bfmu_m$, $\Sigma_m$) that are currently associated with some observations at each iteration. In this subsection, we provide a step-by-step detailed description of the MH algorithm to update $\bfV_j$ in Weibull-DPM model.

First, we update a class membership $m_j$ based on $m_j | \bfm_{(-j)}, \bfV_j$, $j = 1,\cdots,J$. Let $\bfm_{-(j)}$ denote a set of all class memberships from clusters except the cluster $j$. After identifying the ``$n_{m}$" unique classes of $\bfm_{-(j)}$, we compute the following probabilities for each of the unique values $m$.
	\be
		P(m_j = m | \bfm_{(-j)}, \bfV_j) &=& b \frac{n_{-j,m}}{J-1+\tau}\int\textrm{Normal}(\bfV_j | \bfmu_{m_j}, \Sigma_{m_j})dH_{-j, m}(\mu, \Sigma), \label{comp:dp:prob1}\\
		P(m_j \neq m_k, \  \forall k \neq j \  | \bfm_{(-j)}, \bfV_j) &=& b\frac{\tau}{J-1+\tau}\int \textrm{Normal}(\bfV_j | \bfmu, \Sigma) dG_0(\mu, \Sigma) , \label{comp:dp:prob2}
	\ee
where $H_{-j, m}$ is the posterior distribution of $(\mu, \Sigma)$ based on the prior $G_0$ and $\{\bfV_k: k \neq j, m_k = c\}$.  The normalizing constant $b$ makes ``$n_{m}+1$" probabilities above sum to 1. Let $A = \{j: m_j = m\}$ and $H_A$ be the posterior distribution of $(\mu, \sigma)$ based on the prior $G_0$ and $\{\bfV_j : j\in A\}$. It can be shown that the $H_A$ is also Normal-inverse Wishart distribution as $G_0$ is conjugate to multivariate normal distribution:

\noindent 1. we draw a sample of a class membership.
\begin{enumerate}[i)]
	\item For each $m_j$, identify the $n_{m}$ unique values of $\bfm_{(-j)}$.
	\item For each of the unique values $m$, compute the following probabilities:
	\be
		P(m_j = m | \bfm_{(-j)}, \bfV_j) &=& b \frac{n_{-j,m}}{J-1+\tau}\int\textrm{Normal}(\bfV_j | \bfmu_{m_j}, \Sigma_{m_j})dH_{-j, m}(\mu, \Sigma), \label{comp:dp:prob1} \nonumber \\
						\\
		P(m_j \neq m_k, \  \forall k \neq j \  | \bfm_{(-j)}, \bfV_j) &=& b\frac{\tau}{J-1+\tau}\int \textrm{Normal}(\bfV_j | \bfmu, \Sigma) dG_0(\mu, \Sigma) , \label{comp:dp:prob2}
	\ee
	where $H_{-j, m}$ is the posterior distribution of $(\mu, \Sigma)$ based on the prior $G_0$ and $\{\bfV_k: k \neq j, m_k = m\}$.  The normalizing constant $b$ makes $n_{m}+1$ probabilities above sum to 1.0. Let $A = \{j: m_j = m\}$ and $H_A$ be the posterior distribution of $(\mu, \sigma)$ based on the prior $G_0$ and $\{\bfV_j : j\in A\}$. It can be shown that the $H_A$ is also Normal-inverse Wishart distribution as $G_0$ is conjugate to multivariate normal distribution:
	\be
		H_A(\mu, \Sigma | \mu_A, \zeta_A, \Psi_A, \rho_A), \nonumber
	\ee
	where
	\be
		&&\mu_A = \frac{\frac{1}{\zeta_0}\bfmu_0 + |A|\bar{\bfV}_A}{\frac{1}{\zeta_0} + |A|}, \ \ \zeta_A = \left(\frac{1}{\zeta_0} + |A|\right)^{-1}, \ \ \rho_A = \rho_0 + |A|, \nonumber \\
		&&\Psi_A = \Psi_0 + \sum_{j \in A}\left(\bfV_j - \bar{\bfV}_A\right)\left(\bfV_j - \bar{\bfV}_A\right)^{\top} + \frac{\frac{|A|}{\zeta_0}}{\frac{1}{\zeta_0} + |A|} \left(\bar{\bfV}_A - \bfmu_0\right)\left(\bar{\bfV}_A - \bfmu_0\right)^{\top},
	\ee
	with $\bar{\bfV}_A = \frac{1}{|A|}\sum_{k \in A} \bfV_k$. Now we define
	\be
		&&Q(\bfV_j, \mu_0, \zeta_0, \Psi_0, \rho_0) \nonumber \\
		 &=& \int f_{\mathcal{N}_3} (\bfV_j | \mu, \Sigma) dF_{NIW}(\mu, \Sigma | \mu_0, \zeta_0, \Psi_0, \rho_0) \nonumber \\
		&=& \frac{|\Psi_0|^{\frac{\rho_0}{2}}}{\left|\Psi_0 + \bfV_j\bfV_j^{\top} + \frac{1}{\zeta_0}\bfmu_0\bfmu_0^{\top}- \left(1+\frac{1}{\zeta_0}\right)^{-1}\left(\frac{1}{\zeta_0}\bfmu_0 + \bfV_j\right)\left(\frac{1}{\zeta_0}\bfmu_0 + \bfV_j\right)^{\top}\right|^{\frac{\rho_0+1}{2}}} \nonumber  \\
		&& \times \frac{1}{(\pi\sqrt{2(1+\zeta_0)})^3}\times \frac{\Gamma_{\alpha, 3}(\frac{\rho_0+1}{2})}{\Gamma_{\alpha, 3}(\frac{\rho_0}{2})}
	\ee
	It follows that the integrals in (\ref{comp:dp:prob1}) and (\ref{comp:dp:prob2}) are equal to $Q(\bfV_j, \bfmu_A, \zeta_A, \Psi_A, \rho_A)$ and $Q(\bfV_j, \bfmu_0, \zeta_0, \Psi_0, \rho_0)$, respectively.
	
	\item Sample $m_j^{(\textrm{new})}$ based on the probabilities given in (\ref{comp:dp:prob1}) and (\ref{comp:dp:prob2}).
\end{enumerate}

\noindent 2. For all $m \in \{m_1,\ldots, m_J\}$, update $(\mu_m, \Sigma_m)$ using the posterior distribution that is based on $\{\bfV_j :  j \in \{k: m_k = m\}\}$.

\noindent 3. For $j=1,\ldots, J$, update $\bfV_j$ using its full conditional using Metropolis-Hastings algorithm.

\noindent 4. We treat $\tau$ as random and assign gamma prior $\textrm{Gamma}(a_{\tau}, b_{\tau})$ for $\tau$.  Following \cite{escobar1995bayesian}, we update $\tau$ by
	
	\begin{enumerate}[i)]
		\item sampling an $c \in (0, 1)$ from $Beta(\tau+1, J)$,
		\item sampling the new $\tau$ from the mixture of two gamma distributions:
		\be
			p_{c}\textrm{Gamma}(a_{\tau}+n_{m}, b_{\tau} - \log(c)) + (1-p_{c})\textrm{Gamma}(a_{\tau}+n_{m}-1, b_{\tau}-log(c)), \nonumber 
		\ee 
		where the weight $p_{c}$ is defined such that $p_{c}/(1-p_{c}) = (a_{\tau}+n_{m}-1)/\{J(b_{\tau}-log(c))\}$.
	\end{enumerate}
\noindent 5. Finally we calculate the total variance-covariance matrix:
\be
\Sigma_V = \frac{1}{J}\sum_{j=1}^J\big\{\big(\bfmu_{m_j}-\bar{\bfmu}\big)\big(\bfmu_{m_j}-\bar{\bfmu}\big)^{\top} + \Sigma_{m_j}\big\}, 
\ee
where $\bar{\bfmu} =\sum_{j=1}^J\bfmu_{m_j}/J$.

\newpage
\subsection{PEM models} \label{sec:PEM}

Let $\Phi_P = \{\bflambda_{ 1}, \bflambda_{ 2}, \bflambda_{ 3}, \bfbeta_{ 1}, \bfbeta_{ 2}, \bfbeta_{ 3}, \vec{\gamma}, \vec{\bfV}\}$ a set of parameters in the likelihood function of PEM models. The observed data likelihood $L_P(\mathcal{D} | \Phi_P)$ is given by
\be
&&\prod_{j=1}^J\prod_{i=1}^{n_j} \left[ \gamma_{ji}\eta_{ji1} \exp\left\{\sum_{k=1}^{K_1+1}\lambda_{1k}I(s_{1,k-1}<y_{ji1}\leq s_{1,k})\right\} \right]^{\delta_{ji1}(1-\delta_{ji2})}  \nonumber \\
&&\hspace*{0.5in}  \times \left[ \gamma_{ji}^{2}  \eta_{ji1} \eta_{ji3}\exp\left\{ \sum_{k=1}^{K_1+1}\lambda_{1k}I(s_{1,k-1}<y_{ji1}\leq s_{1,k})+ \sum_{k=1}^{K_3+1}\lambda_{3k}I(s_{3,k-1}<y_{ji2}\leq s_{3,k})\right\}  \right]^{\delta_{ji1}\delta_{ji2}} \nonumber \\
&& \hspace*{0.5in} \times \left[ \gamma_{ji} \eta_{ji2}\exp\left\{\sum_{k=1}^{K_2+1}\lambda_{2k}I(s_{2,k-1}<y_{ji2}\leq s_{2,k})\right\} \right]^{\delta_{ji2}(1-\delta_{ji1})}\nonumber \\
&&\hspace*{0.5in} \times  \exp\left\{-r_P(y_{ji1}, y_{ji2})\right\}, \label{LH_M}
\ee
where $\eta_{jig} = \exp\left(\bfx_{jig}^{\top}\bfbeta_g + V_{jg} \right)$ and
\be
 r_P(t_{ji1}, t_{ji2})\hspace*{4.5in}  \nonumber
\ee
\[
=
\begin{cases}
   \gamma_{ji}\left(\eta_{ji1}\sum_{k=1}^{K_1+1}e^{\lambda_{1,k}}\Delta_{jik}^1 + \eta_{ji2}\sum_{k=1}^{K_2+1}e^{\lambda_{2,k}}\Delta_{jik}^2+ \eta_{ji3} \sum_{k=1}^{K_3+1}e^{\lambda_{3,k}}\Delta_{jik}^{*3} \right), & \text{for Markov model}\\
    \gamma_{ji}\left(\eta_{ji1}\sum_{k=1}^{K_1+1}e^{\lambda_{1,k}}\Delta_{jik}^1 + \eta_{ji2}\sum_{k=1}^{K_2+1}e^{\lambda_{2,k}}\Delta_{jik}^2 + \eta_{ji3} \sum_{k=1}^{K_3+1}e^{\lambda_{3,k}}\Delta_{jik}^{*3} \right),  & \text{for semi-Markov model}
\end{cases}
\]
\be
&&\Delta_{jik}^g = \max\Big\{0, \min(y_{ji1}, s_{g,k}) - s_{g, k-1}\Big\}, \nonumber \\
&&\Delta_{jil}^{*g} = \begin{cases} \max\Big\{0, \min(y_{ji2}, s_{g,l}) - \max(y_{ji1}, s_{g, l-1})\Big\}, &\mbox{for Markov model,} \nonumber\\
\max\Big\{0, \min(y_{ji2}-y_{ji1}, s_{g,l}) - s_{g,l-1})\Big\}, & \mbox{for semi-Markov model}. \end{cases}
\ee

\subsubsection{Reversible jump MCMC algorithm}

For PEM models, we use a random scan Gibbs sampling scheme, randomly selecting and updating a (vector of) model parameter at each iteration. Let $BI_g$ and $DI_g$ denote a birth and a death of a new time split for transition $g\in\{1,2,3\}$. The probabilities for the update {$\textrm{BI}_g$} and {$\textrm{DI}_g$} are given by
\be
	\pi_{BI_g}^{K_g} &=& \rho_g \min\Big\{1, ~ \frac{\textrm{Poisson}(K_g+1 | \alpha_{K_g})}{\textrm{Poisson}(K_g | \alpha_{K_g})}\Big\} = \rho_g\min\Big\{1,~ \frac{\alpha_{K_g}}{K_g+1}\Big\}, \nonumber \\
	 \pi_{DI_g}^{K_g}  &=& \rho_g \min\Big\{1, ~ \frac{\textrm{Poisson}(K_g-1 | \alpha_{K_g})}{\textrm{Poisson}(K_g | \alpha_{K_g})}\Big\} = \rho_g\min\Big\{1,~ \frac{K_g}{\alpha_{K_g}}\Big\},\nonumber 
\ee
where $\rho_g$ is set such that $\pi_{BI_g}^{K_g} + \pi_{DI_g}^{K_g} < C_g$ and $\sum_{g=1}^3C_g < 1$ for $K_g = 1,\dots,K_{g,\max}$. $K_{g,\max}$ is the preassigned upper limit on the number of time splits for transition $g$ and we set $\pi_{BI_g}^{K_{g,\max}}$ = 0. The probabilities of updating other parameters are equally specified from remaining probability $1-\sum_{g=1}^3(\pi_{BI_g}^{K_g} + \pi_{DI_g}^{K_g})$.

\subsubsection{Updating $\bfbeta_g$}

The full conditional posterior distribution of $\bfbeta_{ 1}$ can be obtained by
\be
	&&\pi(\bfbeta_{ 1} | \Phi_P^{-(\bfbeta_{ 1})}, \bfmu_{\lambda}, \bfsigma_{\lambda}^2, \theta, \Sigma_V) \nonumber \\
	&\propto& L_{P}(D | \Phi_P) \nonumber\\
	&\propto& \prod_{j=1}^J \prod_{i=1}^{n_j} \exp\left(\delta_{ji1} \bfx_{ji1}^{\top}\bfbeta_{ 1} - \gamma_{ji} e^{\bfx_{ji1}^{\top}\bfbeta_{ 1} + V_{j1}}\sum_{k=1}^{K_1+1}e^{\lambda_{1,k}}\Delta_{jik}^1\right), \nonumber
\ee	
where $\bfmu_{\lambda} = (\mu_{\lambda_1}, \mu_{\lambda_2}, \mu_{\lambda_3})^{\top}$ and $\bfsigma_{\lambda}^2 = (\sigma_{\lambda_1}^2, \sigma_{\lambda_2}^2, \sigma_{\lambda_3}^2)^{\top}$. Analogously, the full conditionals of $\bfbeta_{ 2}$ and $\bfbeta_{ 3}$ are given by
\be
	&&\pi(\bfbeta_{ 2} | \Phi_P^{-(\bfbeta_{ 2})}, \bfmu_{\lambda}, \bfsigma_{\lambda}^2, \theta, \Sigma_V)\nonumber \\
	&\propto& \prod_{j=1}^J \prod_{i=1}^{n_j} \exp\left\{\delta_{ji2}(1-\delta_{ji1}) \bfx_{ji2}^{\top}\bfbeta_{ 2} - \gamma_{ji}e^{\bfx_{ji2}^{\top}\bfbeta_{ 2} + V_{j2}} \sum_{l=1}^{K_2+1}e^{\lambda_{2,l}}\Delta_{jil}^2\right\},\nonumber \\
	&&\pi(\bfbeta_{ 3} | \Phi_P^{-(\bfbeta_{ 3})}, \bfmu_{\lambda}, \bfsigma_{\lambda}^2, \theta, \Sigma_V)\nonumber \\
	&\propto& \prod_{j=1}^J \prod_{i=1}^{n_j} \exp\left(\delta_{ji1}\delta_{ji2} \bfx_{ji3}^{\top}\bfbeta_{ 3} - \gamma_{ji}e^{\bfx_{ji3}^{\top}\bfbeta_{ 3} + V_{j3}}\sum_{m=1}^{K_3+1}e^{\lambda_{3,m}}\Delta_{jim}^{*3}\right).\nonumber 
\ee

As the full conditionals do not have standard forms, we use MH algorithm to update each element of $\bfbeta_g$. A detailed description of the adapted random walk MH algorithm is provided in Section \ref{subsec:beta}.

\subsubsection{Updating $\bflambda_g$}

The full conditional posterior distribution of $\bflambda_{ 1}$ is given by
\be
	&&\pi(\bflambda_{ 1} | \Phi_P^{-(\bflambda_{ 1})}, \bfmu_{\lambda}, \bfsigma_{\lambda}^2, \theta, \Sigma_V) \nonumber \\
	&\propto& L_{P}(D | \Phi_P) \pi(\bflambda_1 | \mu_{\lambda_1}, \sigma_{\lambda_1}^2) \nonumber\\
	&\propto& \prod_{j=1}^J \prod_{i=1}^{n_j} \exp\left\{ \delta_{ji1}\lambda_{1k}I(s_{1,k-1}<y_{ji1}\leq s_{1,k}) - \gamma_{ji}\Delta_{jik}^1e^{\lambda_{1k}}\eta_{ji1}\right\} \nonumber \\
	&&\times \exp\left\{ - \frac{1}{2\sigma_{\lambda_{ 1}}^2}(\bflambda_{ 1}-\mu_{\lambda_{ 1}}\bfone)^{\top}\Sigma_{\lambda_{ 1}}^{-1}(\bflambda_{ 1}-\mu_{\lambda_{ 1}}\bfone\right\}, \nonumber
\ee	
where $\bfone$ denotes a $K_g + 1$ dimensional vector of $1$'s. Analogously, the full conditionals of $\bflambda_{ 2}$ and $\bflambda_{ 3}$ are given by
\be
	&&\pi(\bflambda_{2} | \Phi_P^{-(\bflambda_{ 2})}, \bfmu_{\lambda}, \bfsigma_{\lambda}^2, \theta, \Sigma_V) \nonumber \\
	&\propto& \prod_{j=1}^J \prod_{i=1}^{n_j} \exp\left\{ \delta_{ji2}(1- \delta_{ji1})\lambda_{2k}I(s_{2,k-1}<y_{ji2}\leq s_{2,k}) - \gamma_{ji}\Delta_{jik}^2e^{\lambda_{2k}}\eta_{ji2}\right\} \nonumber \\
	&&\times \exp\left\{- \frac{1}{2\sigma_{\lambda_{ 2}}^2}(\bflambda_{ 2}-\mu_{\lambda_{ 2}}\bfone)^{\top}\Sigma_{\lambda_{ 2}}^{-1}(\bflambda_{ 2}-\mu_{\lambda_{ 2}}\bfone)\right\}, \nonumber\\
	&&\pi(\bflambda_{3} | \Phi_P^{-(\bflambda_{ 3})}, \bfmu_{\lambda}, \bfsigma_{\lambda}^2, \theta, \Sigma_V) \nonumber \\
	&\propto& \prod_{j=1}^J \prod_{i=1}^{n_j} \exp\left\{ \delta_{ji1}\delta_{ji2}\lambda_{3k}I(s_{3,k-1}<y_{ji2}\leq s_{3,k}) - \gamma_{ji}\Delta_{jik}^{*3}e^{\lambda_{3k}}\eta_{ji3}\right\} \nonumber \\
	&&\times \exp\left\{- \frac{1}{2\sigma_{\lambda_{ 3}}^2}(\bflambda_{ 3}-\mu_{\lambda_{ 3}}\bfone)^{\top}\Sigma_{\lambda_{ 3}}^{-1}(\bflambda_{ 3}-\mu_{\lambda_{ 3}}\bfone)\right\}, \nonumber	
\ee

Since the full conditionals do not follow known distributions, MH algorithm is used to update each element of $\bflambda_g$. We follow the adapted random walk MH algorithm described in Section \ref{subsec:beta}.

\subsubsection{Updating $\gamma_{ji}$}
The full conditional posterior distribution of $\gamma_{ji}$ is given by
\be
	&&\pi(\gamma_{ji} |  \Phi_P^{-(\gamma_{ji})}, \bfmu_{\lambda}, \bfsigma_{\lambda}^2, \theta, \Sigma_V) \nonumber \\
	&\propto&  L_{P}(D | \Phi_P) \times \pi(\gamma_{ji} | \theta) \nonumber \\
	&\propto&	\gamma_{ji}^{\delta_{ji1} + \delta_{ji2} + \theta^{-1} - 1} \exp\left[-r_P(y_{ji1}, y_{ji2}) - \theta^{-1}\gamma_{ji}\right]. \nonumber
\ee	
Therefore, we sample $\gamma_{ji}$ from
\be
	\textrm{Gamma}\left(\delta_{ji1} + \delta_{ji2} + \theta^{-1} , \ r_P(y_{ji1}, y_{ji2}; \gamma_{ji} = 1) + \theta^{-1} \right). \nonumber
\ee

\subsubsection{Updating ($\mu_g$, $\sigma_g^2$)}

Full conditional posterior distributions for $\mu_{\lambda_g}$ and $\upsilon_g = 1/\sigma_{\lambda_g}^2$, $g = 1,2,3$ are Normal and Gamma distribution, respectively. Therefore, we use Gibbs sampling to update the parameters. We obtain the posterior samples of $\mu_{\lambda_g}$ from
\be
\textrm{Normal}\left(\frac{\bfone^{\top}\Sigma_{\lambda_g}^{-1}\bflambda_g}{\bfone^{\top}\Sigma_{\lambda_g}^{-1}\bfone}, ~\frac{\sigma_{\lambda_g}^2}{\bf1^{\top}\Sigma_{\lambda_g}^{-1}\bfone}\right), \nonumber
\ee
because the full conditional is given by
\be
	&&\pi(\mu_{\lambda_g} | \Phi_P, \bfmu_{\lambda}^{-(\mu_{\lambda_g})}, \bfsigma_{\lambda}^2, \theta, \Sigma_V) \propto \pi(\bflambda_g | \mu_{\lambda_g}, \sigma_{\lambda_g}^2)\pi(\mu_{\lambda_g}) \nonumber \\
	&\propto& \exp\left\{\frac{\bfone^{\top}\Sigma_{\lambda_g}^{-1}\bfone}{2\sigma_{\lambda_g}^2}											\left(\mu_{\lambda_g}- \frac{\bfone^{\top}\Sigma_{\lambda_g}^{-1}\bflambda_g}												{\bfone^{\top}\Sigma_{\lambda_g}^{-1}\bfone}\right)^2\right\}.  \nonumber
\ee
We update $\upsilon_g = 1/\sigma_{\lambda_g}^{-2}$ from a Gamma distribution given by
\be
\textrm{Gamma}\left(a_{\sigma,g} + \frac{K_g+1}{2}, ~ b_{\sigma,g} + \frac{1}{2}(\mu_{\lambda_g}\bfone - \bflambda_g)^{\top}\bfSigma_{\lambda_g}^{-1}(\mu_{\lambda_g}\bfone - \bflambda_g)\right), \nonumber
\ee
as the full conditional of $\upsilon_g$ is 
 \be
	&&\pi(\upsilon_g |\Phi_P, \bfmu_{\lambda}, (\bfsigma_{\lambda}^{2})^{-(\sigma_{\lambda_g}^2)}, \theta, \Sigma_V) \nonumber \\
	& \propto&  \pi(\bflambda_g | \mu_{\lambda_g}, \sigma_{\lambda_g}^2)\pi(\upsilon_g) \nonumber\\
	& \propto& (\upsilon_g)^{a_{\sigma, g} + \frac{K_g + 1}{2}-1}\exp\left[-\left\{b_{\sigma, g} + \frac{1}{2}(\mu_{\lambda_g}\bfone - \bflambda_g)^{\top}\bfSigma_{\lambda_g}^{-1}(\mu_{\lambda_g}\bfone - \bflambda_g)\right\}\upsilon_g\right]. \nonumber
\ee

\subsubsection{Updating $\theta$}

Updating the precision parameter $\xi = 1/\theta$ in PEM models requires the exactly same step as that in Weibull models. Therefore, the readers are referred to Section \ref{subsec:theta} for the full conditional posterior distribution of $\xi$ and the MH algorithm.

\subsubsection{Updating $\bfV_j$ for PEM-MVN model}

The full conditional posterior distribution of $V_{j1}$ can be obtained by
\be
\pi(V_{j1} |  \Phi_P^{-(V_{j1})}, \bfmu_{\lambda}, \bfsigma_{\lambda}^2, \theta, \Sigma_V) 
	&\propto&  L_{P}(D | \Phi_P) \times  \pi(\bfV_j| \Sigma_{V}). \nonumber \\
	&\propto& \exp\left\{\sum_{i=1}^{n_j} \left(V_{j1}\delta_{ji1} - \gamma_{ji}\eta_{ji1} \sum_{k=1}^{K_1+1}e^{\lambda_{1,k}}\Delta_{jik}^1\right)- \frac{1}{2}\bfV_j^{\top}\Sigma_{V}^{-1}\bfV_j\right\}. \nonumber
\ee
Analogously, the full conditionals of $V_{j2}$ and $V_{j3}$ can be written as
\be
	\pi(V_{j2} |  \Phi_P^{-(V_{j2})}, \bfmu_{\lambda}, \bfsigma_{\lambda}^2, \theta, \Sigma_V)  &\propto& \exp\left\{\sum_{i=1}^{n_j} \left(V_{j2}\delta_{ji2}(1-\delta_{ji1}) - \gamma_{ji}\eta_{ji2}\sum_{l=1}^{K_2+1}e^{\lambda_{2,l}}\Delta_{jil}^2 \right)- \frac{1}{2}\bfV_j^{\top}\Sigma_{V}^{-1}\bfV_j\right\}, \nonumber \\
	\pi(V_{j3} |  \Phi_P^{-(V_{j3})}, \bfmu_{\lambda}, \bfsigma_{\lambda}^2, \theta, \Sigma_V)  &\propto& \exp\left\{\sum_{i=1}^{n_j} \left(V_{j3}\delta_{ji1}\delta_{ji2} - \gamma_{ji}\eta_{ji3}\sum_{m=1}^{K_3+1}e^{\lambda_{3,m}}\Delta_{jim}^{*3} \right)- \frac{1}{2}\bfV_j^{\top}\Sigma_{V}^{-1}\bfV_j\right\}. \nonumber	
\ee
For updating each element of $\bfV_j$, we use the adapted random walk MH algorithm provided in Section \ref{subsec:MVNV}.

\subsubsection{Updating $\Sigma_V$ for PEM-MVN model}
The full conditional posterior distribution of $\Sigma_V$ in PEM-MVN model is the exactly same as that in Weibull-MVN model. Readers are referred to Section \ref{subsec:MVNSigma} for the Gibbs sampling step for updating $\Sigma_V$.

\subsubsection{Updating $\bfV_j$ and $\Sigma_V$ for PEM-DPM model}

Updating $\bfV_j$ and $\Sigma_V$ in PEM-DPM requires the exactly same step as that in Weibull-DPM. Therefore, the readers are referred to Section \ref{subsec:DPM} for detailed algorithm to update $\bfV_j$ and $\Sigma_V$. Note that in step 3 of the algorithm, the full conditional of $\bfV_j$ needs to be obtained based on $L_{P}(D | \Phi_P)$ for PEM-DPM.

\subsubsection{Update BI}

We specify $\log h_0(t)$ = $\sum_{k=1}^{K_g + 1} \lambda_{g,k} I(t\in I_{g,k})$ for the baseline hazard function corresponding to transition $g$ with partition ($K_g$, $\bfs_g$). Updating ($K_g$, $\bfs_g$) requires generating a proposal partition and then deciding whether or not to accept the proposal. For update BI (a birth move), we first select a proposal split time $s^*$ uniformly from among the observed event times which are not included in the current partition. Suppose $s^*$ lies between the $(k-1)^{th}$ and $k^{th}$ split times in the current partition. The proposal partition is then defined as
\be
	&&(0 = s_{g,0}, ..., s_{g,k-1}, s^*, s_{g,k},..., s_{g,K_1+1} = s_{g,\max}) \nonumber \\
	 &\equiv& (0 = s_{g,0}^*,..., s_{g,k-1}^*,s_{g,k}^*,s_{g,k+1}^*,...,s_{g,K_1+2}^*= s_{g,\max}).\nonumber
\ee
A height of the two new intervals created by the split at time $s^*$ also needs to be proposed. In order to make the old height be a compromise of the two new heights, the former is taken to be the weighted mean of the latter on the log scale: 
\be
	(s^{*} - s_{g,k-1})\lambda_{g,k}^* + (s_{g,k} - s^{*})\lambda_{g,k+1}^*
 = (s_{g,k} - s_{g,k-1})\lambda_{g,k}. \nonumber 
 \ee
Defining the multiplicative perturbation $\exp(\lambda_{g,j+1}^*)/\exp(\lambda_{g,j}^*) = (1-U)/U$, where $U\sim \textrm{Uniform}(0, 1)$, the new heights are given by
\be
	\lambda_{g,k}^*\ =\ \lambda_{g,k}\ -\ \frac{s_{g,k} - s^*}{s_{g,k} - s_{g, k-1}}\log\left(\frac{1 - U}{U}\right) \nonumber
\ee
and
\be
	\lambda_{g,k+1}^* \ =\ \lambda_{g,k}\ +\ \frac{s^*-s_{g,k-1}}{s_{g,k} - s_{g, k-1}}\log\left(\frac{1 - U}{U}\right). \nonumber
\ee
The acceptance probability in the Metropolis-Hastings-Green step can be written as the product of the likelihood ratio, prior ratio, proposal ratio, and Jacobian. For $g=1$, they are given by
\be
likelihood~ ratio &=& \frac{L_{P}(D | \Phi_P^*)}{L_{P}(D | \Phi_P)}, \nonumber \\
 prior ~ ratio &=&  \frac{\textrm{Poisson}(K_1+1 | \alpha_{K_1}) \times \textrm{MVN}_{K_1+2}(\bflambda_1^*|\mu_{\lambda_1}\bfone,~ \sigma_{\lambda_1}^2\Sigma_{\lambda_1}^*)}{\textrm{Poisson}(K_1 | \alpha_{K_1}) \times \textrm{MVN}_{K_1+1}(\bflambda_1|\mu_{\lambda_1}\bfone,~ \sigma_{\lambda_1}^2\Sigma_{\lambda_1})}\nonumber \\
 &&\times \frac{(2K_1+3)(2K_1+2)(s^*-s_{1,k-1})(s_{1,k} - s^*)}{s_{1,\max}^2(s_{1,k} - s_{1,k-1})},\nonumber \\
 proposal ~ ratio &=&  \frac{\pi_{DI}\times(1/(K_1+1))}{\pi_{BI}\times (1/\sharp\{y_{ji1}: \delta_{ji1} = 1\})\times \textrm{Uniform}(U|0, 1)} \nonumber \\
 &=& \frac{\rho\min(1, (K_1+1)/\alpha_{K_1})\sharp\{y_{ji1}: \delta_{ji1} = 1\}}{\rho\min(1, \alpha_{K_1}/(1 + K_1))(K_1+1)} = \frac{\sharp\{y_{ji1}: \delta_{ji1} = 1\}}{\alpha_{K_1}},\nonumber \\
 Jacobian &=& \left|
  \begin{array}{cc}
    d\lambda_{1,k}^*/d\lambda_{1,k} 	&  	d\lambda_{1,k}^*/dU    \\
    d\lambda_{1,k+1}^*/d\lambda_{1,k}         & d\lambda_{1,k+1}^*/dU
\end{array} \right| = \frac{1}{U(1-U)},
 \ee
where $\Phi_P^*$ is $\Phi_P$ with $\lambda_1$ replaced by $\lambda_1^*$.

\subsubsection{Update DI}

For update DI (a death or reverse move), we first sample one of the $K_g$ split times, $s_{g, k}$. The proposal for time splits is given by
\be
	&& (0 = s_{g,0}, ..., s_{g,k-1}, s_{g,k+1},..., s_{g,K_g+1} = s_{g,\max}) \nonumber\\
	& \equiv& (0 = s_{g,0}^*,..., s_{g,k-1}^*,s_{g,k}^*,...,s_{g,K_g}^* = s_{g,\max}). \nonumber
\ee
Following \cite{green1995reversible}:
\be
	(s_{g,k} - s_{g,k-1})\lambda_{g,k} + (s_{g,k+1} - s_{g,k})\lambda_{g,k+1}
 = (s_{g,k+1} - s_{g,k-1})\lambda_{g,k}^*, \nonumber \\
perturbation: ~~~~\frac{e^{\lambda_{g,k+1}}}{e^{\lambda_{g,k}}} = \frac{1-U^*}{U^*},~~~~~~~~~~~~~~~~~~~~~~~\nonumber
 \ee
where $U\sim \textrm{Uniform}(0, 1)$. Then the acceptance probability can be obtained as the product of following four components (for $g=1$):
\be
	likelihood~ ratio &=& \frac{L_{P}(D | \Phi_P^*)}{L_{P}(D | \Phi_P)}, \nonumber \\
 prior ~ ratio &=&  \frac{\textrm{Poisson}(K_1-1 | \alpha_{K_1}) \times \textrm{MVN}_{K_1}(\bflambda_1^*|\mu_{\lambda_1}\bfone,~ \sigma_{\lambda_1}^2\Sigma_{\lambda_1}^*)}{\textrm{Poisson}(K_1 | \alpha_{K_1}) \times \textrm{MVN}_{K_1+1}(\bflambda_1|\mu_{\lambda_1}\bfone,~ \sigma_{\lambda_1}^2\Sigma_{\lambda_1})}\nonumber\\
 &&\times \frac{s_{1,\max}^2(s_{1,k+1} - s_{1,k-1})}{(2K_1+1)2K_1(s_{1,k}-s_{1,k-1})(s_{1,k+1}-s_{1,k})},\nonumber \\
 proposal ~ ratio &=&  \frac{\pi_{BI}\times(1/\sharp\{y_{ji1}: \delta_{ji1} = 1\})}{\pi_{DI}\times(1/K_1)} \nonumber \\
 &=& \frac{\rho\min(1,\alpha_{K_1}/K_1)K_1}{\rho\min(1, K_1/\alpha_{K_1})\sharp\{y_{ji1}: \delta_{ji1} = 1\}} = \frac{\alpha_{K_1}}{\sharp\{y_{ji1}: \delta_{ji1} = 1\}},\nonumber \\
Jacobian &=& \left|
  \begin{array}{cc}
    d\lambda_{1k}/d\lambda_{1k}^* 	&  	d\lambda_{1k}/d\lambda_{1,k+1}^*    \\
    dU/d\lambda_{1k}^*         & 		dU/d\lambda_{1,k+1}^* 
\end{array} \right| = U(1-U). \nonumber
\ee

\newpage

\section{The potential use of existing methods}\label{existing}

The methods in the main manuscript were developed specifically for on-going collaboration examining the risk of readmission following a diagnosis of pancreatic cancer. As indicated in the manuscript, the current standard for the analysis of cluster-correlated readmission data is a logisitic-Normal generalized linear mixed model. This model ignores death as a competing risk, however, and, as such, is inappropriate in for the study of pancreatic cancer due to its strong force of mortality.

Viewing the data arising in the pancreatic cancer as \textit{cluster-correlated semi-competing risks data}, the existing literature does have a number of options that could be considered. Here we review two of these options, specifically those proposed in \cite{liquet2012investigating} and \cite{gorfine2011frailty}. For the former, we note that the methods have been implemented in the \texttt{frailtypack} package for \texttt{R}.

For convenience, expressions (4)-(6) from the main manuscript that the describe the key features of the proposed hierarchical model are repeated here:
\be
	h_1(t_{ji1}; \gamma_{ji}, \bfX_{ji1}, V_{j1}) &=& \gamma_{ji}\ h_{01}(t_{ji1})\ \mbox{exp}\{\bfX_{ji1}^T\bfbeta_1\ +\ V_{j1}\}, \hspace*{0.25in}  t_{ji1} > 0 \nonumber \\
	h_2(t_{ji2}; \gamma_{ji}, \bfX_{ji2}, V_{j2}) &=& \gamma_{ji}\ h_{02}(t_{ji2})\ \mbox{exp}\{\bfX_{ji2}^T\bfbeta_2\ +\ V_{j2}\}, \hspace*{0.25in}  t_{ji2} > 0 \nonumber \\
	h_3(t_{ji2}|t_{ji1}; \gamma_{ji}, \bfX_{ji3}, V_{j3}) &=&\gamma_{ji} \ h_{03}(t_{ji2}| t_{ji1})\ \mbox{exp}\{\bfX_{ji3}^T\bfbeta_3\ +\ V_{j3}\}, \hspace*{0.25in}  t_{ji2} > t_{ji1}, \nonumber
\ee

\subsection{\cite{liquet2012investigating}}

The R package \texttt{frailtypack} provides several classes of frailty models for multivariate survival data including shared frailty models, additive frailty models, nested frailty models, joint frailty models \citep{rondeau2012frailtypack}. Among these, the \emph{shared frailty model} and the \emph{joint frailty model} are most relevant the context we consider; additionally, these models form the basis for the analyses presented in \cite{liquet2012investigating}. Here we provide a summary of these two classes using the notation developed in the manuscript, as well as an overview of their drawbacks in regard to the analysis of cluster-correlated semi-competing risks data. 

\subsubsection{The shared frailty model}

In the shared frailty model, the hazard function for the subject $i$ in the cluster $j$ conditional on the cluster-specific shared frailty term $\eta_j$ = ($\eta_{j1}$, $\eta_{j2}$, $\eta_{j3}$) is given by 
\be
	h_1(t_{ji1};  \bfX_{ji1}, \eta_{j1}) &=& \eta_{j1} h_{01}(t_{ji1})\ \mbox{exp}\{\bfX_{ji1}^T\bfbeta_1\}, \hspace*{0.25in}  t_{ji1} > 0 \nonumber \\
	h_2(t_{ji2};  \bfX_{ji2}, \eta_{j2}) &=& \eta_{j2} h_{02}(t_{ji2})\ \mbox{exp}\{\bfX_{ji2}^T\bfbeta_2\}, \hspace*{0.25in}  t_{ji2} > 0 \nonumber \\
	h_3(t_{ji2}|t_{ji1}; \bfX_{ji3}, \eta_{j3}) &=& \eta_{j3} h_{03}(t_{ji2}-t_{ji1})\ \mbox{exp}\{\bfX_{ji3}^T\bfbeta_3\}, \hspace*{0.25in}  t_{ji2} > t_{ji1}, \label{eq:sf}
\ee
Key features of this model, in relation to the proposed framework are:
\begin{itemize}
	\item[$\bullet$] Cluster-specific effects are represented via the ($\eta_{j1}$, $\eta_{j2}$, $\eta_{j3}$), each of which is assigned an independent univariate parametric distribution (either a log-Normal or a Gamma). As such, the model does not permit the characterization of covariation between the cluster-specific random effects. In contrast, the proposed methods provides analysts with two choices for the joint distribution of the $V_j$'s: a parametric MVN or a non-parametric DPM-MVN.
	\item[$\bullet$] There is no patient-specific term, analogous to the $\gamma_{ji}$ in the proposed model. As such a potentially important source of within-subject correlation between $T_1$ and $T_2$ is not accounted for.
	\item[$\bullet$] Similar to the propose methods, however, is that the baseline hazard function for $h_3()$ can be specified non-parametrically (via a spline) or parametrically (using the Weibull distribution).
	\item[$\bullet$] Although not evident from the model specification, estimation of the shared frailty model is based on three separate fits of the three models. In contrast, because the proposed model considers several components of covariation (i.e. covariation among the $V_j$'s and the patient-specific $\gamma_{ji}$'s) we perform estimation/inference using single likelihood. Indeed for the shared frailty model to accommodate these components of covariation, a completely new framework for estimation/inference would need to be developed.
	\item[$\bullet$] Estimation of the ($\eta_{j1}$, $\eta_{j2}$, $\eta_{j3}$) proceeds using empirical Bayes (after estimation of the remaining components via an integrated likelihood). Uncertainty for these estimates are only available when their distributions are taken to be Gamma distributions.
\end{itemize}

\subsubsection{The joint frailty model}

Two variations of a joint frailty model have been implemented in the \texttt{frailtypack} package. The first was developed for the analysis of a recurrent non-terminal event and a terminal event and specifies a single hazard function for each. Specifically, the model is given by:
\be
	h_1(t_{ki1} | \omega_i) &=& \omega_ir_0(t_{ki1})\ \mbox{exp}\{\bfX_{i1}^T\bfbeta_1\}, ~~~ \textrm{for recurrent non-terminal event} \nonumber \\
	 h_2(t_{i2} | \omega_i) &=& \omega_i^{\alpha}h_0(t_{i2})\ \mbox{exp}\{\bfX_{i2}^T\bfbeta_2\},~~~ \textrm{for the terminal event} \nonumber \\
	\omega_{i} &\sim& \textrm{Gamma}(1/\theta, 1/\theta) \label{eq:jf1}.
\ee
where $\omega_{i}$ is a common subject-specific frailty representing unobserved covariates that impact both events. We note that this is specification is similar to the model proposed by \cite{Liu:Wolf:2004}.

The second joint frailty model implemented in the \texttt{frailtypack} package is for modeling two clustered survival outcomes. Specifically, the model posits that the event-specific hazard functions for the $j^{th}$ cluster are structured as follows:
\be
	h(t_{ji1} | \eta_j) &=& h_{01}(t_{ji1})\ \mbox{exp}\{\bfX_{ji1}^T\bfbeta_1 + \eta_j\}, ~~~ \textrm{for any event} \nonumber \\
	h(t_{ji2} | \eta_j) &=& h_{02}(t_{ji2})\ \mbox{exp}\{\bfX_{ji2}^T\bfbeta_2 + \alpha\eta_j\}, ~~~ \textrm{for the terminal event} \nonumber \\
	\eta_{j} &\sim& \textrm{Normal}(0, \sigma^2) \label{eq:jf2}
\ee

In relation to the context we consider, the central limitation of these model is that they only consider a single level of the two-level hierarchy inherent to cluster-correlated semi-competing risks data. Specifically, as applied and described in the \texttt{fraitypack} package, the first model only considers patient-specific effects while the second model only considers cluster-specific effects. As such neither model would be appropriate for our motivating application since (i) ignoring cluster-specific effects means that one cannot address several of our key scientific questions and (ii) ignoring patient-level effects can result in substantial bias (see the simulation studies in Section 5 of the main manuscript).

We also note that a second limitation is that model (\ref{eq:jf2}) does not consider the transition from the non-terminal event to the terminal event; that is there is no analogue for $h_3()$ in the model. This represents a limitation in the sense that information readily available in the data is ignored. In the motivating application in the main manuscript, for example, the fact that the time of death following readmission within 90 days is known for 608 (11.5\%) patients is ignored. Finally, although model (\ref{eq:jf1}) does permit a patient to transition from the non-terminal state to the terminal state, this transition is assumed to occur at the same rate at which a patient who is in the initial state transitions directly into the terminal state; that is, in contrast to the proposed model that distinguishes $h_2()$ from $h_3()$, model (\ref{eq:jf1}) only has a single hazard for the terminal event.

\subsection{\cite{gorfine2011frailty}}

\cite{gorfine2011frailty} explicitly consider the related but distinct problem of analyzing \textit{cluster correlated competing risk data} for which $T_1$ and $T_2$ are both terminal events (i.e. death due to two causes). Towards analyzing such data, they propose the following hierarchical model:
\be
	h(t_{ji1} | \bfX_{ji}, \epsilon_{j1}(t_{ji1})) &=& h_{01}(t_{ji1})\ \mbox{exp}\{\bfX_{ji}^T\bfbeta_1 + \epsilon_{j1}(t_{ji1})\}, ~~~ \textrm{for cause 1} \nonumber \\
	h(t_{ji2} | \bfX_{ji}, \epsilon_{j2}(t_{ji2})) &=& h_{02}(t_{ji2})\ \mbox{exp}\{\bfX_{ji}^T\bfbeta_2 + \epsilon_{j2}(t_{ji2})\}, ~~~ \textrm{for cause 2} \label{eq:gorfin1}
\ee
to describe the risk of transitioning into one of the two terminal states for the $i^{th}$ patient in the $j^{th}$ cluster. As part of their development, \cite{gorfine2011frailty} provide a framework within which the distribution of the cluster-specific $\epsilon_{jg}(t)$ terms can be flexibly specified.

While this flexibility is very appealing, direct application of this model to our motivating application would be subject to a number of limitations mainly because the method was not designed for the cluster-correlated semi-competing risks setting. Specifically,
\begin{itemize}
	\item[$\bullet$] Similar to the joint frailty model given by (\ref{eq:jf2}), the application of model (\ref{eq:gorfin1}) means that one would ignore information in the data on the transition from the non-terminal event to the terminal event; that is, there is not analogue of $h_3()$.
	\item[$\bullet$] Although model (\ref{eq:gorfin1}) includes cluster-specific random effects, it does not include specification of patient-specific terms analogous to $\gamma_{ji}$ in the proposed model. As is clear from the simulations presented in Section 5 of the main manuscript, ignoring this component of variation can lead to substantial bias in estimation and poor inferential properties in the cluster-correlated semi-competing risks setting.
\end{itemize}

\newpage

\section{Simulation Results}\label{addSim}

\begin{figure}[h!]
\centering
\includegraphics[width = 6in]{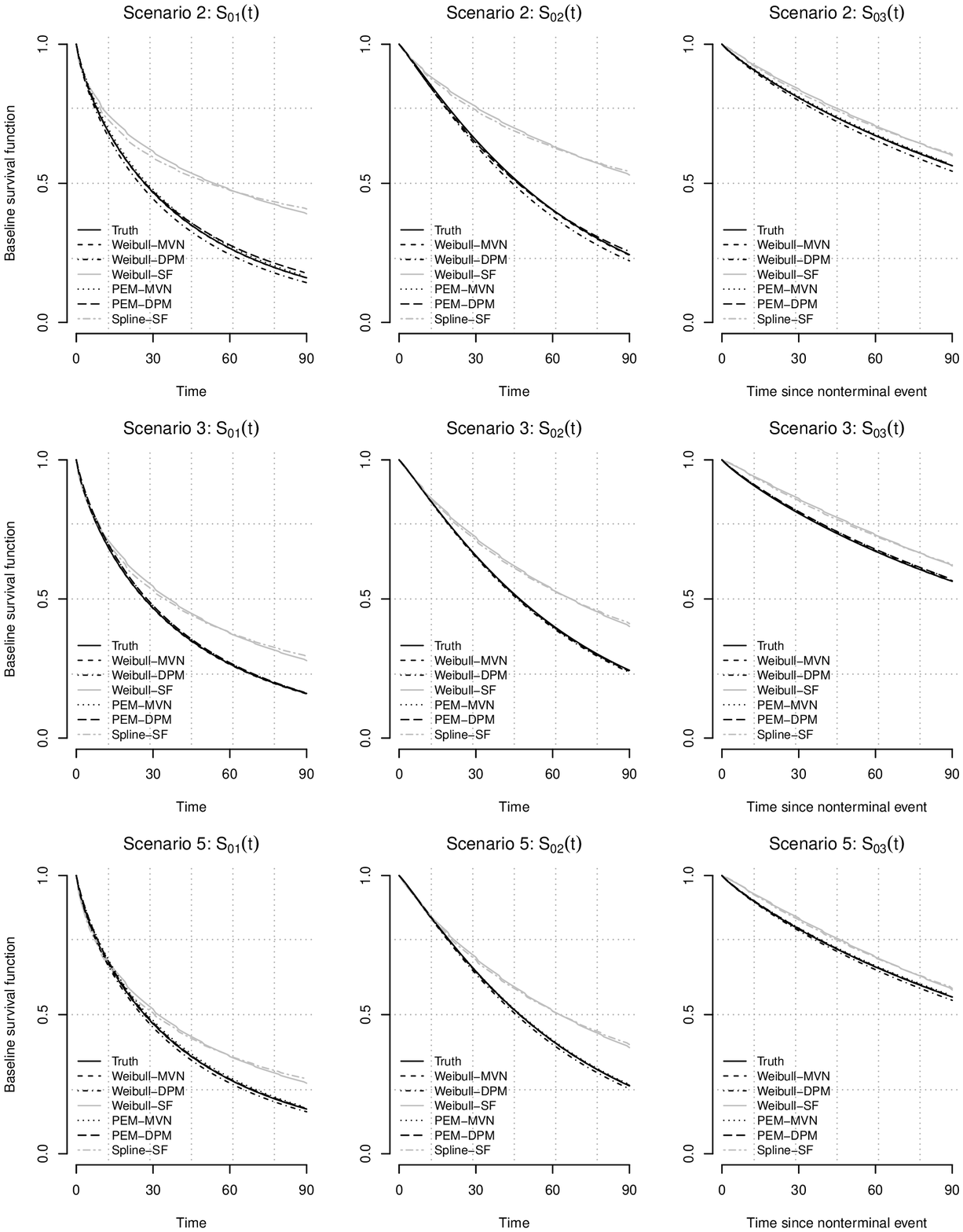}
\caption{Estimated transition-specific baseline survival functions, $S_{0g}(\cdot)$=$\exp(-H_{0g}(\cdot))$, for each six analyses described in Section 5 under simulation scenarios 2, 3 and 5. \label{fig:sim2BS}}
\end{figure}

In order to supplement the results from simulation studies, we provide estimated percent bias, coverage probability, and average relative width of 95\% credible/confidence intervals for $\bfbeta_1$, $\bfbeta_2$, $\bfbeta_3$, and $\theta$ for our four proposed models and four types of SF models of \cite{liquet2012investigating} in Table \ref{tab:simReg1}-\ref{tab:simWidths3}. We also provide estimated transition-specific baseline survival functions for the models under simulation scenarios 2,3, and 5 in Figure \ref{fig:sim2BS}. Note that since results from SF models are almost identical between models that adopt the independent gamma distributions for cluster-specific random effects and those that adopt the independent log-Normal distributions, we only present the results from SF models with the gamma cluster-specific random effects in Figure \ref{fig:sim2BS}. We also present Table \ref{tab:simMSEPall} that augments Table 6 in the main manuscript by additionally presenting results for the \cite{liquet2012investigating}'s models that adopt independent log-Normal distributions for the cluster-specific random effects.

The results presented in this section are generally consistent with the conclusions we drew in the main paper: contrary to the existing SF models, our proposed models yielded a small bias and coverage probability estimated to be close to the nominal 0.95 for regression parameters and $\theta$ (except scenario 4 for which $\theta$=0); all four of the proposed models estimate the three baseline survival functions very well.

\begin{landscape}

\begin{table}[h!]
\centering
\caption{Estimated percent bias and coverage probability for $\bfbeta_1$ and $\theta$ for six analyses described in Section 5.2, across six simulation scenarios given in Table 3. Throughout values are based on results from $R$=500 simulated datasets. \label{tab:simReg1}}
\scalebox{0.75}{
\begin{tabular}{ccrc  cccccccc c  cccccccc}
  \hline
	 	&&   && \multicolumn{8}{c}{Percent Bias} && \multicolumn{8}{c}{Coverage Probability} \\\cline{5-12}\cline{14-21}   
Scenario	 	&& 	\mc{True}	&& Weibull & Weibull & Weibull & Weibull  & PEM & PEM & Spline &Spline &&  Weibull & Weibull & Weibull & Weibull  & PEM & PEM & Spline &Spline \\
			 && \mc{value}	&& -MVN & -DPM 	 &-SF$_{\mathcal{G}}$$^a$ 	& -SF$_{\mathcal{LN}}$$^b$	& -MVN& -DPM & -SF$_{\mathcal{G}}$& -SF$_{\mathcal{LN}}$ && -MVN & -DPM &-SF$_{\mathcal{G}}$ & -SF$_{\mathcal{LN}}$& -MVN& -DPM & -SF$_{\mathcal{G}}$& -SF$_{\mathcal{LN}}$ \\	   
  \hline
	 &$\beta_{11}$ & 0.50 && 0.1 & 0.2 & -19.8 & -21.3 & 0.4 & 0.4 & -21.0 & -20.8 && 0.96 & 0.96 & 0.01 & 0.01 & 0.95 & 0.96 & 0.00 & 0.00 \\ 
 1	 & $\beta_{12}$& 0.80 && 0.2 & 0.3 & -19.7 & -21.3 & 0.5 & 0.4 & -21.0 & -20.8 && 0.95 & 0.95 & 0.00 & 0.00 & 0.96 & 0.97 & 0.00 & 0.00 \\ 
 	& $\beta_{13}$& -0.50 && 0.3 & 0.3 & -19.8 & -18.8 & 0.3 & 0.3 & -21.2 & -20.9 && 0.97 & 0.96 & 0.31 & 0.31 & 0.96 & 0.96 & 0.25 & 0.26 \\ 
 	 & $\theta$	& 0.50 && 1.0 & 1.3 &  &  & 1.4 & 1.2 &  &  				&& 0.95 & 0.95 &  &  & 0.93 & 0.94 &  &  \\ 
   \hline  
	 &$\beta_{11}$ & 0.50 && -0.1 & -0.0 & -31.8 & -33.4 & 0.1 & 0.1 & -32.8 & -32.8 && 0.94 & 0.94 & 0.00 & 0.00 & 0.94 & 0.93 & 0.00 & 0.00 \\ 
 2	 & $\beta_{12}$& 0.80 && 0.1 & 0.2 & -31.7 & -33.3 & 0.4 & 0.3 & -32.7 & -32.7 && 0.97 & 0.97 & 0.00 & 0.00 & 0.94 & 0.95 & 0.00 & 0.00 \\ 
 	& $\beta_{13}$& -0.50 && 1.2 & 1.3 & -31.1 & -29.2 & 1.1 & 1.1 & -32.2 & -32.2 && 0.94 & 0.95 & 0.05 & 0.05 & 0.94 & 0.94 & 0.04 & 0.04 \\ 
 	 & $\theta$	& 1.00 && 0.4 & 0.7 &  &  & 0.7 & 0.6 &  &  				&& 0.94 & 0.95 &  &  & 0.94 & 0.95 &  &  \\ 
   \hline  
	 &$\beta_{11}$ & 0.50 && 0.3 & 0.3 & -19.9 & -20.7 & 0.7 & 0.7 & -21.0 & -20.9 && 0.94 & 0.94 & 0.00 & 0.00 & 0.93 & 0.94 & 0.00 & 0.00 \\ 
 3	 & $\beta_{12}$& 0.80 && 0.4 & 0.4 & -19.8 & -20.7 & 0.8 & 0.8 & -20.9 & -20.8 && 0.94 & 0.94 & 0.00 & 0.00 & 0.94 & 0.94 & 0.00 & 0.00 \\ 
 	& $\beta_{13}$& -0.50 && 0.4 & 0.3 & -20.1 & -19.7 & 0.5 & 0.6 & -21.2 & -21.2 && 0.96 & 0.96 & 0.31 & 0.29 & 0.95 & 0.96 & 0.27 & 0.27 \\ 
 	 & $\theta$	& 0.50 && 2.0 & 2.1 &  &  & 3.2 & 3.2 &  &  				&& 0.96 & 0.96 &  &  & 0.93 & 0.95 &  &  \\ 
   \hline  
	 &$\beta_{11}$ & 0.50 && 3.7 & 3.7 & 0.2 & -2.9 & 4.7 & 4.6 & 0.3 & 0.6 		&& 0.87 & 0.86 & 0.96 & 0.91 & 0.81 & 0.83 & 0.96 & 0.95 \\ 
 4	 & $\beta_{12}$& 0.80 && 3.6 & 3.6 & -0.0 & -3.1 & 4.5 & 4.5 & 0.1 & 0.4 		&& 0.80 & 0.79 & 0.95 & 0.89 & 0.69 & 0.70 & 0.95 & 0.95 \\ 
 	& $\beta_{13}$& -0.50 && 4.0 & 4.0 & 0.2 & 7.3 & 4.8 & 4.7 & 0.2 & 0.6 		&& 0.93 & 0.94 & 0.94 & 0.88 & 0.93 & 0.93 & 0.93 & 0.94 \\ 
 	 & $\theta$	& 0.00 && & &  &  & & &  &  					&&  &  &  &  & & &  &  \\ 
   \hline  
	 &$\beta_{11}$ & 0.50 && -0.3 & 0.1 & -20.3 & -24.6 & 0.0 & 0.3 & -21.1 & -20.9 && 0.94 & 0.95 & 0.00 & 0.00 & 0.96 & 0.96 & 0.00 & 0.00 \\ 
 5	 & $\beta_{12}$& 0.80 && 0.0 & 0.3 & -20.0 & -24.6 & 0.3 & 0.6 & -20.9 & -20.7 && 0.95 & 0.95 & 0.00 & 0.00 & 0.96 & 0.96 & 0.00 & 0.00 \\ 
 	& $\beta_{13}$& -0.50 && -0.2 & 0.2 & -20.4 & -13.7 & -0.2 & 0.2 & -21.3 & -21.1 && 0.94 & 0.94 & 0.29 & 0.26 & 0.94 & 0.94 & 0.25 & 0.26 \\ 
 	 & $\theta$	& 0.50 && -0.2 & 1.0 &  &  & 0.4 & 1.3 &  &  				&& 0.95 & 0.95 &  &  & 0.95 & 0.96 &  &  \\ 
   \hline  
	 &$\beta_{11}$ & 0.50 && 9.3 & 9.4 & -22.1 & -23.6 & 0.4 & 0.3 & -25.9 & -25.1 && 0.58 & 0.57 & 0.00 & 0.00 & 0.94 & 0.94 & 0.00 & 0.00 \\ 
 6	 & $\beta_{12}$& 0.80 && 9.7 & 9.8 & -22.0 & -23.5 & 0.5 & 0.5 & -25.8 & -25.0 && 0.20 & 0.20 & 0.00 & 0.00 & 0.94 & 0.95 & 0.00 & 0.00 \\ 
 	& $\beta_{13}$& -0.50 && 10.2 & 10.2 & -21.6 & -18.2 & 0.8 & 0.7 & -26.1 & -24.9 && 0.81 & 0.80 & 0.21 & 0.21 & 0.93 & 0.94 & 0.10 & 0.10 \\ 
 	 & $\theta$	& 0.50 && 52.8 & 53.0 &  &  & 1.8 & 1.7 &  &  					&& 0.00 & 0.00 &  &  & 0.95 & 0.96 &  &  \\ 
   \hline
   \multicolumn{21}{l}{\footnotesize $^a$ The SF models that adopt the independent gamma distributions for cluster-specific random effects}\\
   \multicolumn{21}{l}{\footnotesize $^b$ The SF models that adopt the independent log-Normal distributions for cluster-specific random effects}   
\end{tabular}}
\end{table}

\begin{table}[h!]
\centering
\caption{Estimated percent bias and coverage probability for $\bfbeta_2$ for six analyses described in Section 5.2, across six simulation scenarios given in Table 3. Throughout values are based on results from $R$=500 simulated datasets. \label{tab:simReg2}}
\scalebox{0.75}{
\begin{tabular}{ccrc  cccccccc c  cccccccc}
  \hline
	 	&&   && \multicolumn{8}{c}{Percent Bias} && \multicolumn{8}{c}{Coverage Probability} \\\cline{5-12}\cline{14-21}   
Scenario	 	&& 	\mc{True}	&& Weibull & Weibull & Weibull & Weibull  & PEM & PEM & Spline &Spline &&  Weibull & Weibull & Weibull & Weibull  & PEM & PEM & Spline &Spline \\
			 && \mc{value}	&& -MVN & -DPM 	 &-SF$_{\mathcal{G}}$$^a$ 	& -SF$_{\mathcal{LN}}$$^b$	& -MVN& -DPM & -SF$_{\mathcal{G}}$& -SF$_{\mathcal{LN}}$ && -MVN & -DPM &-SF$_{\mathcal{G}}$ & -SF$_{\mathcal{LN}}$& -MVN& -DPM & -SF$_{\mathcal{G}}$& -SF$_{\mathcal{LN}}$ \\	   
  \hline
	 &$\beta_{21}$ & 0.50 && -0.1 & -0.0 & -27.9 & -25.9 & -0.2 & -0.3 & -26.1 & -26.1 && 0.93 & 0.93 & 0.00 & 0.00 & 0.94 & 0.94 & 0.00 & 0.00 \\ 
 1	 & $\beta_{22}$& 0.80 && 0.3 & 0.4 & -27.9 & -25.6 & 0.2 & 0.1 & -25.8 & -25.7 && 0.96 & 0.96 & 0.00 & 0.00 & 0.96 & 0.96 & 0.00 & 0.00 \\ 
 	& $\beta_{23}$& -0.50 && 1.3 & 1.4 & -26.5 & -21.1 & 0.9 & 0.9 & -25.2 & -25.1 && 0.94 & 0.93 & 0.34 & 0.35 & 0.95 & 0.95 & 0.34 & 0.34 \\ 
   \hline  
	 &$\beta_{21}$ & 0.50 && -0.1 & 0.1 & -39.2 & -39.2 & -0.2 & -0.2 & -39.9 & -39.8 && 0.96 & 0.96 & 0.00 & 0.00 & 0.96 & 0.96 & 0.00 & 0.00 \\ 
 2	 & $\beta_{22}$& 0.80 && 0.2 & 0.3 & -39.1 & -39.0 & 0.0 & 0.0 & -39.6 & -39.6 && 0.96 & 0.97 & 0.00 & 0.00 & 0.96 & 0.97 & 0.00 & 0.00 \\ 
 	& $\beta_{23}$& -0.50 && 1.3 & 1.4 & -38.5 & -38.3 & 0.8 & 0.8 & -39.2 & -39.2 && 0.93 & 0.94 & 0.07 & 0.07 & 0.93 & 0.93 & 0.06 & 0.06 \\ 
   \hline  
	 &$\beta_{21}$ & 0.50 && 0.3 & 0.3 & -27.4 & -24.9 & 0.3 & 0.3 & -25.7 & -25.6 && 0.95 & 0.95 & 0.01 & 0.01 & 0.95 & 0.96 & 0.00 & 0.00 \\ 
 3	 & $\beta_{22}$ & 0.80 && 0.5 & 0.5 & -27.5 & -24.8 & 0.5 & 0.5 & -25.5 & -25.4 && 0.95 & 0.95 & 0.00 & 0.00 & 0.94 & 0.95 & 0.00 & 0.00 \\ 
 	& $\beta_{23}$& -0.50 && 2.5 & 2.5 & -24.9 & -23.2 & 2.3 & 2.5 & -24.2 & -24.1 && 0.95 & 0.95 & 0.37 & 0.38 & 0.94 & 0.94 & 0.35 & 0.36 \\ 
   \hline  
	 &$\beta_{21}$ & 0.50 && 4.5 & 4.5 & -4.9 & -4.4 & 4.7 & 4.7 & -0.6 & -0.6 && 0.89 & 0.89 & 0.91 & 0.90 & 0.87 & 0.88 & 0.96 & 0.96 \\ 
 4	 & $\beta_{22}$& 0.80 && 4.6 & 4.6 & -5.2 & -4.6 & 4.8 & 4.8 & -0.6 & -0.5 && 0.78 & 0.78 & 0.88 & 0.88 & 0.76 & 0.79 & 0.92 & 0.92 \\ 
 	& $\beta_{23}$& -0.50 && 5.4 & 5.4 & 25.8 & 18.1 & 5.4 & 5.5 & -0.1 & -0.0 && 0.92 & 0.92 & 0.88 & 0.88 & 0.91 & 0.93 & 0.93 & 0.92 \\ 
   \hline  
	 &$\beta_{21}$  & 0.50 && -0.3 & 0.2 & -26.6 & -25.1 & -0.4 & -0.0 & -26.1 & -25.8 && 0.94 & 0.95 & 0.00 & 0.00 & 0.94 & 0.95 & 0.00 & 0.00 \\ 
 5	 & $\beta_{22}$ & 0.80 && -0.4 & 0.1 & -26.8 & -25.3 & -0.5 & -0.1 & -26.2 & -25.9 && 0.96 & 0.96 & 0.00 & 0.00 & 0.96 & 0.96 & 0.00 & 0.00 \\ 
 	& $\beta_{23}$& -0.50 && 0.1 & 0.5 & -27.2 & -24.9 & -0.4 & 0.1 & -26.1 & -25.8 && 0.95 & 0.95 & 0.33 & 0.35 & 0.95 & 0.95 & 0.30 & 0.31 \\ 
   \hline  
	 &$\beta_{21}$ & 0.50 && 9.2 & 9.3 & -25.1 & -23.8 & 0.1 & 0.1 & -25.0 & -25.0 && 0.70 & 0.70 & 0.01 & 0.01 & 0.95 & 0.95 & 0.00 & 0.01 \\ 
 6	 & $\beta_{22}$& 0.80 && 9.7 & 9.7 & -25.4 & -23.8 & 0.4 & 0.3 & -24.8 & -24.8 && 0.40 & 0.39 & 0.00 & 0.00 & 0.94 & 0.95 & 0.00 & 0.00 \\ 
 	& $\beta_{23}$& -0.50 && 10.4 & 10.4 & -26.6 & -13.6 & 0.8 & 0.8 & -24.5 & -24.5 && 0.85 & 0.85 & 0.47 & 0.47 & 0.94 & 0.95 & 0.36 & 0.35 \\ 
   \hline
   \multicolumn{21}{l}{\footnotesize $^a$ The SF models that adopt the independent gamma distributions for cluster-specific random effects}\\
   \multicolumn{21}{l}{\footnotesize $^b$ The SF models that adopt the independent log-Normal distributions for cluster-specific random effects}      
\end{tabular}}
\end{table}

\begin{table}[h!]
\centering
\caption{Estimated percent bias and coverage probability for $\bfbeta_3$ for six analyses described in Section 5.2, across six simulation scenarios given in Table 3. Throughout values are based on results from $R$=500 simulated datasets. \label{tab:simReg3}}
\scalebox{0.75}{
\begin{tabular}{ccrc  cccccccc c  cccccccc}
  \hline
	 	&&   && \multicolumn{8}{c}{Percent Bias} && \multicolumn{8}{c}{Coverage Probability} \\\cline{5-12}\cline{14-21}   
Scenario	 	&& 	\mc{True}	&& Weibull & Weibull & Weibull & Weibull  & PEM & PEM & Spline &Spline &&  Weibull & Weibull & Weibull & Weibull  & PEM & PEM & Spline &Spline \\
			 && \mc{value}	&& -MVN & -DPM 	 &-SF$_{\mathcal{G}}$$^a$ 	& -SF$_{\mathcal{LN}}$$^b$	& -MVN& -DPM & -SF$_{\mathcal{G}}$& -SF$_{\mathcal{LN}}$ && -MVN & -DPM &-SF$_{\mathcal{G}}$ & -SF$_{\mathcal{LN}}$& -MVN& -DPM & -SF$_{\mathcal{G}}$& -SF$_{\mathcal{LN}}$ \\	   
  \hline
	 &$\beta_{31}$ & 1.00 && 0.4 & 0.5 & -21.8 & -12.5 & 0.7 & 0.8 & -13.3 & -13.2 && 0.94 & 0.94 & 0.08 & 0.09 & 0.94 & 0.94 & 0.06 & 0.06 \\ 
 1	 & $\beta_{32}$& 1.00 && 0.2 & 0.3 & -20.5 & -9.0 & 0.6 & 0.6 & -9.7 & -9.7 && 0.96 & 0.96 & 0.28 & 0.32 & 0.93 & 0.94 & 0.27 & 0.28 \\ 
 	& $\beta_{33}$& -1.00 && 0.2 & 0.3 & 44.3 & -12.5 & 0.4 & 0.4 & -13.4 & -13.3 && 0.94 & 0.94 & 0.47 & 0.53 & 0.94 & 0.94 & 0.49 & 0.50 \\ 
   \hline  
	 &$\beta_{31}$ & 1.00 && 0.1 & 0.3 & -24.1 & -25.3 & 0.6 & 0.6 & -24.8 & -24.7 && 0.95 & 0.95 & 0.00 & 0.00 & 0.95 & 0.95 & 0.00 & 0.00 \\ 
 2	 & $\beta_{32}$& 1.00 && 0.4 & 0.5 & -22.8 & -24.1 & 0.8 & 0.8 & -23.3 & -23.3 && 0.94 & 0.94 & 0.00 & 0.00 & 0.95 & 0.95 & 0.00 & 0.00 \\ 
 	& $\beta_{33}$& -1.00 && 0.2 & 0.4 & -23.8 & -22.6 & 0.4 & 0.4 & -24.6 & -24.5 && 0.96 & 0.96 & 0.08 & 0.07 & 0.95 & 0.95 & 0.06 & 0.07 \\ 
   \hline  
	 &$\beta_{31}$ & 1.00 && 0.6 & 0.6 & -21.4 & -14.4 & 1.1 & 1.1 & -12.6 & -12.5 && 0.96 & 0.96 & 0.10 & 0.14 & 0.95 & 0.95 & 0.09 & 0.09 \\ 
 3	 & $\beta_{32}$& 1.00 && 0.7 & 0.8 & -19.2 & -11.2 & 1.2 & 1.2 & -9.0 & -8.9 && 0.95 & 0.95 & 0.35 & 0.40 & 0.94 & 0.94 & 0.37 & 0.38 \\ 
 	& $\beta_{33}$& -1.00 && 0.3 & 0.3 & 13.1 & -10.1 & 0.5 & 0.6 & -12.6 & -12.5 && 0.95 & 0.95 & 0.52 & 0.57 & 0.93 & 0.94 & 0.54 & 0.54 \\ 
   \hline  
	 &$\beta_{31}$ & 1.00 && 3.4 & 3.4 & -0.6 & 11.9 & 4.1 & 3.9 & 11.0 & 11.3 && 0.87 & 0.88 & 0.11 & 0.12 & 0.84 & 0.86 & 0.15 & 0.15 \\ 
 4	 & $\beta_{32}$& 1.00 && 3.4 & 3.5 & 5.6 & 19.2 & 4.2 & 4.0 & 18.3 & 18.6 && 0.88 & 0.88 & 0.00 & 0.00 & 0.83 & 0.86 & 0.00 & 0.00 \\ 
 	& $\beta_{33}$& -1.00 && 3.0 & 3.0 & 20.6 & 11.6 & 3.3 & 3.5 & 10.3 & 10.7 && 0.95 & 0.95 & 0.51 & 0.57 & 0.94 & 0.94 & 0.62 & 0.61 \\ 
   \hline  
	 &$\beta_{31}$ & 1.00 && -0.0 & 0.5 & -22.4 & -16.0 & 0.4 & 0.8 & -14.2 & -14.0 && 0.97 & 0.96 & 0.05 & 0.07 & 0.96 & 0.97 & 0.04 & 0.05 \\ 
 5	 & $\beta_{32}$& 1.00 && 0.0 & 0.6 & -20.2 & -12.4 & 0.4 & 0.9 & -10.4 & -10.1 && 0.96 & 0.95 & 0.27 & 0.31 & 0.95 & 0.95 & 0.25 & 0.26 \\ 
 	& $\beta_{33}$& -1.00 && 0.0 & 0.5 & 36.3 & -11.4 & 0.3 & 0.6 & -14.1 & -13.9 && 0.94 & 0.94 & 0.44 & 0.49 & 0.95 & 0.94 & 0.46 & 0.47 \\ 
   \hline  
	 &$\beta_{31}$ & 1.00 && 8.4 & 8.5 & -28.5 & -28.5 & 0.5 & 0.5 & -30.4 & -30.3 && 0.31 & 0.30 & 0.00 & 0.00 & 0.96 & 0.95 & 0.00 & 0.00 \\ 
 6	 & $\beta_{32}$& 1.00 && 8.9 & 9.0 & -20.3 & -20.3 & 0.6 & 0.6 & -22.0 & -21.8 && 0.28 & 0.27 & 0.00 & 0.00 & 0.95 & 0.95 & 0.00 & 0.00 \\ 
 	& $\beta_{33}$& -1.00 && 8.8 & 8.9 & -27.9 & -27.8 & 0.4 & 0.4 & -30.2 & -30.1 && 0.67 & 0.67 & 0.00 & 0.00 & 0.95 & 0.95 & 0.00 & 0.00 \\ 
   \hline
   \multicolumn{21}{l}{\footnotesize $^a$ The SF models that adopt the independent gamma distributions for cluster-specific random effects}\\
   \multicolumn{21}{l}{\footnotesize $^b$ The SF models that adopt the independent log-Normal distributions for cluster-specific random effects}         
\end{tabular}}
\end{table}

\end{landscape}

\begin{table}[ht]
\centering
\caption{Average relative width of 95\% credible/confidence intervals for $\bfbeta_1$ and $\theta$, with the Weibull-MVN model taken as the referent, across six simulation scenarios given in Table 3. Throughout values are based on results from $R$=500 simulated datasets. \label{tab:simWidths1}}
\scalebox{0.9}{
\begin{tabular}{cc cccccccc}
  \hline
Scenario	&& Weibull & Weibull & Weibull & Weibull & PEM & PEM & Spline & Spline\\
	 	&& -MVN & -DPM &-SF$_{\mathcal{G}}$$^a$ &-SF$_{\mathcal{LN}}$$^b$ & -MVN& -DPM & -SF$_{\mathcal{G}}$ &-SF$_{\mathcal{LN}}$\\	 
  \hline
	& $\beta_{11}$ 	& 1.00 & 1.00 & 0.81 & 0.78 & 1.02 & 1.02 & 0.81 & 0.81 \\ 
 1	& $\beta_{12}$ & 1.00 & 1.00 & 0.77 & 0.74 & 1.04 & 1.04 & 0.77 & 0.77 \\ 
 	& $\beta_{13}$ 	& 1.00 & 1.00 & 0.84 & 0.81 & 1.00 & 1.01 & 0.83 & 0.84 \\ 
 	& $\theta$	 	& 1.00 & 1.00 &  &  & 1.10 & 1.12 &  &  \\ 
    \hline    
	& $\beta_{11}$ 	& 1.00 & 1.00 & 0.73 & 0.70 & 1.02 & 1.02 & 0.73 & 0.73 \\ 
 2	& $\beta_{12}$ & 1.00 & 1.00 & 0.69 & 0.66 & 1.03 & 1.04 & 0.69 & 0.69 \\ 
 	& $\beta_{13}$ 	& 1.00 & 1.00 & 0.76 & 0.72 & 1.00 & 1.00 & 0.76 & 0.76 \\ 
 	& $\theta$	 	& 1.00 & 1.00 &  &  & 1.12 & 1.14 &  &  \\ 
    \hline    
	& $\beta_{11}$ 	& 1.00 & 1.00 & 0.81 & 0.78 & 1.02 & 1.02 & 0.81 & 0.81 \\ 
 3	& $\beta_{12}$ & 1.00 & 1.00 & 0.76 & 0.74 & 1.04 & 1.04 & 0.77 & 0.77 \\ 
 	& $\beta_{13}$ 	& 1.00 & 1.00 & 0.83 & 0.80 & 1.00 & 1.01 & 0.83 & 0.83 \\ 
 	& $\theta$	 	& 1.00 & 1.00 &  &  & 1.10 & 1.13 &  &  \\ 
    \hline    
	& $\beta_{11}$ 	& 1.00 & 1.00 & 0.95 & 0.90 & 1.02 & 1.01 & 0.96 & 0.96 \\ 
 4	& $\beta_{12}$ & 1.00 & 1.00 & 0.94 & 0.88 & 1.03 & 1.03 & 0.95 & 0.95 \\ 
 	& $\beta_{13}$ 	& 1.00 & 1.00 & 0.96 & 0.90 & 1.01 & 1.01 & 0.96 & 0.96 \\ 
 	& $\theta$	 	& 1.00 & 1.00 &  &  & 1.09 & 1.09 &  &  \\ 
    \hline    
	& $\beta_{11}$ 	& 1.00 & 1.00 & 0.81 & 0.74 & 1.02 & 1.02 & 0.81 & 0.81 \\ 
 5	& $\beta_{12}$ & 1.00 & 1.00 & 0.77 & 0.70 & 1.03 & 1.03 & 0.77 & 0.77 \\ 
 	& $\beta_{13}$ 	& 1.00 & 1.00 & 0.83 & 0.76 & 1.00 & 1.00 & 0.83 & 0.83 \\ 
 	& $\theta$	 	& 1.00 & 1.00 &  &  & 1.09 & 1.09 &  &  \\ 
    \hline    
	& $\beta_{11}$ 	& 1.00 & 1.00 & 0.74 & 0.71 & 0.94 & 0.95 & 0.73 & 0.74 \\ 
 6	& $\beta_{12}$ & 1.00 & 1.00 & 0.72 & 0.69 & 0.96 & 0.97 & 0.71 & 0.72 \\ 
 	& $\beta_{13}$ 	& 1.00 & 1.00 & 0.76 & 0.72 & 0.93 & 0.93 & 0.75 & 0.75 \\ 
 	& $\theta$	 	& 1.00 & 1.00 &  &  & 0.89 & 0.90 &  &  \\ 
   \hline
   \multicolumn{10}{l}{\footnotesize $^a$ The SF models that adopt the independent gamma distributions for cluster-specific random effects}\\
   \multicolumn{10}{l}{\footnotesize $^b$ The SF models that adopt the independent log-Normal distributions for cluster-specific random effects}            
\end{tabular}}
\end{table}

\begin{table}[ht]
\centering
\caption{Average relative width of 95\% credible/confidence intervals for $\bfbeta_2$, with the Weibull-MVN model taken as the referent, across six simulation scenarios given in Table 3. Throughout values are based on results from $R$=500 simulated datasets. \label{tab:simWidths2}}
\scalebox{0.9}{
\begin{tabular}{cc cccccccc}
  \hline
Scenario	&& Weibull & Weibull & Weibull & Weibull & PEM & PEM & Spline & Spline\\
	 	&& -MVN & -DPM &-SF$_{\mathcal{G}}$$^a$ &-SF$_{\mathcal{LN}}$$^b$ & -MVN& -DPM & -SF$_{\mathcal{G}}$ &-SF$_{\mathcal{LN}}$\\	
  \hline
	& $\beta_{21}$ 	& 1.00 & 1.00 & 0.80 & 0.82 & 1.01 & 1.01 & 0.84 & 0.84 \\ 
 1	& $\beta_{22}$ & 1.00 & 1.00 & 0.75 & 0.78 & 1.02 & 1.02 & 0.79 & 0.79 \\ 
 	& $\beta_{23}$ & 1.00 & 1.00 & 0.83 & 0.85 & 1.00 & 1.00 & 0.87 & 0.87 \\ 
    \hline    
	& $\beta_{21}$ 	& 1.00 & 1.00 & 0.77 & 0.77 & 1.01 & 1.01 & 0.78 & 0.78 \\ 
 2	& $\beta_{22}$ & 1.00 & 1.00 & 0.72 & 0.72 & 1.02 & 1.02 & 0.73 & 0.73 \\ 
 	& $\beta_{23}$ & 1.00 & 1.00 & 0.80 & 0.80 & 1.00 & 1.00 & 0.80 & 0.80 \\ 
    \hline    
	& $\beta_{21}$ 	& 1.00 & 1.00 & 0.80 & 0.83 & 1.01 & 1.01 & 0.83 & 0.83 \\ 
 3	& $\beta_{22}$ & 1.00 & 1.00 & 0.75 & 0.78 & 1.02 & 1.03 & 0.79 & 0.79 \\ 
 	& $\beta_{23}$ & 1.00 & 1.00 & 0.82 & 0.86 & 1.00 & 1.00 & 0.86 & 0.86 \\ 
    \hline    
	& $\beta_{21}$ 	& 1.00 & 1.00 & 0.90 & 0.90 & 1.01 & 1.01 & 0.96 & 0.96 \\ 
 4	& $\beta_{22}$ & 1.00 & 1.00 & 0.88 & 0.88 & 1.01 & 1.01 & 0.95 & 0.95 \\ 
 	& $\beta_{23}$ & 1.00 & 1.00 & 0.91 & 0.91 & 1.00 & 1.00 & 0.97 & 0.97 \\ 
    \hline    
	& $\beta_{21}$ 	& 1.00 & 1.00 & 0.81 & 0.83 & 1.01 & 1.01 & 0.84 & 0.84 \\ 
 5	& $\beta_{22}$ & 1.00 & 1.00 & 0.77 & 0.78 & 1.02 & 1.02 & 0.79 & 0.79 \\ 
 	& $\beta_{23}$ & 1.00 & 1.00 & 0.84 & 0.86 & 1.00 & 1.00 & 0.86 & 0.86 \\ 
    \hline    
	& $\beta_{21}$ 	& 1.00 & 1.00 & 0.78 & 0.79 & 0.96 & 0.96 & 0.82 & 0.82 \\ 
 6	& $\beta_{22}$ & 1.00 & 1.00 & 0.75 & 0.76 & 0.97 & 0.97 & 0.80 & 0.80 \\ 
 	& $\beta_{23}$ & 1.00 & 1.00 & 0.79 & 0.80 & 0.95 & 0.95 & 0.84 & 0.84 \\ 
   \hline
   \multicolumn{10}{l}{\footnotesize $^a$ The SF models that adopt the independent gamma distributions for cluster-specific random effects}\\
   \multicolumn{10}{l}{\footnotesize $^b$ The SF models that adopt the independent log-Normal distributions for cluster-specific random effects}               
\end{tabular}}
\end{table}

\begin{table}[ht]
\centering
\caption{Average relative width of 95\% credible/confidence intervals for $\bfbeta_3$, with the Weibull-MVN model taken as the referent, across six simulation scenarios given in Table 3. Throughout values are based on results from $R$=500 simulated datasets. \label{tab:simWidths3}}
\scalebox{0.9}{
\begin{tabular}{cc cccccccc}
  \hline
Scenario	&& Weibull & Weibull & Weibull & Weibull & PEM & PEM & Spline & Spline\\
	 	&& -MVN & -DPM &-SF$_{\mathcal{G}}$$^a$ &-SF$_{\mathcal{LN}}$$^b$ & -MVN& -DPM & -SF$_{\mathcal{G}}$ &-SF$_{\mathcal{LN}}$\\	
  \hline
	& $\beta_{31}$ 	& 1.00 & 1.00 & 0.72 & 0.83 & 1.01 & 1.01 & 0.83 & 0.83 \\ 
 1	& $\beta_{32}$ & 1.00 & 1.00 & 0.73 & 0.83 & 1.01 & 1.02 & 0.83 & 0.83 \\ 
 	& $\beta_{33}$ 	& 1.00 & 1.00 & 0.75 & 0.86 & 1.00 & 1.00 & 0.86 & 0.86 \\ 
    \hline    
	& $\beta_{31}$ 	& 1.00 & 1.00 & 0.77 & 0.75 & 1.01 & 1.01 & 0.77 & 0.77 \\ 
 2	& $\beta_{32}$ & 1.00 & 1.00 & 0.77 & 0.75 & 1.01 & 1.02 & 0.77 & 0.77 \\ 
 	& $\beta_{33}$ 	& 1.00 & 1.00 & 0.81 & 0.79 & 1.00 & 1.00 & 0.81 & 0.81 \\ 
    \hline    
	& $\beta_{31}$ 	& 1.00 & 1.00 & 0.73 & 0.81 & 1.01 & 1.01 & 0.84 & 0.84 \\ 
 3	& $\beta_{32}$ & 1.00 & 1.00 & 0.74 & 0.81 & 1.01 & 1.02 & 0.84 & 0.84 \\ 
 	& $\beta_{33}$ 	& 1.00 & 1.00 & 0.76 & 0.84 & 1.00 & 1.00 & 0.87 & 0.87 \\ 
    \hline    
	& $\beta_{31}$ 	& 1.00 & 1.00 & 0.85 & 0.97 & 1.01 & 1.01 & 0.98 & 0.98 \\ 
 4	& $\beta_{32}$ & 1.00 & 1.00 & 0.87 & 0.99 & 1.01 & 1.01 & 1.00 & 1.00 \\ 
 	& $\beta_{33}$ 	& 1.00 & 1.00 & 0.85 & 0.96 & 1.00 & 1.00 & 0.97 & 0.97 \\ 
    \hline    
	& $\beta_{31}$ 	& 1.00 & 1.00 & 0.73 & 0.79 & 1.01 & 1.01 & 0.83 & 0.83 \\ 
 5	& $\beta_{32}$ & 1.00 & 1.00 & 0.73 & 0.80 & 1.01 & 1.01 & 0.83 & 0.84 \\ 
 	& $\beta_{33}$ 	& 1.00 & 1.00 & 0.76 & 0.82 & 1.00 & 1.00 & 0.86 & 0.86 \\ 
    \hline    
	& $\beta_{31}$ 	& 1.00 & 1.00 & 0.69 & 0.69 & 0.96 & 0.96 & 0.69 & 0.69 \\ 
 6	& $\beta_{32}$ & 1.00 & 1.00 & 0.72 & 0.72 & 0.97 & 0.97 & 0.72 & 0.72 \\ 
 	& $\beta_{33}$ 	& 1.00 & 1.00 & 0.73 & 0.73 & 0.94 & 0.94 & 0.73 & 0.73 \\ 
   \hline
   \multicolumn{10}{l}{\footnotesize $^a$ The SF models that adopt the independent gamma distributions for cluster-specific random effects}\\
   \multicolumn{10}{l}{\footnotesize $^b$ The SF models that adopt the independent log-Normal distributions for cluster-specific random effects}               
\end{tabular}}
\end{table}

\begin{table}[ht]
\centering
\caption{Mean squared error of prediction ($\times 10^{-2}$) for cluster-specific random effects based on six analyses described in Section 5.2, across six data scenarios given in Table 3. Throughout values are based on results from $R$=500 simulated datasets. \label{tab:simMSEPall}}
\scalebox{0.80}{
\begin{tabular}{cc r r rr rr r r rr rr}
  \hline
Scenario	 & 		& \mc{Weibull} & \mc{Weibull} & \multicolumn{2}{c}{Weibull}	& \multicolumn{2}{c}{Weibull}& \mc{PEM} 	& \mc{PEM} & \multicolumn{2}{c}{Spline}& \multicolumn{2}{c}{Spline}\\ 
		 & 		& \mc{-MVN} & \mc{-DPM} 	& \multicolumn{2}{c}{-SF$_{\mathcal{G}}$$^a$}	& \multicolumn{2}{c}{-SF$_{\mathcal{LN}}$$^b$}  & \mc{-MVN} 	& \mc{-DPM} &\multicolumn{2}{c}{-SF$_{\mathcal{G}}$}	&\multicolumn{2}{c}{-SF$_{\mathcal{LN}}$}\\ 
		 & 		&		    & 			 	& & \%F 					& & \%F$^{\dag}$ 		&			&			 & & \%F				& & \%F\\ 		 
  \hline
	 &$V_1$ 		& 5.25 & 5.27 & 6.40 &  & 10355.30 &  			& 5.27 & 5.27 & 6.39 &  & 10397.30 &  \\ 
1	 & $V_2$		& 7.66 & 7.70 & 8.70 & 17.8 & 11199.69 & 6.6 		& 7.67 & 7.72 & 8.68 & 0.2 & 11131.71 & 0.8 \\ 
	& $V_3$ 		& 9.91 & 9.95 & 12.13 &  & 11221.04 &  			& 9.91 & 9.96 & 12.11 &  & 11214.38 &  \\ 
\hline  
	 &$V_1$ 		& 6.36 & 6.41 & 8.10 &  & 10535.62 &  			& 6.37 & 6.41 & 8.09 &  & 10476.70 &  \\ 
2	 & $V_2$		& 8.76 & 8.85 & 10.23 & 10.4 & 11328.83 & 7.8	 & 8.77 & 8.86 & 10.20 & 0.0 & 11208.54 & 0.6 \\ 
	& $V_3$ 		& 11.13 & 11.19 & 13.85 &  & 11417.49 &  		& 11.13 & 11.19  & 13.91 &  & 11449.00 &  \\ 
\hline  
	 &$V_1$ 		& 5.03 & 5.04 & 6.27 &  & 10398.49 &  			& 5.04 & 5.04  & 6.22 &  & 10329.00 &  \\ 
3	 & $V_2$		 & 6.34 & 6.34 & 8.28 & 15.8 & 11357.53 & 9.2		 & 6.36 & 6.36 & 8.24 & 0.0 & 11348.16 & 0.6 \\ 
	& $V_3$ 		& 7.55 & 7.49 & 11.66 &  & 10932.18 &  			& 7.57 & 7.55 & 11.69 &  & 10895.77 &  \\ 
\hline  
	 &$V_1$ 		& 3.84 & 3.85 & 4.99 &  & 9798.48 &  			& 3.87 & 3.87 	& 5.01 &  & 9765.63 &  \\ 
4	 & $V_2$		& 6.25 & 6.27 & 7.19 & 12.8 & 11076.72 & 12.4 	& 6.25 & 6.27 & 7.12 & 0.4 & 11102.66 & 5.4 \\ 
	& $V_3$ 		& 7.89 & 7.90 & 9.57 &  & 10893.62 &  			& 7.90 & 7.91 & 9.52 &  & 10886.70 &  \\ 
\hline  
	 &$V_1$ 		& 6.95 & 6.26 & 10.87 &  & 10005.24 &  			& 6.96 & 6.27 & 10.86 &  & 9869.41 &  \\ 
5	 & $V_2$		& 11.52 & 10.50 & 14.95 & 12.8 & 11090.98 & 78.4 	& 11.50 & 10.52 & 14.92 & 0.2 & 10976.78 & 76.4 \\ 
	& $V_3$ 		& 15.46 & 14.66 & 25.04 &  & 11156.06 &  		& 15.46 & 14.72  & 24.94 &  & 11073.46 &  \\ 
\hline  
	 &$V_1$ 		& 5.05 & 5.01 & 6.34 &  & 9670.25 & 			 & 4.89 & 4.85  & 6.26 &  & 9804.42 &  \\ 
6	 & $V_2$		& 7.58 & 7.55 & 8.60 & 5.4 & 11259.19 & 9.2 		& 7.41 & 7.39 & 8.49 & 1.4 & 11272.07 & 0.4 \\ 
	& $V_3$ 		& 6.72 & 6.65 & 13.42 &  & 10095.61 &  			& 6.44 & 6.40 & 13.70 &  & 10233.22 &  \\ 
   \hline
   \multicolumn{14}{l}{\footnotesize $^a$ The SF models that adopt the independent gamma distributions for cluster-specific random effects}\\
   \multicolumn{14}{l}{\footnotesize $^b$ The SF models that adopt the independent log-Normal distributions for cluster-specific random effects}\\               
\multicolumn{14}{l}{\footnotesize $^\dag$ \% of times SF models yield at least one of $\hat{\bfV}_j$ being $-\infty$, resulting in MSEP being $\infty$}
\end{tabular}}
\end{table}

\clearpage

\section{Application to Medicare data from New England}\label{additional}

In our main paper, posterior summaries for hazard ratio (HR) parameters for readmission, $\exp(\bfbeta_1)$, from models for which a semi-Markov specification was adopted for $h_{03}(\cdot)$ are presented. In Table \ref{tab:results:smBeta2}-\ref{tab:results:Beta3}, we provide posterior summaries for HR parameters for death without readmission, $\exp(\bfbeta_2)$, and those for death following readmission, $\exp(\bfbeta_3)$, from both Markov and semi-Markov models. We also provide the posterior estimates of $\exp(\bfbeta_1)$ from Markov models in Table \ref{tab:results:Beta1}. 

From Table 4 (in the main paper) and Table \ref{tab:results:smBeta2}-\ref{tab:results:Beta1}, we see the little difference in posterior estimates of HR parameters between Markov and semi-Markov models in this particular application. Therefore, our analyses in this document mainly focus on HR parameters for death ($\exp(\bfbeta_2)$ and $\exp(\bfbeta_3)$) under semi-Markov models. As seen in Table \ref{tab:results:smBeta2} and Table \ref{tab:results:smBeta3}, our proposed framework show how risk for death changes depending on whether or not a patient experiences the readmission. For example, whereas there is evidence of an increased risk of death for long hospital stay among individuals who have not been readmitted (HR 1.10; 95\% CI 1.04, 1.18 in PEM-DPM), the same conclusion cannot be drawn for individuals who have been readmitted (HR 0.98; 95\% CI 0.87, 1.09 in PEM-DPM). In addition, the association between death and two of covariates (age and Charlson/Deyo score) is stronger in this magnitude (i.e. farther from zero) while the association between death and some other covariates (sex, source of entry to initial hospitalization, length of stay, discharged location, and whether or not patients underwent a procedure during the hospitalization) is weakened among patient who have been readmitted. In general, our analyses show evidence of increased risk for death for patients with male gender, older age, initially hospitalized via some route other than ER, higher comorbidity score, a procedure during the hospitalization, a discharge to a place other than home (without care).

\begin{table}[ht]
\centering
\caption{Posterior medians (PM) and 95\% credible intervals (CI) for hazard ratio parameters for death without readmission ($\exp(\bfbeta_2)$) from semi-competing risks data analyses based on semi-Markov models. \label{tab:results:smBeta2}}
\bigskip
\scalebox{0.8}{
\begin{tabular}{ l r r r r}
  \hline
  				&	 Weibull-MVN 	& Weibull-DPM 	&  PEM-MVN 		& PEM-DPM \\
	 			& PM (95\%CI) 		& PM (95\%CI)  	& PM (95\%CI)  	& PM (95\%CI)  \\ 
  \hline
Sex 				&					&					&					& \\
~~~Male 			& 1.00 				& 1.00 				& 1.00				& 1.00\\
~~~Female 		& 0.69 (0.60, 0.79) 		& 0.69 (0.61, 0.78) 		& 0.75 (0.67, 0.83) 		& 0.75 (0.67, 0.83) \\ \\ 
Age$^*$			& 1.09 (1.04, 1.13) 		& 1.09 (1.04, 1.14) 		& 1.07 (1.03, 1.11) 		& 1.07 (1.03, 1.11) \\ \\   
Race 			&					&					&					&\\
~~~White			& 1.00				& 1.00 				& 1.00 				& 1.00\\  
~~~Non-white		& 0.93 (0.70, 1.22) 		& 0.93 (0.70, 1.22) 		& 0.94 (0.74, 1.17) 		& 0.94 (0.75, 1.18) \\  \\
Source of entry to initial &&&&\\
hospitalization & &&&\\
~~~Emergency room& 1.00& 1.00 & 1.00 & 1.00\\
~~~Other facility	& 1.61 (1.41, 1.85) 		& 1.61 (1.41, 1.86) 		& 1.50 (1.33, 1.70) 		& 1.49 (1.32, 1.68) \\  \\
Charlson/Deyo score& &&&\\
~~~$\leq 1$ 		& 1.00& 1.00 & 1.00 & 1.00\\
~~~$> 1$ 			& 1.40 (1.12, 1.71) 		& 1.39 (1.13, 1.73) 		& 1.26 (1.08, 1.50) 		& 1.27 (1.06, 1.51) \\ \\ 
Procedure during hospitalization			&&&&\\
~~~No 		 	& 1.00& 1.00 & 1.00 & 1.00\\
~~~Yes 			& 0.09 (0.07, 0.12) 		& 0.09 (0.07, 0.12) 		& 0.13 (0.10, 0.16) 		& 0.13 (0.10, 0.16) \\    \\
Length of stay$^{**}$& 1.15 (1.07, 1.24) 		& 1.15 (1.06, 1.24) 		& 1.10 (1.04, 1.18) 		& 1.10 (1.04, 1.18) \\  \\
Care after discharge 	& &&&\\
~~~Home 			& 1.00& 1.00 & 1.00 & 1.00\\
~~~Home with care 		& 2.41 (2.00, 2.91) 		& 2.45 (2.02, 2.94) 		& 2.22 (1.85, 2.63) 		& 2.21 (1.90, 2.61) \\ 
~~~Hospice 			& 22.99 (18.08, 30.16) 	& 23.71 (18.28, 31.20) 	& 13.94 (11.22, 17.43) 	& 13.85 (11.33, 17.08) \\ 
~~~ICF/SNF	 		& 5.22 (4.29, 6.39) 		& 5.33 (4.32, 6.45) 		& 4.25 (3.57, 5.06) 		& 4.25 (3.66, 5.01) \\ 
~~~Other 				& 4.81 (3.48, 6.70) 		& 4.93 (3.58, 6.84) 		& 3.79 (2.91, 4.98) 		& 3.81 (2.94, 4.91) \\ 
\hline
   	\multicolumn{5}{l}{$^*$ standardized so that 0 corresponds to an age of 77 years and so that one unit increment corresponds to 10 years}\\
   	\multicolumn{5}{l}{$^{**}$ standardized so that 0 corresponds to 10 days and so that one unit increment corresponds to 7 days}	 \\
\end{tabular}}
\end{table}

\begin{table}[ht]
\centering
\caption{Posterior medians (PM) and 95\% credible intervals (CI) for hazard ratio parameters for death following readmission ($\exp(\bfbeta_3)$) from semi-competing risks data analyses based on semi-Markov models. \label{tab:results:smBeta3}}
\bigskip
\scalebox{0.8}{
\begin{tabular}{ l r r r r}
  \hline
  				&	 Weibull-MVN 	& Weibull-DPM 	&  PEM-MVN 		& PEM-DPM \\
	 			& PM (95\%CI) 		& PM (95\%CI)  	& PM (95\%CI)  	& PM (95\%CI)  \\ 
  \hline
Sex 				&					&					&					& \\
~~~Male 			& 1.00 				& 1.00 				& 1.00				& 1.00\\
~~~Female 		& 0.81 (0.66, 1.00) 		& 0.81 (0.66, 0.98) 		& 0.84 (0.70, 1.00) 		& 0.84 (0.71, 1.01) \\ \\
Age$^*$			& 1.10 (1.02, 1.19) 		& 1.10 (1.02, 1.19) 		& 1.09 (1.02, 1.17) 		& 1.09 (1.02, 1.17) \\ \\ 
Race 			&					&					&					&\\
~~~White			& 1.00				& 1.00 				& 1.00 				& 1.00\\  
~~~Non-white		& 1.15 (0.77, 1.67) 		& 1.14 (0.79, 1.65) 		& 1.12 (0.79, 1.54) 		& 1.11 (0.78, 1.54) \\  \\
Source of entry to initial &&&&\\
hospitalization & &&&\\
~~~Emergency room& 1.00& 1.00 & 1.00 & 1.00\\
~~~Other facility	& 1.54 (1.25, 1.90) 		& 1.55 (1.25, 1.91) 		& 1.42 (1.17, 1.72) 		& 1.42 (1.16, 1.72) \\  \\
Charlson/Deyo score& &&&\\
~~~$\leq 1$ 		& 1.00& 1.00 & 1.00 & 1.00\\
~~~$> 1$ 			& 1.51 (1.11, 2.06) 		& 1.52 (1.11, 2.07) 		& 1.41 (1.06, 1.85) 		& 1.40 (1.05, 1.84) \\  \\
Procedure during hospitalization			&&&&\\
~~~No 		 	& 1.00& 1.00 & 1.00 & 1.00\\
~~~Yes 			& 0.21 (0.15, 0.29) 		& 0.21 (0.15, 0.29) 		& 0.28 (0.20, 0.39) 		& 0.28 (0.21, 0.39) \\  \\
Length of stay$^{**}$& 1.01 (0.89, 1.13) 		& 1.01 (0.89, 1.13) 		& 0.98 (0.88, 1.09) 		& 0.98 (0.87, 1.09) \\  \\
Care after discharge 	& &&&\\
~~~Home 			& 1.00& 1.00 & 1.00 & 1.00\\
~~~Home with care 		& 1.44 (1.13, 1.81) 		& 1.44 (1.13, 1.82) 		& 1.35 (1.08, 1.68) 		& 1.34 (1.08, 1.65) \\ 
~~~Hospice 			& 10.23 (4.66, 22.01) 	& 10.43 (4.83, 22.33) 	& 6.46 (3.33, 12.58) 		& 6.35 (3.30, 12.29) \\ 
~~~ICF/SNF	 		& 2.54 (1.87, 3.45) 		& 2.57 (1.87, 3.46) 		& 2.08 (1.52, 2.77) 		& 2.07 (1.56, 2.76) \\ 
~~~Other 				& 2.78 (1.64, 4.49) 		& 2.72 (1.61, 4.44) 		& 2.24 (1.40, 3.48) 		& 2.25 (1.41, 3.43) \\ 
\hline
   	\multicolumn{5}{l}{$^*$ standardized so that 0 corresponds to an age of 77 years and so that one unit increment corresponds to 10 years}\\
   	\multicolumn{5}{l}{$^{**}$ standardized so that 0 corresponds to 10 days and so that one unit increment corresponds to 7 days}	 \\
\end{tabular}}
\end{table}

\begin{table}[ht]
\centering
\caption{Posterior medians (PM) and 95\% credible intervals (CI) for hazard ratio parameters for death without readmission ($\exp(\bfbeta_2)$) from semi-competing risks data analyses based on Markov models. \label{tab:results:Beta2}}
\bigskip
\scalebox{0.8}{
\begin{tabular}{ l r r r r}
  \hline
  				&	 Weibull-MVN 	& Weibull-DPM 	&  PEM-MVN 		& PEM-DPM \\
	 			& PM (95\%CI) 		& PM (95\%CI)  	& PM (95\%CI)  	& PM (95\%CI)  \\ 
  \hline
Sex 				&					&					&					& \\
~~~Male 			& 1.00 				& 1.00 				& 1.00				& 1.00\\
~~~Female 		& 0.69 (0.60, 0.79) 		& 0.69 (0.60, 0.79) 		& 0.75 (0.67, 0.84) 		& 0.75 (0.67, 0.83) \\ \\
Age$^*$			& 1.08 (1.04, 1.13) 		& 1.08 (1.04, 1.13) 		& 1.07 (1.03, 1.11) 		& 1.07 (1.03, 1.11) \\ \\
Race 			&					&					&					&\\
~~~White			& 1.00				& 1.00 				& 1.00 				& 1.00\\  
~~~Non-white		& 0.92 (0.69, 1.21) 		& 0.92 (0.70, 1.22) 		& 0.94 (0.75, 1.18) 		& 0.94 (0.75, 1.16) \\ \\
Source of entry to initial &&&&\\
hospitalization & &&&\\
~~~Emergency room& 1.00& 1.00 & 1.00 & 1.00\\
~~~Other facility	& 1.61 (1.42, 1.84) 		& 1.62 (1.42, 1.86) 		& 1.49 (1.32, 1.67) 		& 1.49 (1.33, 1.69) \\ \\
Charlson/Deyo score& &&&\\
~~~$\leq 1$ 		& 1.00& 1.00 & 1.00 & 1.00\\
~~~$> 1$ 			& 1.40 (1.12, 1.73) 		& 1.39 (1.12, 1.72) 		& 1.26 (1.06, 1.49) 		& 1.27 (1.06, 1.51) \\ \\
Procedure during hospitalization			&&&&\\
~~~No 		 	& 1.00& 1.00 & 1.00 & 1.00\\
~~~Yes 			& 0.09 (0.07, 0.12) 		& 0.09 (0.07, 0.12) 		& 0.13 (0.10, 0.16) 		& 0.13 (0.11, 0.17) \\ \\
Length of stay$^{**}$& 1.15 (1.07, 1.23) 		& 1.15 (1.07, 1.23) 		& 1.10 (1.04, 1.17) 		& 1.10 (1.04, 1.17) \\ \\
Care after discharge 	& &&&\\
~~~Home 			& 1.00& 1.00 & 1.00 & 1.00\\
~~~Home with care 		& 2.44 (2.04, 2.91) 		& 2.42 (2.02, 2.93) 		& 2.20 (1.86, 2.62) 		& 2.20 (1.90, 2.61) \\ 
~~~Hospice 			& 23.52 (17.98, 30.13) 	& 23.55 (18.05, 30.53) 	& 13.72 (11.22, 17.49) 	& 13.78 (11.10, 17.1) \\ 
~~~ICF/SNF	 		& 5.30 (4.36, 6.43) 		& 5.29 (4.38, 6.47) 		& 4.23 (3.61, 5.16) 		& 4.25 (3.62, 5.02) \\ 
~~~Other 				& 4.88 (3.59, 6.72) 		& 4.87 (3.54, 6.73) 		& 3.78 (2.93, 4.99) 		& 3.82 (2.92, 4.97) \\ 
\hline
   	\multicolumn{5}{l}{$^*$ standardized so that 0 corresponds to an age of 77 years and so that one unit increment corresponds to 10 years}\\
   	\multicolumn{5}{l}{$^{**}$ standardized so that 0 corresponds to 10 days and so that one unit increment corresponds to 7 days}	 \\
\end{tabular}}
\end{table}

\begin{table}[ht]
\centering
\caption{Posterior medians (PM) and 95\% credible intervals (CI) for hazard ratio parameters for death following readmission ($\exp(\bfbeta_3)$) from semi-competing risks data analyses based on Markov models. \label{tab:results:Beta3}}
\bigskip
\scalebox{0.8}{
\begin{tabular}{ l r r r r}
  \hline
  				&	 Weibull-MVN 	& Weibull-DPM 	&  PEM-MVN 		& PEM-DPM \\
	 			& PM (95\%CI) 		& PM (95\%CI)  	& PM (95\%CI)  	& PM (95\%CI)  \\ 
  \hline
Sex 				&					&					&					& \\
~~~Male 			& 1.00 				& 1.00 				& 1.00				& 1.00\\
~~~Female 		& 0.81 (0.66, 0.98) 		& 0.81 (0.67, 0.99)	& 0.84 (0.71, 1.01) 	& 0.85 (0.71, 1.02) \\ \\
Age$^*$			& 1.11 (1.02, 1.19) 		& 1.10 (1.03, 1.20)	& 1.10 (1.03, 1.18) 	& 1.09 (1.02, 1.18) \\ \\
Race 			&					&					&					&\\
~~~White			& 1.00				& 1.00 				& 1.00 				& 1.00\\  
~~~Non-white		& 1.14 (0.78, 1.64) 		& 1.15 (0.79, 1.67) 	& 1.13 (0.81, 1.55) 	& 1.14 (0.79, 1.58) \\ \\
Source of entry to initial &&&&\\
hospitalization & &&&\\
~~~Emergency room& 1.00& 1.00 & 1.00 & 1.00\\
~~~Other facility	& 1.58 (1.28, 1.97) 		& 1.58 (1.28, 1.97) 	& 1.44 (1.18, 1.75) 	& 1.46 (1.21, 1.77) \\ \\
Charlson/Deyo score& &&&\\
~~~$\leq 1$ 		& 1.00& 1.00 & 1.00 & 1.00\\
~~~$> 1$ 			& 1.53 (1.12, 2.11) 		& 1.53 (1.12, 2.11) 	& 1.40 (1.02, 1.84) 	& 1.40 (1.06, 1.86) \\ \\
Procedure during hospitalization			&&&&\\
~~~No 		 	& 1.00& 1.00 & 1.00 & 1.00\\
~~~Yes 			& 0.20 (0.14, 0.28) 		& 0.20 (0.14, 0.28) 	& 0.27 (0.19, 0.37) 	& 0.27 (0.19, 0.36) \\ \\
Length of stay$^{**}$& 1.00 (0.89, 1.13) 		& 1.01 (0.89, 1.13) 	& 0.98 (0.88, 1.09) 	& 0.98 (0.88, 1.08) \\ \\
Care after discharge 	& &&&\\
~~~Home 			& 1.00& 1.00 & 1.00 & 1.00\\
~~~Home with care 		& 1.44 (1.15, 1.82) 		& 1.44 (1.13, 1.81) 	& 1.32 (1.06, 1.63) 	& 1.33 (1.07, 1.66) \\ 
~~~Hospice 			& 11.81 (5.18, 25.66) 	& 11.6 (5.08, 24.49) 	& 6.95 (3.49, 12.75) 	& 6.79 (3.24, 13.28) \\ 
~~~ICF/SNF	 		& 2.70 (1.96, 3.68) 		& 2.69 (1.99, 3.61) 	& 2.12 (1.59, 2.81) 	& 2.17 (1.63, 2.87) \\ 
~~~Other 				& 2.92 (1.74, 4.77) 		& 2.89 (1.74, 4.68) 	& 2.32 (1.46, 3.65) 	& 2.36 (1.47, 3.67) \\ 
\hline
   	\multicolumn{5}{l}{$^*$ standardized so that 0 corresponds to an age of 77 years and so that one unit increment corresponds to 10 years}\\
   	\multicolumn{5}{l}{$^{**}$ standardized so that 0 corresponds to 10 days and so that one unit increment corresponds to 7 days}	 \\
\end{tabular}}
\end{table}

\begin{table}[ht]
\centering
\caption{Posterior medians (PM) and 95\% credible intervals (CI) for hazard ratio parameters for readmission ($\exp(\bfbeta_1)$) from semi-competing risks data analyses based on Markov models. \label{tab:results:Beta1}}
\bigskip
\scalebox{0.8}{
\begin{tabular}{ l r r r r}
  \hline
  				&	 Weibull-MVN 	& Weibull-DPM 	&  PEM-MVN 		& PEM-DPM \\
	 			& PM (95\%CI) 		& PM (95\%CI)  	& PM (95\%CI)  	& PM (95\%CI)  \\ 
  \hline
Sex 				&					&					&					& \\
~~~Male 			& 1.00 				& 1.00 				& 1.00				& 1.00\\
~~~Female 		& 0.79 (0.70, 0.91) 		& 0.80 (0.70, 0.91) 		& 0.85 (0.76, 0.95) 		& 0.85 (0.76, 0.95) \\ \\
Age$^*$			& 0.90 (0.86, 0.95) & 0.90 (0.86, 0.95) & 0.91 (0.87, 0.94) & 0.91 (0.87, 0.95) \\ \\
Race 			&					&					&					&\\
~~~White			& 1.00				& 1.00 				& 1.00 				& 1.00\\  
~~~Non-white		& 1.11 (0.86, 1.45) & 1.11 (0.86, 1.44) & 1.12 (0.89, 1.40) & 1.11 (0.89, 1.38) \\ \\
Source of entry to initial &&&&\\
hospitalization & &&&\\
~~~Emergency room& 1.00& 1.00 & 1.00 & 1.00\\
~~~Other facility	& 1.18 (1.03, 1.35) & 1.19 (1.03, 1.36) & 1.12 (0.99, 1.26) & 1.12 (0.99, 1.27) \\ \\
Charlson/Deyo score& &&&\\
~~~$\leq 1$ 		& 1.00& 1.00 & 1.00 & 1.00\\
~~~$> 1$ 			& 1.49 (1.19, 1.84) & 1.50 (1.19, 1.85) & 1.40 (1.15, 1.68) & 1.39 (1.15, 1.68) \\ \\
Procedure during hospitalization			&&&&\\
~~~No 		 	& 1.00& 1.00 & 1.00 & 1.00\\
~~~Yes 			& 0.45 (0.37, 0.53) & 0.45 (0.37, 0.53) & 0.57 (0.49, 0.66) & 0.57 (0.48, 0.66) \\ \\
Length of stay$^{**}$& 1.15 (1.07, 1.23) & 1.15 (1.07, 1.23) & 1.12 (1.05, 1.19) & 1.12 (1.05, 1.19) \\ \\
Care after discharge 	& &&&\\
~~~Home 			& 1.00& 1.00 & 1.00 & 1.00\\
~~~Home with care 		& 0.95 (0.82, 1.11) & 0.95 (0.82, 1.11) & 0.89 (0.78, 1.02) & 0.89 (0.78, 1.01) \\ 
~~~Hospice 			& 0.39 (0.23, 0.62) & 0.38 (0.22, 0.64) & 0.27 (0.16, 0.42) & 0.27 (0.16, 0.43) \\ 
~~~ICF/SNF	 		& 0.88 (0.73, 1.06) & 0.88 (0.73, 1.07) & 0.76 (0.63, 0.90) & 0.76 (0.64, 0.90) \\ 
~~~Other 				& 1.05 (0.77, 1.43) & 1.04 (0.77, 1.45) & 0.89 (0.68, 1.18) & 0.89 (0.67, 1.18) \\ 
\hline
   	\multicolumn{5}{l}{$^*$ standardized so that 0 corresponds to an age of 77 years and so that one unit increment corresponds to 10 years}\\
   	\multicolumn{5}{l}{$^{**}$ standardized so that 0 corresponds to 10 days and so that one unit increment corresponds to 7 days}	 \\
\end{tabular}}
\end{table}

\newpage

\clearpage

\begin{figure}[h]
\centering
\includegraphics[width = 6in]{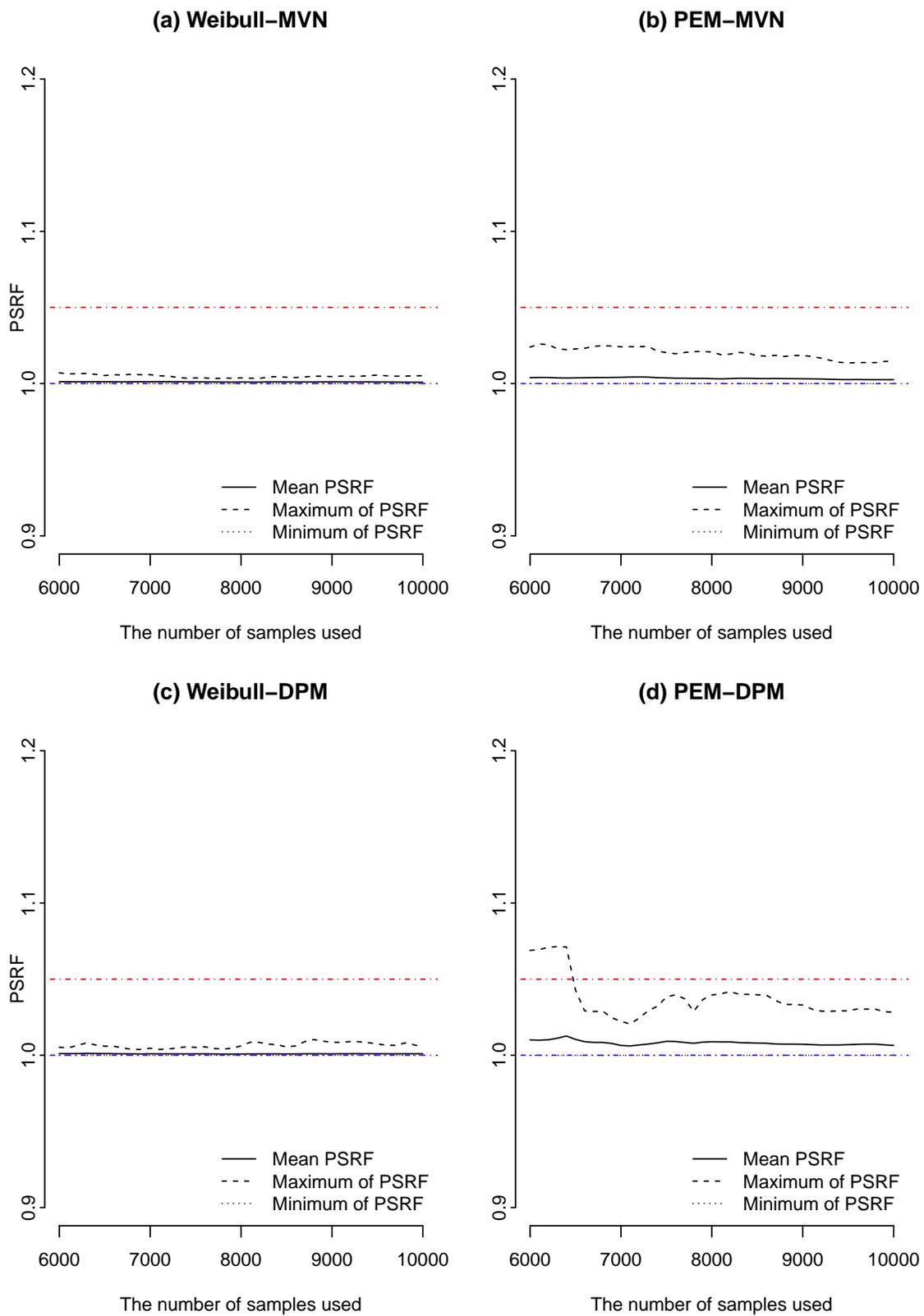}
\caption{The mean, maximum, and minimum value of the potential scale reduction factor (PSRF) of all model parameters from the analysis of Medicare data. \label{fig:conv}}
\end{figure}

For our proposed models, we assess the convergence of our MCMC scheme by evaluating the potential scale reduction factor (PSRF) of \cite{gelman2003bayesian}. The potential problem with PSRF is that it has not converged but happens to be close to 1 by chance even though the PSRF is actually fluctuating. Therefore, for each parameter, the PSRF was calculated at several points in time with the first half discarded as burn-in. Then, we summarize the results using mean, maximum, and minimum value of PSRF for all model parameters at different iterations. The results are shown in Figure \ref{fig:conv}. As the number of MCMC iterations increases, the mean PSRF converges toward 1 and the maximum of PSRF is less than 1.05 indicating that all model parameters have converged well.

\clearpage
\bibliographystyle{apalike}
\bibliography{ref.bcscr}